\documentclass[11pt]{article}
\pdfoutput=1
\usepackage{jheppub} 
\usepackage{enumerate}
\usepackage{amsmath,amssymb,epsfig,cancel}
\usepackage{color}
\usepackage{subfig}
\usepackage{graphicx}
\usepackage{caption}
\usepackage{braket}
\usepackage{hyperref}
\usepackage{float}
\usepackage[english]{babel}

\title{Integrability and Holographic Aspects of Six-Dimensional ${\cal N}=(1,0)$ Superconformal Field Theories} 
\author{Kostas Filippas,}
\author{Carlos N\'u\~nez,}
\author{and Jeroen van Gorsel}

\affiliation{Department of Physics, Swansea University, Swansea SA2 8PP, United Kingdom}

\emailAdd{kphilippas@hotmail.com, c.nunez@swansea.ac.uk, jeroen.van.gorsel@gmail.com} 

\abstract{ In the framework of six-dimensional conformal field theories with ${\cal N}=(1,0)$ supersymmetry we develop the map between the holographic description, the field
theoretical description and the associated Hanany-Witten set-ups. General expressions that calculate various observables are presented. The study of string solitons singles out a special background of Massive IIA on which we show (by explicitly finding a Lax pair)
that the Neveu-Schwarz part of the string sigma model is classically integrable. We study the particular dual conformal field theory and compute some of its observables.
\\[10pt]
 } 

\subheader{}
\begin{document}
\def\Tr{{\textrm{Tr}}}

%%%%%%%%%%%%%%%%%%%%%%%%%%%%%%%%%%%%%%%%%%%%%%%%%%%%%%%%%%%%%%%%
%\hfill { CERN-TH-}
%%%%%%%%%%%%%%%%%%%%%%%%%%%%%%%%%%%%%%%%%%%%%%%%%%%%%%%%%%%%%%%%

\maketitle 

\newpage
%%%%%%%%%%%%%%%%%%%%%%%%%%%%%%%%%%%%%%%%%%%%%%%%%%%
\section{Introduction and General Idea}
Quantum field theories are usually thought of in the context of renormalisation group flows from short distances in the UV to long distances in the IR. 
The endpoints of these flows called fixed points are scale invariant. In $d$-dimensional relativistic field theories, it is common to assume that the fixed-point theory is a conformal field theory (CFT), whose spacetime symmetry  $SO(1,d-1)$ is enhanced to the conformal algebra $SO(2,d)$. Aside from free CFTs, there is compelling evidence for a vast landscape of interacting CFTs in diverse dimensions, with many of these theories being non-Lagrangian (CFTs that do not have a known representation in terms of fields and a Lagrangian). In such cases, one typical route to learn about the dynamics of these CFTs is to use algebraic methods. These are `kinematic' in nature and hence limited for the goal of calculating detailed dynamical information. In this sense, the best way to learn about these field theories is to use  Maldacena's Conjecture \cite{Maldacena:1997re} and find the holographic dual background to calculate field theory correlators with it.
\\
\\
Geometrical Engineering, that is the interplay between the geometry of the string theory's extra dimensions, and the resulting low energy field theories on branes probing this geometry, is by now a well-established research tool. In the case of geometries with singularities, such methods have led to the construction and study of conformal field theories in diverse dimensions. Of particular significance for this paper, are the conformal field theories in six dimensions, which resist a  Lagrangian description. The key ingredients of these theories are tensionless strings coupled to dynamical tensor modes.
Examples of such theories include the ADE ${\cal N}= (2,0)$ theories \cite{Witten:1995ex}, realised on a stack of M5-branes. Less is known about the ${\cal N}= (1,0)$ theories.
These CFTs do not have a weakly coupled UV Lagrangian. However, one can always go to the tensor branch of these theories, which corresponds to giving vacuum expectation values (VEVs) to scalars in tensor multiplets. In such cases, one can find an effective Lagrangian description for ${\cal N}=(1, 0)$ theories in terms of a weakly coupled quiver gauge theory, where the scalars in the tensor multiplets controlling the coupling constants of the corresponding gauge groups,  are promoted to a set of dynamical fields. Moving to the origin of the tensor branch typically leads to a strongly coupled 6d SCFT with ${\cal N}=(1, 0)$ supersymmetry.  Some of these systems have a realisation in string theory \cite{Seiberg:1996vs},\cite{Blum:1997mm}. The description of these CFTs is well advanced, see for example \cite{Seiberg:1996qx}-\cite{Cremonesi:2015bld}, for a sample of papers that deal with the system from the field-theoretical, brane picture, or holographic perspectives.
\\
\\
The general idea of this paper is to develop important new formal tools in the mapping between six-dimensional CFTs with ${\cal N}=(1,0)$ SUSY and their holographic description. In particular, we will discuss aspects of the CFTs that are not suitable to be understood with either Geometrical Engineering or algebraic methods, hence rely heavily on the holographic description. 

In more detail, the contents of this work are the following: in Section \ref{holodescr}, we summarise the known field theoretical and holographic description, we describe the mapping between these descriptions and provide new holographic expressions for some of the characteristic quantities that follow from a Hanany-Witten description \cite{Hanany:1996ie} of the CFTs. For example, we give new expressions that calculate the number of branes, linking numbers and entanglement entropy of a given quiver in terms of the holographic description. For the benefit of the readers new to the topic, we present detailed examples of the mapping between the quiver CFT and the holographically dual background. In Section \ref{sectionintegrability}, we fully rely on the holographic description and discuss the possibility of finding a particularly special background in Massive IIA on which the string theory is classically integrable. The treatment uses semiclassical string solitons and studies its dynamical evolution. A particular background is then singled out as special, on which the non-integrable characteristics of the soliton are absent. In Section \ref{integrablespecial}, we present the Lax pair from which the equations of motion of a bosonic string on such particular background are derived. We connect this Lax pair with the so-called $\lambda$-deformation of a Wess-Zumino-Witten (WZW) model. In Section \ref{CFTsmeared}, we study some of the field theoretical observables associated with this special background. These observables can only be accessed through the holographic description and provide a {\it definition} of the particular integrable ${\cal N}=(1,0)$ six-dimensional CFT. In fact, an alternative to giving the precise colour and flavour groups at the origin of the tensor branch is to display the observables of the conformal field theory. We summarise our findings and conclude in Section \ref{concl}, where we also lay-out some ideas to work in the future.
Our work is complemented by very extensive and dense appendixes for the benefit of the readers willing to work on these topics.

\section{Six dimensional SCFTs  and their holographic description}\label{holodescr}
Let us start with a summary of six-dimensional ${\cal N}=(1,0)$ conformal field theories and their holographic description.

It is useful to remind the reader of the main issue afflicting higher dimensional ($d>4$) field theories. Consider a simple interacting field theory in six dimensions with action,
\begin{equation}
S=\int d^6x \left[-\frac{1}{2}(\partial_\mu \phi)^2 - V(\phi)\right].\nonumber
\end{equation}
Here $\phi$ represents a real scalar field with classical dimension $[\phi]=m^2$. The potential can be a mass term $V=\frac{m^2}{2}\phi^2$ or more interestingly a classically marginal interaction term, like $V=g\phi^3$, but this would lead to a system without ground state (for $\phi<0$). On the other hand, a potential like $V=\lambda \phi^4$ has a well-defined vacuum, but the interaction is irrelevant, hence the theory is not well defined without a UV completion.
The  Wilsonian logic, according to which we start from a  conformal (not necessarily weakly coupled) field theory and deform it by inserting relevant operators into the Lagrangian, flowing to interesting field theories at low energies, does not seem to apply here. 

Nevertheless, different string theoretic constructions have suggested that supersymmetric field theories of scalars coupled to gauge fields have an interacting UV fixed point. In fact, for a Lagrangian like
\begin{equation}
\mathcal{L}\sim -\frac{1}{2}(\partial_\mu \phi)^2 -c~\phi F_{\mu\nu}^2 +\textrm{fermions},\label{kakit}
\end{equation}
 when $\langle \phi\rangle \to0$ we are dealing with the strong coupling limit of a gauge field theory (since the scalar $\phi$ takes the role of the inverse coupling of the gauge theory). The presence of fermions in the supersymmetric theory implies the possible existence of gauge anomalies that need to be cancelled.
This cancellation is possible if the scalar  $\phi$ belongs to a tensor multiplet \cite{Seiberg:1996qx}, \cite{Danielsson:1997kt} (see below for a description of the relevant multiplets) and a certain tuning between the amount of adjoint and fundamental matter must be imposed.

This picture was realised in brane constructions. The relevant Hanany-Witten set-ups  \cite{Hanany:1996ie} were presented in \cite{Hanany:1997gh}. The associated field theories preserve eight Poincare supercharges, have $SO(1,5)$ Lorentz and  $SU(2)$ R-symmetries. In more detail, the field theories with ${\cal N}=(1,0)$ SUSY are constructed in terms of the following multiplets:
\begin{itemize}
\item{Tensor multiplets with field content   $(B_{\mu\nu},\lambda_1,\lambda_2,\phi)$. A two form with self-dual curvature $H_3=dB_2$, two fermions and a real scalar.}
\item{Vector multiplets with field content  $(A_\mu,\hat{\lambda}_1, \hat{\lambda}_2)$, a six-dimensional vector and two fermions.}
\item{Hypermultiplets with field content  $(\varphi_1,\varphi_2,\psi_1,\psi_2)$, two scalars and two fermions.}
\item{Linear multiplets with field content  $(\vec{\pi}, c,\tilde{\xi})$ an $SU(2)$ triplet and a singlet, together with a fermion.}
 \end{itemize}
The field theories have a `tensor branch' when the scalar $\phi$ gets a non-zero VEV. In this case, the $SU(2)_R$ symmetry is preserved. On the other hand, when the scalars inside the hyper or the linear multiplet get VEVs, we explore the Higgs branch breaking the R-symmetry. In what follows we will be concerned with the tensor branch only.

To reproduce the Lorentz and R-symmetry mentioned above, the authors of \cite{Hanany:1997gh} distributed D6, NS5, and D8 branes according to Table \ref{table:BraneSetup}.

\begin{table}[h!]
\centering
\begin{tabular}{c||c c c c c c |c|c c c}
  & $t$ &  $x_1$ & $x_2$ & $x_3$ & $x_4$ & $x_5$ & $x_6$ & $x_7$ & $x_8$ & $x_9$ \\ [0.5ex] 
 \hline\hline %-----------------------------------------------------------------------------------------------
 NS5 & $\bullet$ & $\bullet$ & $\bullet$ & $\bullet$ & $\bullet$ & $\bullet$ & $\cdot$ & $\cdot$ & $\cdot$ & $\cdot$ \\ 
 \hline\hline %-----------------------------------------------------------------------------------------------
  D6 & $\bullet$ & $\bullet$ & $\bullet$ & $\bullet$ & $\bullet$ & $\bullet$ & $\bullet$ & $\cdot$ & $\cdot$ & $\cdot$ \\
  \hline
  D8 & $\bullet$ & $\bullet$ & $\bullet$ & $\bullet$ & $\bullet$ & $\bullet$ & $\cdot$ & $\bullet$ & $\bullet$ & $\bullet$ 
\end{tabular}
\caption{The generic brane set-ups. All the branes are extended on the Minkowski $R^{1,5}$ directions. The D6-branes also extend over $x_6$ where they have finite size extension between NS5-branes. The D8-branes also extend along the $x_7,x_8$ and $x_9$ directions, preserving the $SO(3)_R$ symmetry.}
\label{table:BraneSetup}
\end{table}
There are some key differences with Hanany-Witten set-ups in lower dimensions,
\begin{itemize}
\item{The dimension of the field theory on the NS5-branes is the same as that on the bounded D6-branes. The non-decoupling of the five-branes dynamics adds the dynamical tensor multiplets to the field theories. These are absent in lower dimensional set-ups.}
\item{The bending of the NS5-branes due to other $p$-branes ending on them leads to a Laplace equation in $6-p$ dimensions. In this case, where $p=6$, there is no-bending. The field content is always such that anomalies are cancelled, namely 
\begin{equation}
N_{D6,R} + N_{D6,L} + N_{D8}= 2 N_{D6,c},\label{anomalycancellation}
\end{equation}
being $N_{D6,R/L}$ the number of sixbranes to the right/left of a given stack with $N_{D6,c}$ branes. }
\item{We can consider D2-branes on $(t,x_1, x_6)$ that end on the NS5-branes. These branes represent one dimensional magnetically charged defects identified with the instantonic strings charged under the self-dual $H_3$.}
\item{
When the system is in the tensor branch (the difference between the scalars in different tensor multiplets  $\langle\phi_i-\phi_{i-1}\rangle $ is non-zero) the instantonic strings are massive and the field theory can be described by an anomaly-free quiver. When   $\langle\phi_i-\phi_{i-1}\rangle\to  0$, the theory is proposed  \cite{Seiberg:1996qx} to flow to a strongly coupled six dimensional CFT with $(1,0)$ SUSY. These are the theories that we study in this paper. }
\end{itemize}

\subsection{Holographic description}\label{secthol}
Let us now discuss the holographic description of the CFTs that appear when we move to the origin of the tensor branch. This description 
was developed in a set of papers, most notably \cite{Gaiotto:2014lca}-\cite{Cremonesi:2015bld}. We adopt the notation of \cite{Cremonesi:2015bld}. 

The six dimensional SCFTs have $SO(2,6)\times SU(2)_R$ bosonic symmetries, see for example \cite{Nahm:1977tg}. They are realised 
as the isometries of a Massive Type IIA background of the form,
\begin{eqnarray}\label{eq:TomasielloGeometryGeneral}
& & ds^2=f_1(z) ds^2_{AdS_7}+f_2(z) dz^2+f_3(z)\;d \Omega_2^2(\chi, \xi) ,\nonumber\\
& & B_2=f_4(z)  \mathrm{Vol}_{\Omega_2},\;\;\; F_2= f_5(z)  \mathrm{Vol}_{\Omega_2}, \;\;\;e^{\phi}=f_6(z),\;\;\; F_0= F_0(z).\label{backgroundads7xm3}
\end{eqnarray}
We have defined $d \Omega_2^2(\chi, \xi)=d \chi^2+ \sin^2 \chi\; d \xi^2$ and $\mathrm{Vol}_{\Omega_2}=\sin\chi\;d\chi\wedge d\xi$. 

 If we impose that  ${\cal N}=(1,0)$ SUSY is preserved by the background, we need the functions $f_i(z)$  to satisfy some first order and nonlinear differential equations.
 These  BPS equations are solved if the functions $f_i(z)$ in eq.(\ref{eq:TomasielloGeometryGeneral}) are all defined in terms of a function $\alpha(z)$ and its derivatives---see \cite{Apruzzi:2014qva}-\cite{Cremonesi:2015bld} for the details,
\begin{eqnarray}\label{eq:TomasielloGeometriesFunctions}
& & f_1(z)= 8 \sqrt{2} \pi  \sqrt{-\frac{\alpha}{{\alpha''}}},\;\;\; f_2(z)= \sqrt{2} \pi \sqrt{-\frac{{\alpha''}}{{\alpha}}},\;\;\;
f_3(z)= \sqrt{2} \pi \sqrt{-\frac{{\alpha''}}{{\alpha}}}\left( \frac{\alpha^2}{{\alpha'}^2-2 \alpha {\alpha''}}\right),\nonumber\\
& &  f_4(z)=\pi \left(-z +\frac{\alpha {\alpha'}}{{{\alpha'}}^2-2 \alpha {\alpha''}}\right),\qquad f_5(z)=\left( \frac{{\alpha''}}{162 \pi^2}+ \frac{\pi F_0 \alpha {\alpha'}}{ {\alpha'}^2-2 \alpha {\alpha''}}  \right),\\
& &f_6(z)=2^{\frac{5}{4}} \pi^{\frac{5}{2}}3^4 \frac{(-\alpha/ {\alpha''})^{\frac{3}{4}}}{\sqrt{{\alpha'}^2-2 \alpha {\alpha''}}}.\nonumber
\end{eqnarray}
Where $\alpha(z)$ has to satisfy the differential equation 
\begin{equation}\label{eq:AlphaThird}
{\alpha'''}=-162 \pi^3 F_0.
\end{equation}
The function $\alpha(z)$ must be piece-wise continuous, this implies that $F_0$ can be piece-wise constant and discontinuous. The internal space $\mathcal{M}_3=(z,\Omega_2)$ is a two-sphere `fibered' over the $z$-interval. The warp factor  $f_3(z)$ must vanish at the beginning  and  at the end of the $z$-interval  ($z=0$ and $z=z_f$ by convention), in such a way that the two-sphere  shrinks smoothly at those  points.
\\
\\
For a piece-wise constant and possibly discontinuous $F_0(z)$, the general solution to eq.(\ref{eq:AlphaThird}) in each interval of constant $F_0$ is, 
\begin{equation}
\alpha(z)=a_0+ a_1 z +\frac{a_2}{2}z^2 -\frac{162\pi^3 F_0}{6}z^3.
\nonumber
\end{equation}
As we observed above, the function $\alpha(z)$ is in general piece-wise continuous and generically a polynomial solution like the one above should be proposed for each interval $[z_i, z_{i+1}]$. Imposing that the two-sphere
shrinks smoothly at $z=0$ and $z=z_f$ implies that $\alpha(0)=\alpha(z_f)=0$. 
We shall discuss a generic solution below. Before that, let us find general expressions for the brane-charges associated with the backgrounds in eq.(\ref{eq:TomasielloGeometryGeneral}).  

\subsubsection{Page charges}
We define the conserved Page charges,
\begin{eqnarray}
& & Q_{Dp}=\frac{1}{(2\pi)^{7-p} g_s( \alpha')^{\frac{(7-p)}{2}}} \int F_{8-p}- B_2\wedge F_{6-p},\;\;\;\; Q_{NS5}=\frac{1}{4\pi^2 g_s^2 \alpha'}\int H_3.\label{cargas}
\end{eqnarray}
In what follows we set $g_s=\alpha'=1$. Calculating explicitly for the NS5-brane charge. Using that $\alpha(0)=\alpha(z_f)=0$ we find,
\begin{equation}
Q_{NS5}=\frac{1}{4\pi^2} \int_{z,\Omega_2} \partial_z f_4= \frac{1}{\pi}\int_{z=0}^{z=z_f} \partial_z f_4= \frac{f_4(z_f)-f_4(0) }{\pi} = -z_f.\label{chargeNS5good}
\end{equation}
Up to an orientation-related sign, the size of the $z$-interval equals the number of fivebranes. Hence we need to choose $z_f$ to be a positive integer. We shall take $Q_{NS5}=z_f=N_5$ in what follows. 

Calculating  the charge of D6-branes, we find
\begin{eqnarray}\label{chargeD6}
& & Q_{D6}=\frac{1}{2\pi} \int_{ (\chi,\xi)} F_2-F_0 B_2= \left[\frac{\alpha''+162\pi^3 F_0 z}{81\pi^2}
\right] =\frac{\alpha''-z \alpha'''}{81\pi^2}.
\end{eqnarray}
The charge in eq.(\ref{chargeD6}) computes the charge of D6-branes but also includes the charge of D6-brane induced on the D8-branes. To avoid this `overcounting', note that we can perform a large gauge transformation in any interval $[k,k+1]$ such that,
 \begin{equation}
\hat{ B}_2\to B_2+ k\pi\;d\Omega_2.
 \end{equation}
 We then find that in the interval $[k,k+1]$ the Page charge reads,
 \begin{equation}
 Q_{D6}=\frac{1}{2\pi} \int_{S^2} F_2- F_0 \hat{B}_2= \frac{1}{2\pi}\times \frac{\alpha''-\alpha'''(z-k)}{162\pi^2} \times 4\pi.
 \end{equation}
Using that on the $[k,k+1]$ interval the function
  $\alpha''(z)=- 81\pi^2 \left[N_k +(N_{k+1}-N_k)(z-k)\right]$, we find that
   \begin{equation}
N_{D6}=\frac{1}{2\pi}\times \frac{\alpha''-\alpha'''(z-k)}{162\pi^2} \times 4\pi= -N_k.
 \end{equation}
The sign can be attributed to a choice of orientation for the two-sphere. The expression above indicates that in the $[k,k+1]$ interval,  there are $N_k$ D6-branes. Notice that the expression in eq.(\ref{chargeD6}) also counts the 
charge of D6's induced on the D8's. We are subtracting these, by performing the large gauge transformation above.\\

We thus find that the number of only the D6-branes in the associated Hanany-Witten set-up is given by,
\begin{equation}
N_{D6}=-\frac{1}{81\pi^2}\int_0^{z_f} \alpha''(z) dz.\label{chargeD6good}
\end{equation}
This can be verified by explicitly performing this integral for a generic function $\alpha''(z)$, observing that it counts the sum of the ranks of the gauge groups (see eq.(\ref{eq:genericquiver4Nodes}) for an example of a function $\alpha(z)$ for a generic quiver with four nodes and four flavour groups). On each interval $[k, k+1]$ this gives,
 \begin{equation}
 -\frac{1}{81\pi^2}\int_{k}^{k+1} \alpha'' dz= -\int_{k}^{k+1} \left[ N_k +(N_{k+1}-N_k)(z-k)\right] dz
=\frac{N_k+N_{k+1}}{2}.
\end{equation}
Summing over all the intervals (using that $N_0=N_{P+1}=0$), gives the total quantity,
\begin{equation}
N_{D6}=\sum_{k=0}^{P} \frac{N_k+N_{k+1}}{2}= N_1+N_2+....+N_P.
\end{equation}

We also present an expression that calculates the number of D8-branes in any given Hanany-Witten set-up. Our proposed  new expression reads,
\begin{equation}
N_{D8}=\frac{1}{81\pi^2}\left[\alpha'''(0)-\alpha'''(z_f)\right].\label{chargeD8good}
\end{equation}
In other words, the jumps in $\alpha'''(z)$ across any interval counts D8-branes according to eq.(\ref{eq:AlphaThird}). Adding these jumps leads to eq.(\ref{chargeD8good}).
 
These expressions are analogous to those derived in \cite{Nunez:2019gbg}, for the case of Hanany-Witten set-ups associated with four dimensional ${\cal N}=2$ SCFTs. In Section \ref{connect} we  test the new expressions in eqs.(\ref{chargeD6good}),(\ref{chargeD8good}) on some examples.

\subsubsection{Linking numbers}
One interesting quantity characterising the Hanany-Witten set-ups are the linking numbers. For the case at hand, with D6, D8 and NS5-branes these topological invariants (unchanged under Hanany-Witten moves) are defined for the j-th D8-brane ($L_j$ is the linking number) and the i-th Neveu-Schwarz fivebrane ($K_i$ being the corresponding linking number) by counting the number of the other branes to the left and right. More precisely, we have
\begin{eqnarray}
& & L_j= N_{D6}^{right} - N_{D6}^{left} + N_{NS}^{left},\nonumber\\
& & K_i= N_{D6}^{right} - N_{D6}^{left} - N_{D8}^{right}.\label{linkings}
\end{eqnarray}
They must satisfy a consistency relation
\begin{equation}
\sum_{j=1}^{N_8} L_j+\sum_{i=1}^{N5}K_i =0.\label{consistency}
\end{equation}
In the Hanany-Witten set-ups that are relevant for the CFTs we study in this paper,  all the linking numbers for the different Neveu-Schwarz fivebranes are equal. 
We have found that they can be holographically calculated by very simple expressions. Our proposal is that for these CFTs we calculate the linking numbers as,
\begin{eqnarray}
& & K_1=K_2=....=K_{N5}=\frac{1}{81\pi^2}\alpha'''(z_f)\to \sum_{i=1}^{N_5}K_i
=\frac{1}{81\pi^2} \alpha'''(z_f) z_f.\label{linkingNS}\\
& & L_i= z_i\to \sum_{j=1}^{N_8} L_j=-\frac{1}{81\pi^2}\alpha'''(z_f) z_f.\label{linkingD8}
\end{eqnarray}
These expressions satisfy eq.(\ref{consistency}) and are analogous to those presented in the case of four dimensional CFTs with eight supercharges \cite{Nunez:2019gbg}. In Section \ref{connect} we  test these expressions in a couple of examples. The reader is invited to apply the expressions of eqs.(\ref{chargeD6good}),(\ref{chargeD8good}),(\ref{linkingNS}),(\ref{linkingD8}) to the examples of the paper \cite{Nunez:2018ags}.

\subsubsection{Entanglement Entropy}
We briefly discuss this interesting observable characterising CFTs. Our treatment is an extension of that presented in  \cite{Macpherson:2014eza} for backgrounds of the form in eq.(\ref{eq:TomasielloGeometryGeneral}). The entanglement entropy $S_{EE}$ for a rectangular region is calculated by solving a minimisation problem for an eight manifold hanging from infinity in the AdS-radial direction. The two regions of the space are separated by a distance $L_{EE}$ (calculated in terms of the background functions as shown below). A regularisation is needed analogously to what happens when computing Wilson loops, see  \cite{Macpherson:2014eza} for the details. In particular for the backgrounds of the form in eq.(\ref{eq:TomasielloGeometryGeneral}). The eight surface is parametrised by the coordinates
\begin{equation}
\Sigma_8=[x_1,x_2,x_3,x_4,x_5,z,\chi,\xi],\;\;\;\;\; R=R(x_1).\nonumber
\end{equation}
Using Poincar\'e coordinates for the AdS$_7$ space, the induced metric of the eight-surface is
\begin{equation}
ds_{8,ind}^2= f_1 \left[ R^2 d\vec{x}_4^2 + dx_1^2\left(R^2+\frac{R'^2}{R^2} \right)     \right] + f_2 dz^2 + f_3(d\chi^2+ \sin^2\chi d\xi^2).\nonumber
\end{equation}
The entanglement entropy is
\begin{eqnarray}
 & & S_{EE}= \frac{1}{4G_N}\int d^8\sigma ~e^{-2\phi}\sqrt{\det g_{8,ind}}, \;\;\;\;\; G_N=8\pi^6 g_s^2\alpha'^4=8\pi^6\label{eexx}\\
 & & S_{EE}= \frac{128 V_4}{6561 G_N}\left(\!\int_0^{z_f} \!\alpha''(z)\alpha(z) dz\!\right) \int dx_1 R^5\sqrt{1+\frac{R'^2}{R^4}}, \;\;\; V_4=\int\!\! dx_2dx_3dx_4dx_5.\nonumber
\end{eqnarray}
Following the formalism of the works \cite{Macpherson:2014eza}, we find the regularised version of the entanglement entropy, $S_{EE}^{reg}$ and the separation between the regions $L_{EE}$ to be,
\begin{eqnarray}
 & & S_{EE}^{reg}= \frac{V_4}{2G_N} \left[\int_1^\infty  dy\left(\frac{y^8}{\sqrt{y^{10}-1}}-y^3   \right)   \right] {\cal N} R_0^4= \mu_1 {\cal N}R_0^4,\nonumber\\
 & & L_{EE}= \left[2\int_{1}^{\infty} \frac{dy}{\sqrt{y^4(y^{10}-1)}}  \right] \frac{1}{R_0}=\frac{\mu_2}{R_0},\nonumber\\
 & & S_{EE}^{reg}={\cal N}\left( \frac{\mu_1\mu_2^4}{L^4}\right),\;\;\;\; {\cal N}=-\frac{512}{6561}\int_0^{z_f} \alpha(z)\alpha''(z) dz.\label{eecc}
\end{eqnarray}
The factors $\mu_1\mu_2^4$ are common to all six-dimensional conformal field theories. The power $L^{-4}$ is the only possible one given conformality and the dimension of the CFT. All the information about the particular CFT in consideration is in the factor ${\cal N}\sim\int \alpha\alpha''$. Notice that this factor also appears when computing the central charge  of the CFT, see \cite{Nunez:2018ags}. This is not a surprise as both quantities measure the number of degrees of freedom.
 
\subsection{Connecting  the holographic background with the CFT}\label{connect}
Let us discuss the connection between a quiver field theory, and the geometry in eq.(\ref{eq:TomasielloGeometryGeneral}). 

The problem can be organised as follows: first, we consider a non-anomalous quiver with bifundamental matter, gauge and flavour groups satisfying the relation in eq.(\ref{anomalycancellation}). Then, we define the function $R(z)$,  a piecewise continuous linear function such that at $z=j$ (with $j$ being a positive integer number) the value $R(j)=N_j$ is the rank of the $j$-th gauge group. It was shown in \cite{Cremonesi:2015bld} that this rank-function must be convex to satisfy the anomaly cancellation condition in eq.(\ref{anomalycancellation}). 

The link with the holographic description is given by the identification,
\begin{equation}
R(z)=-\frac{1}{81\pi^2}\alpha''(z).\label{identification}
\end{equation}
Finally, after this identification, we need to determine the function $\alpha(z)$ by imposing boundary conditions and continuity of $\alpha$ and $\alpha'$.

\begin{figure}[h!]
    \centering
    {{\includegraphics[width=9cm]{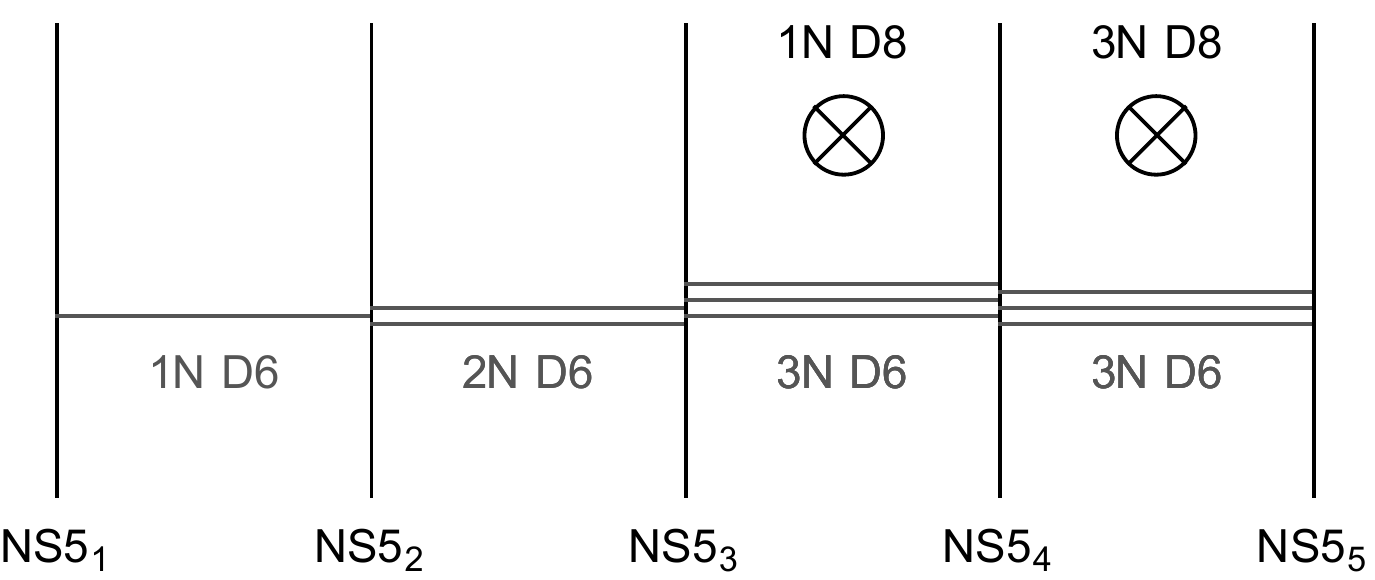} }}
\caption{The  Hanany-Witten set-up for the field theory. The vertical lines denote individual Neveu-Schwarz branes extended on the $(x_4,x_5)$ space. The horizontal ones D6-branes, that extend on $x_6$, in between fivebranes. The crossed-circles represent D8-branes, that extend on the $(x_7,x_8,x_9)$ directions. All the branes share the Minkowski directions. This realises the isometries $SO(1,5) \times SO(3) $.}
\label{figure1x}
\end{figure}

\begin{figure}[h!]
    \centering
    {{\includegraphics[width=9cm]{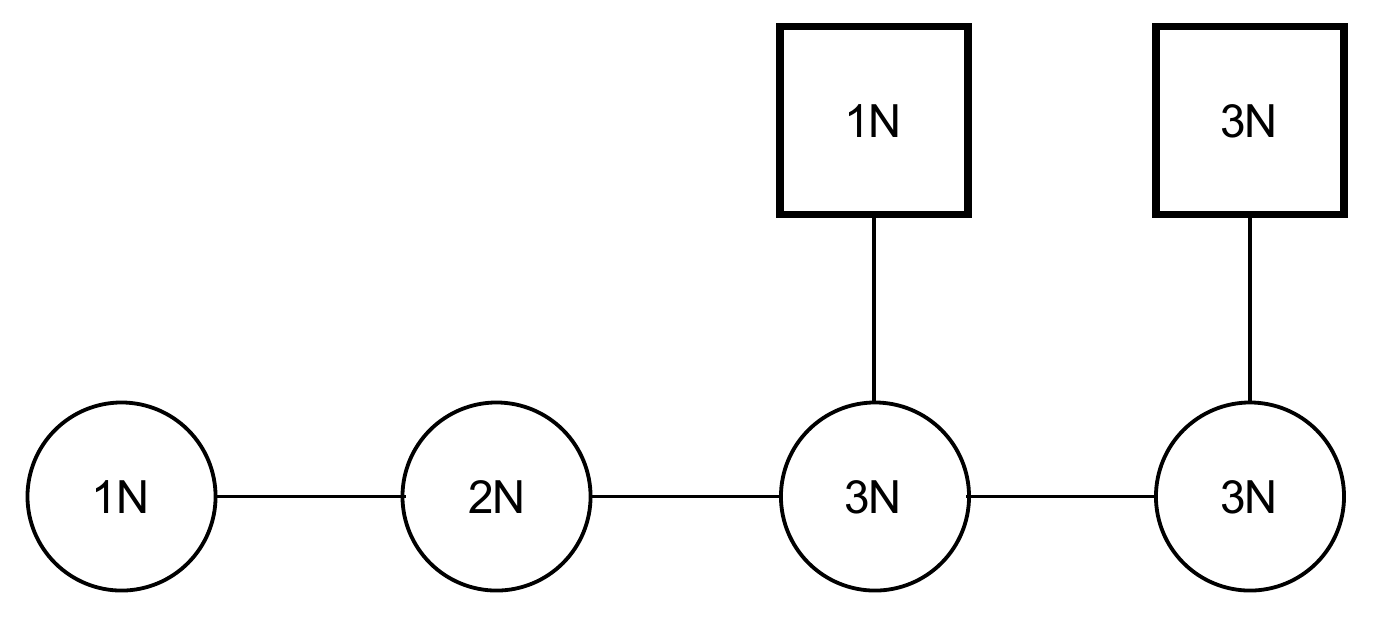} }}
\caption{The quiver corresponding to the Hanany-Witten set-up above.}
\label{figure2x}
\end{figure}

\begin{figure}[h!]
    \centering
    {{\includegraphics[width=9cm]{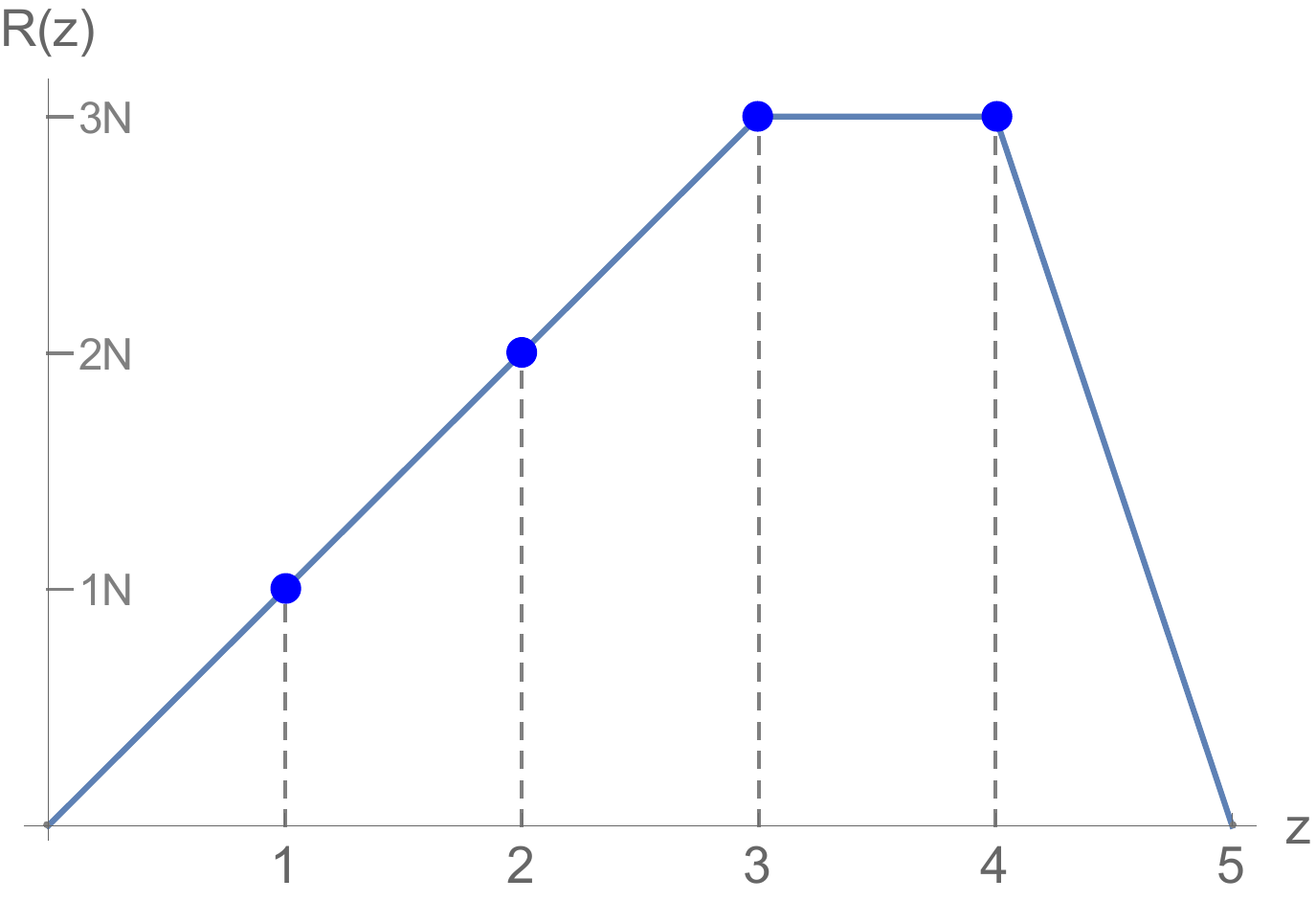} }}
\caption{The rank function $R(z)$ corresponding to the field theory.}
\label{figure3x}
\end{figure}

Working out examples is possibly the clearest way to explain the procedure to the reader not acquainted with this formalism. We first present full details for a simple example and then we consider a more generic situation.   The interested reader can consult the examples in Section  2.1.1 of the paper \cite{Nunez:2018ags}
\footnote{ In order for the background to capture faithfully the CFT dynamics one should work with long linear quivers and with large ranks. In this sense the examples of eqs.(2.6) and (2.8) of \cite{Nunez:2018ags} are rigorously trustable. Our examples in this section should be taken as illustrative of the procedure.}.
\\
\\
\underline{{\bf A simple example}}\\
Consider the  Hanany-Witten set-up, quiver and Rank function $R(z)$ in Figures \ref{figure1x}-\ref{figure3x}.

In this example, the rank function and the function $\alpha''(z)$ are given  by,
\[ R(z)=-\frac{1}{81\pi^2} \alpha''(z)= N \begin{cases} 
      z  &  0 \leq z\leq 1\\
      (z-1)+1 & 1\leq z \leq 2 \\
      (z-2)+2 &  2\leq z \leq 3\\
      3  & 3\leq z \leq 4\\
     3-3 (z-4)  & 4\leq z \leq 5. 
   \end{cases}
\]
This implies that the generic function $\alpha(z)$ for this example is,
\[ \alpha(z)=-{81\pi^2}  N \begin{cases} 
  a_0+ a_1 z + \frac{z^3}{6}     &  0 \leq z\leq 1\\
     b_0+ b_1 (z-1) +\frac{1}{2}(z-1)^2 +\frac{1}{6}(z-1)^3 & 1\leq z \leq 2 \\
c_0+ c_1 (z-2)+2\frac{(z-2)^2}{2} +\frac{1}{6} (z-2)^3&  2\leq z \leq 3\\
     d_0+d_1(z-3)+ 3\frac{(z-3)^2}{2}  & 3\leq z \leq 4\\
    p_0+p_1(z-4) + 3\frac{(z-4)^2}{2}-3\frac{ (z-4)^3}{6}  & 4\leq z \leq 5. 
   \end{cases}
\]
To determine the ten integration constants, we need to impose:
\begin{itemize}
\item{That $\alpha(0)=\alpha(5)=0$. This  is to have an internal space that shrinks smoothly at the beginning and end of the $z$-interval. These conditions imply
\begin{equation}
a_0=0,\;\;\;\; p_0+p_1+\frac{3}{2} -\frac{3}{6}=0\nonumber
\end{equation}}
\item{That the function $\alpha(z)$ is continuous, this implies the equations,
\begin{eqnarray}
& &a_1+\frac{1}{6}=b_0,\;\;\; b_0+b_1+\frac{1}{2}+\frac{1}{6}=c_0,\;\;c_0+c_1+1+\frac{1}{6}= d_0,\;\;\;\; d_0+d_1+\frac{3}{2}=p_0.\nonumber
\end{eqnarray}}
\item{That the function $\alpha'(z)$ is continuous. This implies,
\begin{eqnarray}
& & a_1+\frac{1}{2}
=b_1,\;\;\; b_1+1+\frac{1}{2}=c_1,\;\; c_1+2+\frac{1}{2} =d_1,\;\;\; d_1+3 =p_1.\nonumber
\end{eqnarray}}
\end{itemize}
Solving these equations we find,
\begin{eqnarray}
& & a_0=0,\;\; -5 a_1=19,\;\;\; -30 b_0=109,\;\;\; -10 b_1=33,\;\;\; -15c_0=94,\;\; -5c_1=9,\nonumber\\
& & -10 d_0=69,\;\;\; 10 d_1=7,\;\;\;\; -10p_0=47,\;\;\;\; 10p_1=37.\nonumber
\end{eqnarray}
In this way, we have a well defined function $\alpha(z)$. 

We can apply our expressions for the number of NS, D6 and D8-branes and linking numbers. Using eqs.(\ref{chargeNS5good}), (\ref{chargeD6good}) and (\ref{chargeD8good}) we find,
\begin{eqnarray}
& &N_{NS5}=z_f=5,\;\;\; N_{D8}=\frac{1}{81\pi^2}\left[\alpha'''(0)-\alpha'''(z_f)  \right]=4 N,\nonumber\\
& & N_{D6}=-\frac{1}{81\pi^2}\int_0^{z_f}\alpha''(z) dz=9 N.
\end{eqnarray}
Notice that this coincides with the numbers we count from the Hanany-Witten set-up in Figure \ref{figure1x}.

We can also calculate the linking numbers using eqs.(\ref{linkingNS}),(\ref{linkingD8}). We find,
\begin{eqnarray}
& & K_1=K_2=....=K_5=\frac{1}{81\pi^2}\alpha'''(z_f)=-3N\to \sum_{i=1}^{N_5}K_i=-15 N.\nonumber\\
& & L_1=L_2=....=L_N=3,\;\;\;\hat{L}_{1}=....=\hat{L}_{3N}= 4\to \sum_{i=1}^{N_8}L_i +\hat{L}_i= 3 N + 4 \times 3 N=15 N.\nonumber
\end{eqnarray}
We have denoted by $L_i,\hat{L}_i$ the linking numbers of the D8-branes in the first and second stacks. These numbers are also obtained by simple inspection of the Hanany-Witten diagram. The relation in eq.(\ref{consistency}) is satisfied.  The entanglement entropy can be calculated straightforwardly using eq.(\ref{eecc}).

\begin{figure}[h!]
    \centering
    {{\includegraphics[width=9cm]{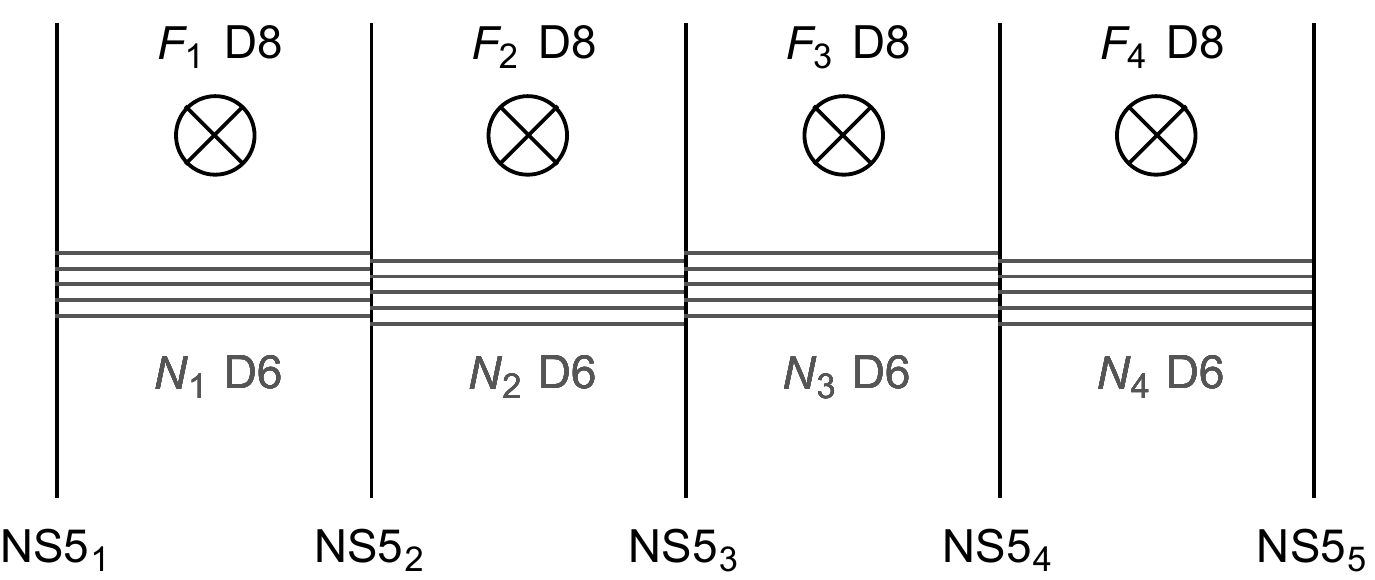} }}
\caption{The Hanany-Witten set-up corresponding to the generic field theory studied here.}
\label{Figures2Branes}
\end{figure}

\begin{figure}[h!]
    \centering
    {{\includegraphics[width=9cm]{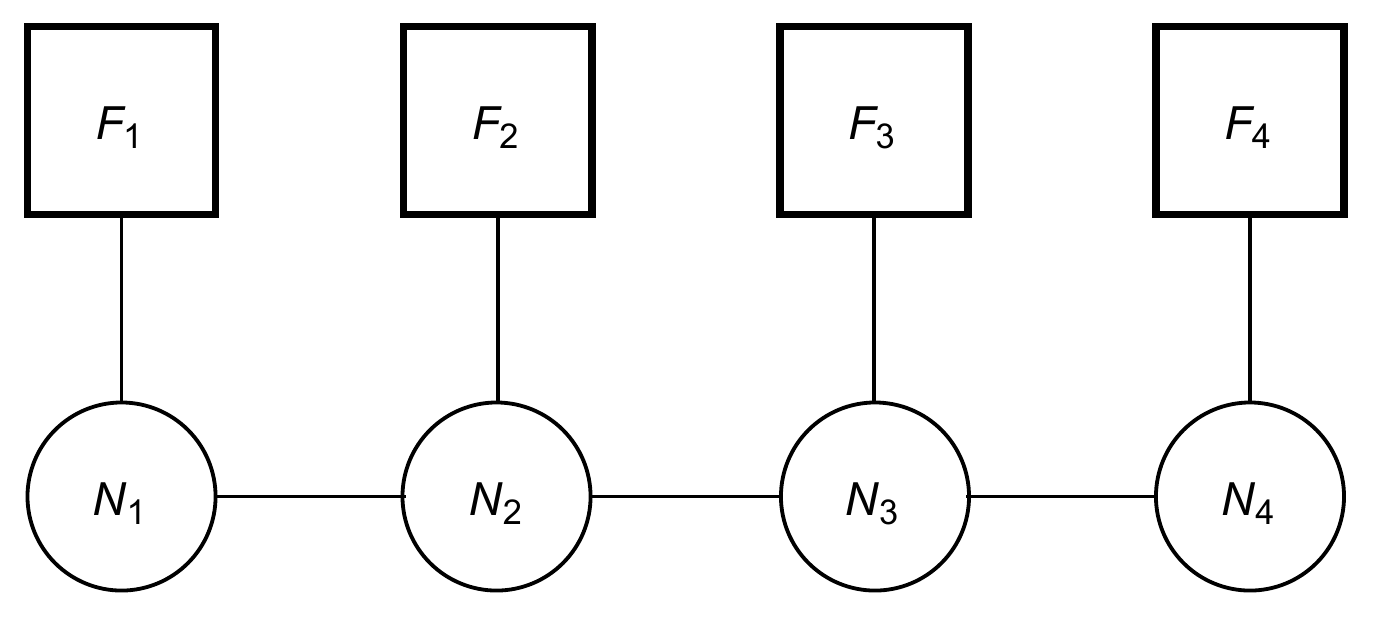} }}
\caption{The quiver corresponding to the generic field theory studied here.}\label{Figures2Quiver}
\end{figure}

Let us now study a more generic example.\\
\\
\underline{{\bf A more generic example}}\\
Following the same logic, we can work out a slightly more generic situation. Consider the field theory represented by the Hanany-Witten set-up in Figure \ref{Figures2Branes} or  equivalently, the quiver in Figure \ref{Figures2Quiver} .

For the gauge anomalies to cancel, we need
\begin{eqnarray}
& & 2N_1\!- \!N_2=\!F_1,\;\;\;\;\;\; 2 N_2-\! N_1-N_3 \!=\!F_2,\nonumber\\
& &  2 N_3- \! N_2-N_4=\!F_3,\;\;\;\;\;\; 2N_4-\!N_3\!=\! F_4.\label{rangosvv}
\end{eqnarray}  
We construct the rank-function
\[ R(z)= \begin{cases} 
      N_1 z  &  0 \leq z \leq 1\\
      (N_2-N_1)(z-1)+N_1 & 1\leq z \leq 2 \\
      (N_3-N_2)(z-2)+N_2 &  2\leq z \leq 3\\
      (N_4-N_3)(z-3) + N_3 & 3\leq z \leq 4\\
     -N_4 (z-4) + N_4 & 4\leq z \leq 5. 
   \end{cases}
\]
The function $\alpha(z)$ reads,
\begin{equation} \label{eq:genericquiver4Nodes} 
\alpha(z)=-81\pi^2  \begin{cases} 
      a_0+a_1z+N_1\frac{ z^3}{6}  &  0 \leq z \leq 1\\
b_0+ b_1 (z-1) +N_1\frac{(z-1)^2}{2} +    (N_2-N_1)\frac{(z-1)^3}{6} & 1\leq z \leq 2 \\
     c_0+c_1(z-2)+ N_2\frac{(z-2)^2}{2}+ (N_3-N_2)\frac{(z-2)^3}{6} &  2\leq z \leq 3\\
 d_0+ d_1(z-3) + N_3\frac{(z-3)^2}{2} +   (N_4-N_3)\frac{(z-3)^3}{6}  & 3\leq z \leq 4\\
  p_0+ p_1(z-4)+ N_4\frac{(z-4)^2}{2}   -N_4\frac{ (z-4)^3}{6}  & 4\leq z \leq 5. 
   \end{cases}
\end{equation} 
We determine the ten coefficients by imposing that $\alpha(0)=\alpha(5)=0$, the continuity of $\alpha(z)$ and $\alpha'(z)$. The resolution of the system is straightforward, we do not quote the result here.

We calculate the number of branes in the Hanany-Witten set-up. We find,
\begin{eqnarray}
& & N_{NS5}=z_f=5,\;\;\;\; N_{D8}= \frac{1}{81\pi^2}\left[\alpha'''(0)-\alpha'''(5)\right]=N_1+N_4=F_1+F_2+F_3+F_4,\nonumber\\
& & N_{D6}=-\frac{1}{81\pi^2}\int_0^{z_f}\alpha'' dz= N_1+N_2+N_3+N_4,\nonumber
\end{eqnarray}
this coincides with the counting from the Hanany-Witten set-up. Similarly, for the linking numbers we obtain
\begin{equation}
K_1=K_2=....=K_5=-N_4\to \sum_{i=1}^{N_5}K_i=\frac{1}{81\pi^2}\alpha'''(z_f) z_f=- 5 N_4.\nonumber
\end{equation} 
For the i-th stack of D8-branes we find the linking number $L_{(i)}$,
\begin{eqnarray}
& & L_{(1)}=1\to \sum_{1}^{F_1} L_{(1)}= F_1,\;\;\;L_{(2)}=2\to \sum_{1}^{F_2} L_{(2)}=2 F_2,\nonumber\\
& & L_{(3)}=3\to \sum_{1}^{F_3} L_{(3)}=3 F_3,\;\;\;\;
L_{(4)}=4\to \sum_{1}^{F_4} L_{(4)}= 4F_4,\nonumber\\
& &\sum_{1}^{N_8}L_i= F_1+2F_2+3F_3+4F_4=5 N_4.\nonumber
\end{eqnarray}
Again, we find agreement between the holographic calculation and the direct inspection of the Hanany-Witten set-up, validating our proposed expressions (\ref{chargeD6good}),(\ref{chargeD8good}),(\ref{linkingNS}),(\ref{linkingD8}).

In what follows, we analytically study the integrability of strings in generic backgrounds dual to six-dimensional ${\cal N}=(1,0)$ CFTs.

\section{Study of the integrability of strings on generic  string backgrounds}\label{sectionintegrability}
In this section we discuss the non-integrability of the classical string sigma model on the backgrounds given by eqs.(\ref{eq:TomasielloGeometryGeneral}),(\ref{eq:TomasielloGeometriesFunctions}). The strategy we adopt is the following: suppose the sigma model on the background is integrable, then any subsector of the sigma model (in particular, any string soliton) must also be integrable. The aim is then to find a contradiction by finding a subsector that is {\it not} (Liouville) integrable. If such a subsector is found, the whole sigma model can {\it not} be integrable. This strategy was applied to a large variety of examples, see the papers \cite{Basu:2011di}, \cite{Basu:2011fw} and citations to them. Of course, not finding such a subsector does not guarantee the integrability of the sigma model.

In the papers \cite{Nunez:2018ags}, \cite{Nunez:2018qcj} the authors presented particular string solitons for which the classical equations of motion were solvable and the non-integrability was shown under generic circumstances. We would like to revisit the treatment in those works. Indeed, we present a generalisation of \cite{Nunez:2018ags}, \cite{Nunez:2018qcj} that leads to a condition on the background for which the string soliton fails to detect non-integrability. 
In other words, the string soliton proposed in the papers \cite{Nunez:2018ags}, \cite{Nunez:2018qcj} and its generalisation presented below are very `efficient' at detecting non-integrable subsectors of the sigma model. Imposing that the soliton fails to detect a non-integrable subsector singles out a particular function $\alpha(z)$.
As we discuss further in Section \ref{integrablespecial}, the particular background satisfying this condition turns out to be integrable.

To proceed, we write the relevant part of the Neveu-Schwarz sector of the background in eq.(\ref{eq:TomasielloGeometryGeneral}) choosing
global coordinates for $AdS_7$. We have
\begin{eqnarray}
 ds^2&=& f_1(z)\left[ -dt^2 \cosh\rho + d\rho^2+\sinh^2\rho (d\varphi^2+\cos^2\varphi d\theta^2 +\sin^2\varphi d\Omega_3) \right] + f_2(z)dz^2\nonumber\\
& &  +f_3(z)\left(d\chi^2+\sin^2\chi d\xi^2\right), \;\;\;\;\;\;\;\;~~B_2= f_4(z)\sin\chi d\chi\wedge d\xi.\label{nspartbackground}
\end{eqnarray}
We propose an embedding for the string soliton of the form,
\begin{eqnarray}\label{eq:StringEmbedding}
 t = t(\tau), \quad \rho = \rho(\tau), \quad \varphi = \varphi(\tau), \quad \theta = \mu\sigma, \quad z = z(\tau), \quad \chi = \chi(\tau), \quad \xi = \kappa\sigma.
\end{eqnarray}
Here  the integers $\kappa$ and $\mu$ indicate how many times the string wraps  around the $\xi$ and $\theta$-directions respectively. 

We study the equations of motion of this soliton derived from the Polyakov action, supplemented by the Virasoro constraint,
\begin{eqnarray}
\label{PolyakovAction}
& & S_P=-\frac{1}{4\pi\alpha'}\int_\Sigma d^2\sigma\left(\eta^{ab}G_{\mu\nu}+\epsilon^{ab}B_{\mu\nu}\right)\partial_a X^\mu\partial_b X^\nu, \\
& & T_{ab}=\partial_a X^\mu\partial_b X^\nu G_{\mu\nu}-\frac{1}{2}\eta_{ab}\eta^{cd}\partial_c X^\mu\partial_d X^\nu G_{\mu\nu}=0.\nonumber
\end{eqnarray}
The equations of motion are,
\begin{eqnarray}\label{eq:EulerLagrangeEquations}
& &f_1(z) \dot{t} = \frac{E}{\cosh^2 \rho} \nonumber\\
& &f_1(z) \ddot{\rho} = - \frac{E^2}{f_1(z)} \frac{\sinh \rho}{\cosh^3 \rho}+  f_1(z) \sinh\rho \cosh \rho \left( \dot{\varphi}^2 - \mu^2 \sin^2 \varphi \right) - f_1'(z)\dot{\rho}\dot{z} \nonumber\\
& &f_1(z) \ddot{\varphi} = -f_1(z) \left( 2\frac{\cosh\rho}{\sinh\rho}\dot{\varphi}\dot{\rho} + \mu^2 \cos\varphi\sin\varphi \right) - f_1'(z)\dot{z}\dot{\varphi} \\
& &f_3(z) \ddot{\chi} = \kappa f_4'(z) \dot{z} \sin \chi -  f_3'(z) \dot{z}\dot{\chi} -\kappa^2 f_3(z)\sin\chi \cos\chi \nonumber\\
& &2f_2(z) \ddot{z} = f_1'(z) \left( - \frac{E^2}{f_1(z)^2 \cosh^{2}\rho} + \dot{\rho}^2 + \sinh^2\rho \left( \dot{\varphi}^2 - \mu^2 \sin^2 \varphi \right) \right) - f_2'(z) \dot{z}^2  \nonumber\\
& &\qquad\qquad  + f_3'(z) \left( \dot{\chi}^2 - \kappa^2 \sin^2 \chi \right) - 2 \kappa \dot{\chi} \sin \chi f_4'(z). \nonumber
\end{eqnarray}
Here the dots indicate derivatives with respect to $\tau$ and the primes indicate derivatives with respect to $z$. 
We have used the first equation above,  to replace for $\dot{t}$ in the other four equations.
Notice that when we set $\rho = \varphi =  0$ this system of equations reduces to the system that was studied in eq.(3.5) of the paper \cite{Nunez:2018ags}.

The Virasoro constraints for the string soliton are,
\begin{eqnarray}\label{eq:VirasoroConstraint}
& &T_{\tau\tau} = 0,\;\;\;T_{\sigma\sigma} =0,\;\;\;T_{\sigma\tau}=0\to\\
& & f_1(z) \left( -\cosh^2\rho\;\dot{t}^2 + \dot{\rho}^2 + \sinh^2 \rho \left( \dot{\varphi}^2 + \mu^2 \sin^2 \varphi \right) \right)  + f_2(z)\dot{z}^2 + f_3(z)\left(\dot{\chi}^2 + \kappa^2 \sin^2 \chi \right)=0. \nonumber
\end{eqnarray}
The reader can check that the equations of motion imply that  the string soliton satisfies the Virasoro constraints  by making an appropriate choice for the integration constant $E$.

We proceed with the  strategy outlined above. The reader unfamiliar with the technical details should consult the papers  \cite{Nunez:2018ags}, \cite{Nunez:2018qcj} for a clear explanation of the procedure. First, we find a simple solution by solving the equation for
 $\ddot{z}(\tau)$, choosing configurations with,
 \begin{equation}
 \ddot{\varphi}(\tau) = \dot{\varphi}(\tau) = \varphi(\tau) = \ddot{\chi}(\tau) = \dot{\chi}(\tau) = \chi(\tau) = \ddot{\rho}(\tau) = \dot{\rho}(\tau) = \rho(\tau) = 0.\nonumber
 \end{equation}
In fact, the equations of motion in (\ref{eq:EulerLagrangeEquations}) are automatically solved, aside from the equation for $\ddot{z}$  that reads,
\begin{equation}
2 f_2(z) \ddot{z} = - \frac{f_1'(z)}{f_1(z)^2}E^2 - \dot{z}^2 f_2'(z).
\end{equation}
After inserting the expressions for the functions $f_1(z)$ and $f_2(z)$ in terms of $\alpha(z)$, the above equation for $\ddot{z}$ reads,
\begin{equation}
\ddot{z} -  \left( \frac{\alpha \alpha''' - \alpha'\alpha''}{4\alpha^2} \right) \left(\frac{\alpha}{\alpha''}\right) \left( \frac{E^2}{16\pi^2} - \dot{z}^2 \right) = 0,
\end{equation}
which has a simple solution
\begin{equation}\label{eq:solutionForZ}
z_{sol}(\tau) = \frac{E}{4\pi} \tau.
\end{equation}
This  also solves the constraint in eq.(\ref{eq:VirasoroConstraint}) using the first of eqs.(\ref{eq:EulerLagrangeEquations}) for $\dot{t}$ and the expressions for $f_1(z), f_2(z)$.

We then have a `base solution' around which we perturb the other variables. This leads to writing the Normal Variational Equation (NVE) for the different coordinates $\varphi,\rho,\chi$. These NVEs are second order linear equations with variable coefficients, for which there are well developed criteria for the existence of Liouville integrable solutions---see \cite{Nunez:2018ags}, \cite{Nunez:2018qcj} for a summary of these criteria. Below, we study the NVE for the $\rho$-variable. The detailed study of this NVE and those for the other coordinates together with the analysis of the Liouville integrability are given in Appendix \ref{sec:StringSolitons}. 

\subsection{NVE for the $\rho$-coordinate}
We allow for small fluctuations in $\rho(\tau) = 0 + \epsilon r(\tau)$ and insert the  $z_{sol}(\tau)$ in (\ref{eq:solutionForZ}) into the equation of motion for $\ddot{r}(\tau)$, we find for the NVE, at leading order in the small parameter $\epsilon$,
\begin{eqnarray}\label{eq:NVEequationRb}
& & \ddot{r}(\tau) + \mathcal{B}_r(\tau) \dot{r}(\tau) + \mathcal{A}_r(\tau) r(\tau) = 0 \nonumber\\
& & \qquad\qquad \mathcal{B}_r(\tau) = \left.\frac{f_1'(z)}{f_1(z)} \frac{E}{4\pi} \right|_{z_{sol}} = \left.\frac{E}{8\pi} \left( \frac{\alpha'}{\alpha} - \frac{\alpha'''}{\alpha''} \right) \right|_{z_{sol}} \\
& & \qquad\qquad \mathcal{A}_r(
\tau) = \left.\frac{E^2}{f_1(z)^2} \right|_{z_{sol}} = \left.\frac{-E}{128\pi^2} \frac{\alpha''}{\alpha} \right|_{z_{sol}} \nonumber
\end{eqnarray}
A detailed analysis of the Liouville integrability of this equation is relegated to Appendix \ref{sec:StringSolitons}. 
\\
Here, we make a simple observation:
 if the warp factor $f_1(z)$ is equal to a constant,  then $\mathcal{B}_r = 0$, and the above differential equation is that of a harmonic oscillator, which admits a Liouvillian solution of the form $r(\tau) = \exp (iE \tau )$. 
On the other hand, if $f_1(z)$ is not equal to a constant, we  show  in Appendix \ref{sec:StringSolitons} that the NVE in eq.(\ref{eq:NVEequationRb}) does not admit Liouvillian solutions.

In summary, the analysis above strongly suggests that the situation with constant AdS$_7 $ warp factor is quite special. This implies that the background-defining function $\alpha(z)=A \sin(\omega z)$, for which the functions $f_1(z),f_2(z)$ are constant, is {\it special}. This solution does not fall within the class of solutions studied in Section \ref{secthol}. In spite of this, we study below the background and the integrability of the string sigma model for strings moving on this special solution. We postpone to Section \ref{CFTsmeared} a more detailed analysis of the  CFT dual to this special solution.

\section{Integrability of strings on the special background}\label{integrablespecial}
As pointed out above, there is only one particular function $\alpha(z)=A \sin(\omega z)$, that makes constant the warp factor of AdS$_7$. We gave reasons to suspect that strings might be integrable on the resulting background. We analyse this in what follows.

Let us first write the complete Massive IIA solution and then show that the equations of motion of the sigma model can be derived from a Lax pair.

When $\alpha(z)= A \sin(\omega z)$, the $z$-coordinate varies in the interval $0\leq z\leq \frac{\pi}{\omega}$. We choose $\omega=\frac{\pi}{N_5}$, being $N_5$ a large integer number. The full background in eqs.(\ref{eq:TomasielloGeometryGeneral}), (\ref{eq:TomasielloGeometriesFunctions}) reads,
\begin{eqnarray} 
& & {ds^2  =\frac{\sqrt{2}\pi}{\omega} \left( 8 AdS_7 + \omega^2\;dz^2+\left(\frac{\sin^2\omega z}{1+\sin^2\omega z}\right)d\Omega_2 \right)},\label{eq:SinAlphaZGeometryMetric}\\
& &{e^{-2\phi}=e^{-2\phi_0}(1+\sin^2\omega z),\qquad\quad B_2=\pi\left(-z +\frac{\sin\omega z\cos\omega z}{\omega(1+\sin^2 \omega z)}\right)d\Omega_2},\label{eq:SinAlphaZGeometryDilatonB2}\\
& & 
{F_0= \frac{A\omega^3 \cos\omega z}{162\pi^3}, \qquad\qquad~~~~~ F_2=- \frac{A \omega^2}{81 \pi^2}  \left(\frac{\sin^3\omega z}{1+\sin^2 \omega z}\right) d\Omega_2}. \label{eq:SinAlphaZGeometryRR}
\end{eqnarray}
The expression for $F_0$ suggests that we have a continuous distribution of D8-branes. Indeed, in contrast with the results of the examples discussed in Section \ref{secthol},  the $F_0$ in eq.(\ref{eq:SinAlphaZGeometryRR}) is a continuous function, instead  of a piece-wise constant and discontinuous one that is characteristic of localised D8-branes. Postponing to Section \ref{CFTsmeared} the discussion of  the dual field theoretical understanding of the Massive IIA background in eqs.(\ref{eq:SinAlphaZGeometryMetric})-(\ref{eq:SinAlphaZGeometryRR}), we focus here on the integrability of the string sigma model in this solution.

As it is well-known, the way to show classical integrability of the string sigma model is to find a Lax pair that encodes the classical equations of motion. We present a Lax connection for the Polyakov action  of strings moving on a background with the Neveu-Schwarz sector  of eq.(\ref{eq:SinAlphaZGeometryMetric})-(\ref{eq:SinAlphaZGeometryDilatonB2}).

As discussed above, the warp-factor $f_1(z)$ as defined in eq.(\ref{eq:TomasielloGeometryGeneral}) is constant for this background, making the metric in  eq.(\ref{eq:SinAlphaZGeometryMetric}) a direct product of the $\mathrm{AdS}_7$ and $\mathcal{M}^3$ spaces. This will simplify proving integrability of the string on this background considerably, as the oscillations of the string on these different spaces decouple. We can write the Polyakov action in eq.(\ref{PolyakovAction}), for the string on this background as the sum of the action for a string on a seven-dimensional $\mathrm{AdS}_7$ geometry, and the action for the string on the three-dimensional internal space $\mathcal{M}^3$, with a $B_2$-field. This reads,
\begin{eqnarray}\label{eq:PolyakovActionSmearedD8Branes}
& S_P &= S_P^{\mathrm{AdS}_7} + S_P^{\mathcal{M}^3} \\
& &= -\frac{1}{4\pi \alpha'}\int_\Sigma d^2\sigma\;\eta^{ab}G_{\alpha\beta}^{\mathrm{AdS}_7} \partial_a X^\alpha \partial_b X^\beta - \frac{1}{4\pi \alpha'}\int_\Sigma d^2\sigma\; \left(\eta^{ab}G_{\mu\nu}^{\mathcal{M}^3} +\epsilon^{ab}B_{\mu\nu}^{\mathcal{M}^3}\right)\partial_a X^\mu\partial_b X^\nu \nonumber
\end{eqnarray}
where the Latin indices range over the worldsheet coordinates, and the Greek indices range over the target space. In particular, $\alpha, \beta$ range over the $\mathrm{AdS}_7$ directions, and $\mu, \nu$ range over the coordinates $z$, $\chi$ and $\xi$ of the internal space $\mathcal{M}^3$.  We reinstated the constant $\alpha'=1$. Let us study in turn the Lax connection for each part of the action.

\subsubsection*{\underline{Lax pair for $\mathrm{AdS}_7$}}
The Polyakov action on an $\mathrm{AdS}_n$ target space without a $B_2$-field, is known to be integrable (as is the action of the string on the other symmetric space dS$_n$ and the Euclidean counterparts H$^n$ and S$^n$, see Appendix \ref{appendix:IntegrabilitySymmetricSpaces} for a more detailed explanation and references). We can think of these symmetric spaces as cosets $F=G/H$ of a Lie group $G$ by a Lie subgroup $H$.

Therefore, we first introduce the Principal Chiral Model (PCM) on a semisimple Lie group $G$,
\begin{equation}
S_{PCM} = -\frac{\kappa^2}{2\pi}\int d^2\sigma\;\mathrm{Tr}\left[\partial_a g \partial^a g^{-1} \right], \qquad\qquad g\in G.
%\end{equation}
\end{equation}
which exhibits a $G_L\times G_R$ global symmetry and can be written in terms of the Maurer-Cartan form $j_a$, a Lie algebra valued connection on the group manifold,
\begin{equation}\label{eq:PCMAction}
S_{PCM}=\frac{\kappa^2}{2\pi} \int d^2\sigma\;\mathrm{Tr}\left[j_a j^a\right],\qquad\qquad j_a=g^{-1}\partial_a g\in\mathfrak{g}.
\end{equation}
This Maurer-Cartan form is by construction flat. The flatness condition together with the equations of motion for the action in eq.(\ref{eq:PCMAction}) read
\begin{equation}
\begin{split}
\partial_+ j_- + \partial_- j_+ = 0,\\
\partial_- j_+ - \partial_+ j_- - \left[ j_+, j_- \right] = 0.
\end{split}\label{eq:PCMeom+flat}
\end{equation}
Here we used lightcone coordinates on the string worldsheet. The above eqs.(\ref{eq:PCMeom+flat}) combine to construct a Lax connection
\begin{equation}
\mathcal{L}_\pm=\frac{1}{1\mp Z}j_\pm,
\end{equation}
where $Z\in\mathbb{C}$ is the spectral parameter, such that the flatness of the Lax connection
\begin{equation}
d\mathcal{L} + \mathcal{L}\wedge\mathcal{L} = 0,
\end{equation}
is  equivalent to the equations of motion obtained from the action in eq.(\ref{eq:PCMAction}).

However, our case at hand is not a group manifold $G$, but a symmetric coset $F=G/H$. That means that there is a $\mathbb{Z}_2$ automorphism of the algebra of $G$, under which the latter decomposes as $\mathfrak{g}=\mathfrak{f}\oplus\mathfrak{h}$. Thus, the right action of $H$ is realized as a gauge symmetry and, by introducing a $\mathfrak{h}$-valued gauge field $B_a$, the new gauge invariant PCM action reads
\begin{equation}
S_{PCM}=\frac{\kappa^2}{2\pi} \int d^2\sigma\;\mathrm{Tr}\left[J_aJ^a\right],\qquad\qquad J_a=j_a-B_a,
\end{equation}
where we have defined the projection $J_a=P_{\mathfrak{f}}(j_a)$. The resulting equations of motion are
\begin{equation}
D_aJ^a=0,\qquad\qquad D_a=\partial_a+\left[B_a,\cdot\:\right],\label{eq:PCMcosetEOM}
\end{equation}
while the new flatness condition
\begin{equation}
\partial_aB_b-\partial_bB_a+\left[B_a,B_b\right]+D_aJ_b-D_bJ_a+\left[J_a,J_b\right]=0,
\end{equation}
uses the commutation relations $\left[\mathfrak{h},\mathfrak{h}\right]\subset\mathfrak{h},\left[\mathfrak{h},\mathfrak{f}\right]\subset\mathfrak{f}$ and $\left[\mathfrak{f},\mathfrak{f}\right]\subset\mathfrak{h}$ to decompose into two separate projections on the algebras $\mathfrak{h},\mathfrak{f}$ as
\begin{equation}
\begin{split}
\partial_aB_b-\partial_bB_a+\left[B_a,B_b\right]+\left[J_a,J_b\right]=0,\\
D_aJ_b-D_bJ_a=0.\label{eq:PCMcosetflat}
\end{split}
\end{equation}
As before, the flatness eq.(\ref{eq:PCMcosetflat}) together with eq.(\ref{eq:PCMcosetEOM}) combine in a Lax connection for the coset space,
\begin{equation}
\mathcal{L}_\pm=B_\pm+Z^{\pm 1}J_\pm,
\end{equation}
whose flatness condition is equivalent the equations of motion of the symmetric PCM. This demonstrates that the string on a symmetric space is classically integrable in the absence of a $B_2$-field.

In Appendix \ref{appendix:IntegrabilitySymmetricSpaces} we introduce a more natural environment to realise the symmetric PCM. The reader unfamiliar with these technical details should consult this Appendix.  We now construct the Lax pair on the $\mathcal{M}^3$ part of the space.\\

\subsubsection*{\underline{Lax  pair for $\mathcal{M}^3$}}
Here we state the Lax connection whose flatness condition gives the equations of motion for  the three coordinates in $\mathcal{M}^3$ derived from the action $S_P^{\mathcal{M}^3}$ 
in eq.(\ref{eq:PolyakovActionSmearedD8Branes}). We will elaborate more on the derivation of this Lax connection in Section \ref{sec:RelationToLambdaDeformation}.
The Lax connection is of the form
\begin{equation}\label{laxm3}
\mathcal{L}_\pm = 2\left(1 + \sqrt{2}\right) \frac{A_\pm}{1 \mp Z},
\end{equation}
where
\begin{eqnarray}\hspace*{-0.5cm}
A_{\pm} = \begin{pmatrix}
\pm \sin \chi \sin \xi z_\pm \pm \frac{\sin 2z}{2\left(1 + \sin^2 z\right)} \left( \cos\chi \sin\xi\;\chi_\pm + \sin\chi \cos\xi\;\xi_\pm \right) - \frac{\sin^2 z}{\sqrt{2}\left(1 + \sin^2 z\right)} \left( 2\cos\xi\;\chi_\pm - \sin 2\chi \sin\xi\;\xi_\pm \right) \\
\mp \sin \chi \cos \xi z_\pm \pm \frac{\sin 2z}{2\left(1 + \sin^2 z\right)} \left( \cos\chi \cos\xi\;\chi_\pm - \sin\chi \sin\xi\;\xi_\pm \right) - \frac{\sin^2 z}{\sqrt{2}\left(1 + \sin^2 z\right)} \left( 2\sin\xi\;\chi_\pm + \sin 2\chi \cos\xi\;\xi_\pm \right) \\
\mp \cos\chi\; z_\pm + \frac{\sin\chi}{2\left(1+\sin^2 z\right)} \left( \pm \sin 2z\;\chi_\pm + 2\sqrt{2} \sin^2 z \sin \chi\;\xi_\pm \right) \nonumber
\end{pmatrix}.\nonumber
\end{eqnarray}
We denoted by $Z$ the spectral parameter, and use the notation $f_+ = \partial_+ f$, $f_- = \partial_- f$ and $f_{+-} = f_{-+} = \partial_+\partial_- f$. 
We wrote $A_{\pm}$ in the adjoint representation of $SU(2)$.
One can check that the flatness condition for this connection is equivalent to the equations of motion for the Polyakov action on the internal space $\mathcal{M}^3$ with $B_2$-field. These equations of motion take the form,
\begin{equation}\hspace*{-0.5cm}
z_{+-} = \frac{1}{2\sqrt{2} \left(\sin ^2 z +1 \right)^2}\left(  \sqrt{2}  \sin 2 z \left(\chi_+ \chi_- + \sin ^2 \chi\;\xi_+ \xi_- \right) - \left(5 + \sin ^2 z \right)\sin ^2 z \sin \chi \left(\chi_+ \xi_- - \chi_- \xi_+ \right) \right), \nonumber
\end{equation}
\begin{equation}\hspace*{-0.8cm}
\chi_{+-} = \frac{1}{2} \sin 2 \chi\; \xi_+ \xi_- + \frac{1}{2\sqrt{2}}\left(1 + \frac{4}{1+\sin^2 z}\right) \sin \chi \left( z_+ \xi_- - z_- \xi_+ \right) -\frac{\cot z}{1 + \sin^2 z} \left( z_+ \chi_- + z_- \chi_+ \right), \nonumber
\end{equation}
\begin{equation}\hspace*{-0.5cm}
\xi_{+-} = -\frac{1}{2 \sqrt{2}} \left(1 + \frac{4}{1+\sin^2 z}\right) \sin^{-1} \chi \left( z_+ \chi_- - z_- \chi_+ \right) - \frac{\cot z}{1+ \sin^2 z} \left( z_+ \xi_- + z_- \xi_+ \right) - \frac{\cos \chi}{\sin \chi} \left( \chi_+ \xi_- +\chi_- \xi_+ \right). \nonumber
\end{equation}
\\
It is illuminating to discuss the reason why this background is integrable, or more precisely, where the form of the Lax pair on $\mathcal{M}^3$ in eq.(\ref{laxm3}) comes from.

\subsection{Relation with $\lambda$-deformation}\label{sec:RelationToLambdaDeformation}
We explain more intuitively the reason why the bosonic sector of the string worldsheet on the background (\ref{eq:SinAlphaZGeometryMetric})-(\ref{eq:SinAlphaZGeometryRR}) is integrable. 

It turns out that the Neveu-Schwarz sector part of the internal space $\mathcal{M}^3$ for this background---eqs.(\ref{eq:SinAlphaZGeometryMetric})-(\ref{eq:SinAlphaZGeometryDilatonB2}) is exactly equal to that of the $\lambda$-deformed Wess–Zumino–Witten (WZW) model on SU(2). The $\lambda$-deformation is an integrable deformation of the WZW model proposed by Sfetsos in  \cite{Sfetsos:2013wia}. The WZW model is given by an action of the form,
\begin{equation}\label{eq:WZWAction}
S_{WZW, k} = \frac{k}{2\pi} \int_{\partial\mathcal{B}} \mathrm{Tr}\left[ j_a j^a \right] + \frac{k}{6\pi} \int_{\mathcal{B}} \epsilon_{abc} \mathrm{Tr} \left[ j^a j^b j^c\right]
\end{equation}
Here the first term is  the action of the PCM and describes the fluctuations of the string on a group manifold. Just like the PCM, the WZW model on any Lie group G is integrable. 

The $\lambda$-deformation is an integrable deformation of the WZW model. Namely, a deformation term added to the action in eq. (\ref{eq:WZWAction}) that preserves the integrability. The action of the $\lambda$-deformed WZW model is given by
\begin{equation}\label{eq:WZWActionLambdaDeformation}
S_\lambda = S_{WZW, k} + \frac{k}{\pi}   \int_{\partial\mathcal{B}}  \tilde{\j}^A_+ \left( \lambda^{-1} - D^T \right)_{AB}^{-1} j_-^B
\end{equation}
where $j_a$ as before in eq. ({\ref{eq:WZWAction}}) and ({\ref{eq:PCMAction}) is the left invariant current, and $\tilde{\j}_a = \partial_a gg^{-1}$ is the  right invariant current. Notice these currents are algebra valued and the indices $A$ and $B$ range over the components of the algebra of the group on which we study this action. The matrix $D_{AB} = \mathrm{Tr}\left[ T_A g T_B g^{-1} \right]$  relates the left and right invariant currents as $j_a^A = D^A_{\;B} \tilde{\j}_a^B$. Here the $T_A$ are the generators of the group.

If we analyse the $\lambda$-deformed WZW model on the Lie group SU(2), the action (\ref{eq:WZWActionLambdaDeformation}) is equivalent to the Polyakov action of the string on a target space of the form \cite{Driezen:2018glg},
\begin{eqnarray}
&ds^2_\lambda =& 2k \left( \frac{1 + \lambda}{1 - \lambda} dz^2 + \frac{1 - \lambda^2}{\Delta} \sin^2 z d \Omega_2^2\right), \nonumber\\
&B_2^\lambda =& -2k \left( z - \frac{\left(1 - \lambda\right)^2}{\Delta} \cos z \sin z \right) \text{vol} \Omega_2,\label{ccxx} \\
&e^{-2\Phi^\lambda} =& e^{-2 \Phi_0} \Delta,  \nonumber
\end{eqnarray}
where $\Delta = 1 + \lambda^2 - 2 \lambda \cos 2z$ and $\lambda \in \left[0, 1\right]$. For $\lambda = 0$ we obtain the original WZW model. The action we obtain for $\lambda \to 1$ is related to the non-Abelian T-dual of the WZW model in eq. (\ref{eq:WZWAction}). See \cite{Driezen:2018glg} for a detailed explanation.\\

\subsubsection*{The $\lambda$-deformation for $\lambda = 3 - 2\sqrt{2}$}
The connection between the $\lambda$-deformation of the WZW model on SU(2) in eq.(\ref{ccxx}) and the background solution in eqs. (\ref{eq:SinAlphaZGeometryMetric})-(\ref{eq:SinAlphaZGeometryRR}) is made by noting that for $\lambda = 3 - 2\sqrt{2}$, which implies $\Delta = 4 \lambda \left( 1 + \sin^2 z \right)$, the  geometry of eq.(\ref{ccxx}) reads,
\begin{eqnarray}
ds^2_{\lambda} =& 2\sqrt{2}k \left( dz^2 + \frac{\sin^2 z}{1 + \sin^2 z} d\Omega_2^2 \right), \nonumber\\
B_2^\lambda =& -2k \left( z - \frac{\sin z \cos z}{1 + \sin^2 z} \right) \text{vol} \Omega_2, \\
e^{-2\Phi^\lambda} =& e^{-2 \Phi_0^\lambda} \left( 12 - 8\sqrt{2} \right) \left(1 + \sin^2 z \right). \nonumber
\end{eqnarray}
This is identical to the Neveu-Schwarz sector of the internal space $\mathcal{M}^3$ in eqs. (\ref{eq:SinAlphaZGeometryMetric}), (\ref{eq:SinAlphaZGeometryDilatonB2}) if we identify
 $\omega =\frac{\pi}{2k}$ and choose conveniently $e^{-2 \Phi_0^\lambda}$.  The holographic limit $\omega\to 0$ associated to long quivers, corresponds to $k\to\infty$, the semi-classical limit of the WZW model. We write the metric, dilaton and $B_2$-field for this solution as,
\begin{equation}\label{eq:WriteGeometryAsAdS7LambdaS3}
ds^2_{10} = 8\sqrt{2}\pi ds^2_{\mathrm{AdS}_7} + \frac{\pi}{2}ds_{\lambda}^2,\;\;\; e^{-2\Phi}= e^{-2 \Phi^\lambda}, \quad B_2=\pi B_2^\lambda.
\end{equation}

In summary, for the function $\alpha(z)=A \sin(\omega z)$, the geometry becomes a direct product of AdS$_7\times \mathcal{M}^3$. The sigma model for the string factorises into a sigma model on AdS$_7$ times a sigma model on $\mathcal{M}^3$ coupled to a $B_2$-field. The first is integrable, and a Lax pair can be written as explained in detail in Appendix \ref{appendix:IntegrabilitySymmetricSpaces}. The sigma model on $\mathcal{M}^3$ is the $\lambda$-deformation of the WZW model on $S^3$---see \cite{Driezen:2018glg}--- for a particular value of the parameter $\lambda=3-2\sqrt{2}$. This implies the existence of a Lax pair, given in eq.(\ref{laxm3}), for this part of the space. As a consequence, the Neveu-Schwarz sector of the string sigma model on the whole solution of eqs.(\ref{eq:SinAlphaZGeometryMetric}),(\ref{eq:SinAlphaZGeometryRR}) is integrable.

There are other examples in the literature of integrable supergravity backgrounds where it is observed that those geometries are a direct product of integrable sub-spaces, with constant warp factors. {Examples of this are the Sfetsos-Thompson solution \cite{Sfetsos:2010uq} in the Gaiotto-Maldacena class of supergravity backgrounds, the Lunin-Maldacena  real $\beta$-deformations
\cite{Lunin:2005jy},  etc.} It would be interesting to derive a no-go theorem for the integrability of a string background with non-trivial warp factors. Some similar ideas have been
presented in \cite{Wulff:2017lxh}.
%Since the internal space in eq. (\ref{eq:WriteGeometryAsAdS7LambdaS3}) is given by the $\lambda$-deformation of the three-sphere and $B_2$-field in (\ref{eq:WZWModelTargetSpace}) for a particular value of $\lambda$, one might wonder if we can think of the complete background in eq. (\ref{eq:SinAlphaZGeometryMetric}-\ref{eq:SinAlphaZGeometryRR}) as somehow being a $\lambda$-deformed version of $\mathrm{AdS}_7 \times S^3$. However, there is no supergravity solution of this form.

%Considering the Polyakov action on a target space $\mathrm{AdS}_7 \times S^3$ would break the conformality of the action. The same would be true when we consider the Polyakov action on the $\lambda$-deformed target space $\mathrm{AdS}_7 \times S^3_\lambda$ for any other value of $\lambda$. Only for $\lambda = 3 - 2\sqrt{2}$ we find a point where the action becomes conformal. It would be interesting to see - from a quantum field theoretical point of view on the worldsheet - if this corresponds to a renormalisation flow, where $\lambda$ will change change until we reach the conformally fixed point at $\lambda = 3 - 2\sqrt{2}$, and if so what would be special about this particular value of $\lambda$ from a renormalisation point of view. 
To complement this analytical proof of integrability, in Appendix \ref{numericsD8smeared}, we perform a careful numerical treatment of the string soliton
in eq.(\ref{eq:StringEmbedding}), analysing its dynamics and finding  results in agreement with the integrability of the solitons, like trajectories in phase space, Poincar\'e sections, power spectrum, and Lyapunov exponents.

We will now present a short analysis of the background in eqs.(\ref{eq:SinAlphaZGeometryMetric})-(\ref{eq:SinAlphaZGeometryRR}) from the point of view of the dual 
${\cal N}=(1,0)$ six-dimensional CFT.

\section{Field theory interpretation of the special background}\label{CFTsmeared}
In this section we present a first approach to the conformal field theory dual to the background in eqs.(\ref{eq:SinAlphaZGeometryMetric})-(\ref{eq:SinAlphaZGeometryRR}).
Since the function $R'(z)=-\frac{\alpha'''(z)}{81\pi^2}$ is not piece-wise discontinuous and constant, a description in terms of  well defined six-dimensional gauge and flavour groups as that given in Section \ref{holodescr} is not the most suitable. Instead, we will {\it define} the CFT by calculating some of its observables. The use of the background to  calculate  these observables defining the CFT, is the main message of this section.

It is illustrative to first present a different way to arrive to the function $\alpha(z)=A \sin(\omega z)$,  than that  of our presentation of Section \ref{sectionintegrability} was purely based on integrability of the sigma model. In the paper \cite{Nunez:2018ags}, the authors gave a way to write solutions to the equation of motion (\ref{eq:AlphaThird})---see Section 2.3 in  \cite{Nunez:2018ags}. The idea was to choose a quiver, write the rank function $R(z)$ and  the function $F_0$ (typically piece-wise constant and discontinuous). An even periodic extension of $F_0$ was proposed and a Fourier series expansion of $F_0$ found. By integration, the function $\alpha(z)$ was written as,
\begin{equation}
\alpha(z)=\sum_{n=1}^{\infty} c_n \sin\left(\frac{n\pi }{N_5}z \right).\nonumber
\end{equation}
While the infinite sum of harmonics reproduces the piecewise continuous  function $\alpha(z)$ made out of cubic polynomials in each interval, it is natural to wonder what is the physical content of each harmonic in the sum (since the dynamical equation (\ref{eq:AlphaThird}) is linear). As we discussed around eqs.(\ref{eq:SinAlphaZGeometryMetric})-(\ref{eq:SinAlphaZGeometryRR}), this leads to a background that can be interpreted as if the D8 sources are smeared all along the $z$-coordinate, instead of sharply localised as the expression for $\alpha'''$ suggests in the examples of Section \ref{secthol}. A situation of this sort was  also suggested  (though not analysed in detail) in the work of Cremonesi and Tomasiello \cite{Cremonesi:2015bld}. These authors observed that a possible scaling under which the backgrounds of the form in eq.(\ref{eq:TomasielloGeometryGeneral}) are trustable representations of ${\cal N}=(1,0)$ SCFTs, involved taking the number of D8-branes to infinity and creating a continuous distribution. The authors of   \cite{Cremonesi:2015bld} emphasise that their treatment of anomalies still holds true in this case. Below, we analyse the different observables discussed in Section \ref{secthol}  for the particular solution of eqs.(\ref{eq:SinAlphaZGeometryMetric})-(\ref{eq:SinAlphaZGeometryRR}).
\\
\\
We  consider the solution derived from $\alpha(z)=A \sin(\omega z)$. We choose $\omega=\frac{n \pi}{N_5}$ which makes the coordinate range in $0\leq z\leq \frac{N_5}{n}$. We work with $n=1$ only (the first harmonic) in what follows. The expression of eq.(\ref{chargeNS5good}) indicates that $N_{NS5}= N_5$. We can calculate the number of D6 and D8-branes in this background. Using eqs.(\ref{chargeD6good}),(\ref{chargeD8good}) we find,
\begin{eqnarray}
& & N_{D6}=-\frac{1}{81\pi^2}\int_0^{N_5}\alpha''(z)dz= \frac{2A}{81\pi N_5},\label{smearedsolD6}\\
& & N_{D8}=\frac{1}{81\pi^2}\left[ \alpha'''(0)-\alpha'''(N_5) \right]=-\frac{2A\pi}{81N_5^3}.\label{smearedsolD8}
\end{eqnarray}
In absolute value, these expressions imply relations among the quantities,
\begin{equation} 
A=\frac{81\pi}{2}N_5 N_{D6},\;\;\; A=\frac{81}{2\pi}N_5^3 N_{D8},\;\;\;\;\; \pi^2N_{D6}=N_{D8} N_{5}^2.\label{relationsA}
\end{equation}
We can use the expressions for the linking numbers---eqs.(\ref{linkingNS}),(\ref{linkingD8}) and the expression for $A$ in terms of the number of flavour D8-branes, 
\begin{equation}
\sum_{i=1}^{N_5} K_i=\frac{1}{81\pi^2}\alpha'''(N_5) N_5=\frac{A\pi}{81N_5^2}= - \frac{N_5 N_{D8}}{2}.\label{linkingsmeared}
\end{equation}
In the paper  \cite{Nunez:2018ags}, the authors found an expression for the central charge of the conformal quiver---see eq.(2.14) in  \cite{Nunez:2018ags}. This coincides
with the holographic central charge $a$ found in \cite{Cremonesi:2015bld}, derived by field theoretical means. Let us apply this expression for the case at hand. For $\alpha(z)=A \sin(\omega z)$, we find
\begin{eqnarray}
c=-\frac{2^8}{3^8 \times 16 \times G_N}\int_{0}^{z_f}\alpha(z)\alpha''(z) dz=\frac{8}{3^8 \times G_N}A^2\omega^2N_5= \frac{N_{D6}^2 N_5}{4\pi^2}.\label{centralsmeared}
\end{eqnarray}
We have used the expression for $A$ in terms of the number of colour D6-branes and that in our conventions $G_N=8\pi^6$. 

We can compute the entanglement entropy. Using the expression in eq.(\ref{eecc}), we find that for this particular CFT
\begin{equation}
S_{EE}^{reg}=\left(\frac{\mu_1\mu_2^4}{L^4}\right) \times 64\pi^4 N_{D6}^2 N_5,\label{xxxcccvvv}
\end{equation}
that, as anticipated, has the same scaling with $N_5$ and $N_{D6}$ as the central charge.

An interesting observation is that these expressions for the linking numbers,  central charge and entanglement entropy in eqs.(\ref{linkingsmeared}),(\ref{centralsmeared}),(\ref{xxxcccvvv}) have the same scaling with $N_{D6}$ and $N_5$ as  a {\it four-dimensional} ${\cal N}=2$ quiver that starts with a flavour group of rank $N_{D6}$, continues with $N_5-1$ colour groups of rank $N_{D6}$ and closes with a flavour group of rank $N_{D6}$. See around eq.(3.16) of the paper \cite{Nunez:2019gbg}.
\\
\\
Another interesting observable in all CFTs is the Wilson loop, in particular those for which the non-dynamical quarks transform under the internal symmetries. We find  it interesting to study a fundamental string on a generic background of the form in eq.(\ref{eq:TomasielloGeometryGeneral}), parametrised by
\begin{equation}
t=\tau,\;\; x=\sigma, \;\;\;\; R=R(\sigma),\;\;\;\; z=z(\sigma).\label{parameterstring}
\end{equation} 
We use Poincar\'e coordinates for the AdS$_7$ space, parametrised by $(t,\vec{x},R)$. The Nambu-Goto action of the fundamental string on a generic background is,
\begin{equation}
S_{NG}=\frac{1}{2\pi}\int d\tau d\sigma \sqrt{f_1^2 R^4 + f_1^2 R'^2+ f_1 f_2 R^2 z'^2}=\frac{T}{2\pi} \int d\sigma \sqrt{f_1^2 R^4 + f_1^2 R'^2+ f_1 f_2  R^2 z'^2}.\label{ngstring}
\end{equation}
This action does not depend explicitly on the 'time variable' $\sigma$ and this implies the conserved `Hamiltonian',
\begin{equation}
\frac{f_1^2 R^4}{ \sqrt{f_1^2 R^4 + f_1^2 R'^2+ f_1 f_2 R^2 z'^2}}=C.\label{conserved}
\end{equation}
At this point, it is interesting to analyse three situations:
\begin{itemize}
\item{The situation for which the coordinate $z(\sigma)$ is constant. In this case, we are back to the usual Wilson loop calculation in strongly coupled CFTs \cite{Maldacena:1998im}, that gives $E_{QQ}\sim \frac{\sqrt{\lambda}}{L_{QQ} }$}

\item{The situation in which $R(\sigma)=R_0$ is constant. In this case we find the Nambu-Goto action, 
\begin{equation}
S_{NG}=\frac{T}{2\pi}\int d\sigma \sqrt{f_1^2(z) R_0^4 + f_1(z) f_2(z) R_0^2 z'^2}.\label{stringzz}
\end{equation}
That leads to more a conventional minimisation problem, equivalent to the calculation of the `usual' rectangular Wilson loop in a background of the form $ds^2\sim R_0^2 f_1(z)\left[dx_{1,p}^2\right] + f_2(z) dz^2$. Using eq.(\ref{eq:TomasielloGeometriesFunctions}) we find that $f_1(z) f_2(z)=16\pi^2$. The main difference with the situations  calculated previously in the bibliography is that the $z$-coordinate is bounded.
}
\item{
More interesting than the general study presented above is to consider the action in eq.(\ref{ngstring}) for the case of our special background in eq.(\ref{eq:SinAlphaZGeometryMetric}), for which $f_1(z)=\frac{8\sqrt{2}\pi}{\omega}$ and $f_2(z)=\sqrt{2}\pi \omega$. 
Using these values, the action in eq.(\ref{ngstring}) reads,
\begin{equation}
S_{NG}=\frac{\sqrt{32} T}{\omega}\int d\sigma \sqrt{R^4(\sigma) +R'^2(\sigma) +\frac{\omega^2}{8} R^2(\sigma) z'(\sigma)^2}.\label{ngspecial}
\end{equation}}
\end{itemize}
We can redefine the variable $\tilde{z}=\frac{\omega z}{\sqrt{8}}$ and from the action in eq.(\ref{ngspecial}) we find two conserved quantities,
\begin{equation}
\frac{R^4}{\sqrt{R^4 +R'^2+ R^2 \tilde{z}'^2}}=E, \quad \frac{R^2 \tilde{z}'}{\sqrt{R^4 +R'^2+ R^2 \tilde{z}'^2}}=J, \quad R^4\frac{J^2}{E^2}=\tilde{z}'^2. \label{conservnges}
\end{equation}
Following the usual procedure to write the separation of the external quarks in the $x$-direction $L_{QQ,x}$ and in the $\tilde{z}$-direction $L_{QQ,\tilde{z}}$,
\begin{eqnarray}
& &L_{QQ,x}=\frac{E}{R_0^3}\int_{1}^{\infty} dy\frac{1}{y^2\sqrt{y^4 -\frac{J^2}{R_0^2} y^2-\frac{E^2}{R_0^4}}},\nonumber\\
& & L_{QQ,\tilde{z}}=\frac{J}{R_0}\int_{1}^{\infty} dy\frac{1}{\sqrt{y^4 -\frac{J^2}{R_0^2} y^2-\frac{E^2}{R_0^4}}}.\label{Lqq}
\end{eqnarray}
The Energy of the quark-antiquark pair $E_{QQ}$ is after  regularisation,
\begin{eqnarray}
& & E_{QQ}=\frac{\sqrt{32} R_0}{\omega}\left[  \int_{1}^\infty dy \left( \frac{y^2}{\sqrt{y^4 -\frac{J^2}{R_0^2} y^2-\frac{E^2}{R_0^4}}    } -1    \right) -1      \right].\label{Eqq}
\end{eqnarray}
After redefining $\tilde{J}R_0=J$ and $\tilde{E}R_0^2=E$, we observe that these expressions in eqs.(\ref{Lqq})-(\ref{Eqq}) are the same as those obtained by Maldacena in \cite{Maldacena:1998im}
 when considering quarks that are charged under the R-symmetry. In our background the $z$-coordinate is not isometric, but the fundamental string in the configuration of eq.(\ref{parameterstring}) does see it as part of the R-symmetry. 

We close this section hoping to have given the reader a flavour of the many things that can be holographically computed with the background of eqs.(\ref{eq:SinAlphaZGeometryMetric})-(\ref{eq:SinAlphaZGeometryRR}). These observables serve as a definition of the six-dimensional ${\cal N}=(1,0)$ CFT, even when the precise ranks of the colour and flavour groups are not easy to determine. Let us present some summary and conclusions.

\section{Conclusions and Future Directions}\label{concl}
In this paper, we have studied six-dimensional superconformal field theories with ${\cal N}=(1,0)$ SUSY. The main goal was to learn about these non-Lagrangian, strongly coupled, field theories using holography. 

In particular, we have found new expressions calculating (in holographic language) 
the number of NS5, D6, and D8-branes, and their linking numbers, that characterise the Hanany-Witten set-ups associated with the CFTs. We also found a closed expression calculating the entanglement entropy of a rectangular region, explicitly dependent on the matter content of the CFT.

Interestingly, we have found a particular background in Massive Type IIA on which the NS sector of the string sigma model is classically integrable. We have written the Lax pair from which the sigma model equations of motion are derived. We related this special background to a $\lambda$-deformation of a WZW model. Our study was complemented with an intensive numerical analysis and a careful discussion of the Liouville integrability of string solitons. Various explicitly worked out examples and detailed appendixes complement our study.

Let us comment on the natural lines of investigation suggested by this work.
It is interesting to understand in detail the character of our special background in eqs.(\ref{eq:SinAlphaZGeometryMetric})-(\ref{eq:SinAlphaZGeometryRR}). In fact, as we commented, D8-branes are smeared in this solution. Finding the precise smearing form and fitting the solution in the framework developed in the past, see for example the works  \cite{Nunez:2010sf},
may be illuminating and useful for further progress.
 
It would be good to exploit the integrable background of eqs.(\ref{eq:SinAlphaZGeometryMetric})-(\ref{eq:SinAlphaZGeometryRR}), by repeating various of the studies that in the paradigmatic case of AdS$_5\times S^5$ gave insightful results.  It would also be interesting to learn about the applicability of the formalism that we presented for the situation in which a flow from the six-dimensional CFT to a lower dimensional field theory is realised by a background solution.

This work and previous experience suggest that when the pre-factor in front of the $AdS$-space is independent of the coordinates of the internal space, the study of the string sigma model decouples between an AdS$_p$ part and a $\Sigma_{10-p}$ one. We suggest that searching for backgrounds with these characteristics is a good guide to find integrable solutions. Along these lines, it should be interesting to understand the conditions that allow to formulate a no-go theorem for integrability.

The study of these issues is fascinating and we hope to report on them in the near future.

\section*{Acknowledgments:} We wish to thank various colleagues for their input
that helped to sharpen the ideas presented here. In particular,  Riccardo Borsato, Stefano Cremonesi, Sibylle Driezen, Tim Hollowood, Linus Wulff, Kostas Sfetsos, Daniel Thompson (very specially!) and Alessandro Tomasiello.

KF acknowledges the support of  an STFC scholarship. CN is Wolfson Merit Research Fellow of the Royal Society. JvG acknowledges the support of  an STFC scholarship.

\appendix
\section{Non-integrability of Strings on General AdS$_7$ Backgrounds}
\label{sec:StringSolitons}
In this Appendix, we analyse in more detail the eqs. (\ref{eq:EulerLagrangeEquations}) for a string soliton on the $\mathrm{AdS}_7$-backgrounds defined in eqs. (\ref{eq:TomasielloGeometryGeneral})-(\ref{eq:AlphaThird}), where we take for $\alpha(z)$ a general third order polynomial of the form $a_3 z^3 + a_2 z^2 + a_1 z + a_0$. A similar analysis was originally performed in \cite{Nunez:2018ags}, but we will generalise the results found there in two ways: 
\begin{itemize}
\item{First we will derive two relations between the coefficients $a_0$, $a_1$, $a_2$ and $a_3$. When either of these two relations - (\ref{eq:NonIntegrabilityRelationC}) and (\ref{eq:NonIntegrabilityRelationD}) - are met, the string soliton will be non-integrable for these coefficients. }
\item{Second, from these relations we then derive that the string soliton will always be non-integrable at the beginning and end of the $z$-interval for any generic quiver.}
\end{itemize}

We start from the equations of motion for the string soliton that are listed in eq. (\ref{eq:EulerLagrangeEquations}). As is explained in Section \ref{sectionintegrability}, we can solve the equations of motion for $\ddot{z}(\tau)$ by first choosing the solutions $\ddot{\varphi}(\tau) = \dot{\varphi}(\tau) = \varphi(\tau)$ = 0, $\ddot{\chi}(\tau) = \dot{\chi}(\tau) = \chi(\tau) = 0$, and $\ddot{\rho}(\tau) = \dot{\rho}(\tau) = \rho(\tau) = 0$. These solutions simplify the equation for $\ddot{z}$ to a new expression that has the solution $z_{sol}(\tau) = \frac{E}{4\pi} \tau$.

\subsection*{NVE for $\rho$}
If we now allow for small fluctuations in $\rho(\tau) = 0 + \epsilon r(\tau)$ and insert the solution for $z_{sol}(\tau)$, we find for the NVE
\begin{eqnarray}\label{eq:NVEequationR}
& & \ddot{r}(\tau) + \mathcal{B}_r(\tau) \dot{r}(\tau) + \mathcal{A}_r(\tau) r(\tau) = 0 \nonumber\\
& & \qquad\qquad \mathcal{B}_r(\tau) = \left.\frac{f_1'(z)}{f_1(z)} \frac{E}{4\pi} \right|_{z_{sol}} = \left.\frac{E}{8\pi} \left( \frac{\alpha'}{\alpha} - \frac{\alpha'''}{\alpha''} \right) \right|_{z_{sol}} \\
& & \qquad\qquad \mathcal{A}_r(
\tau) = \left.\frac{E^2}{f_1(z)^2} \right|_{z_{sol}} = \left.\frac{-E}{128\pi^2} \frac{\alpha''}{\alpha} \right|_{z_{sol}} \nonumber
\end{eqnarray}
When only considering a string that moves along the $z$ and $\rho$-directions, it is now easy to see that if the warp factor $f_1(z)$ is equal to a constant, $\mathcal{B}_r = 0$, and the above differential equation admits a Liouvillian solution of the form $r(\tau) = \exp (iE \tau )$. 

When we allow for a warp factor between the $\mathrm{AdS}_7$ and $\mathcal{M}^3$ spaces, such that $f_1(z)$ is no longer equal to a constant, we can use Kovacic's algorithm \cite{Kovacic:jsc} to try to determine if the resulting differential equation (\ref{eq:NVEequationR}) will still admit any Liouvillian solutions. This can be done by combining the coefficients $\mathcal{A}(\tau)$ and $\mathcal{B}(\tau)$ of a second order differential equation into a new function $V(\tau)$, defined below. By applying Galois theory to second order differential equations, one can find if the solutions to the differential equation will be Liouvillian by studying the pole structure and the behaviour at infinity of this function $V(\tau)$. For the differential equation (\ref{eq:NVEequationR}) to admit Liouvillian solutions, the function $V(\tau)$ has to satisfy at least one of following \emph{necessary but not sufficient} conditions:
\begin{itemize}
\item  The poles of $V(\tau)$ are all either of order 1 or of even order. At infinity, the function $V(\tau)$ is of even order, or of order greater than two.
\item The function $V(\tau)$ has at least one single pole that is either of odd order greater than 2, or of order 2.
\item The order of the poles does not exceed 2, and the order of $V(x)$ at infinity is at least 2.
\end{itemize} Here the order of $V(\tau)$ at infinity is equal to the degree of the denominator minus the degree of the numerator. For a more detailed summary of Kovacic's procedure, see Appendix B of \cite{Nunez:2018ags}. In this case the NVE for $\rho$, in eq. (\ref{eq:NVEequationR}), has a corresponding function $V_r(\tau)$ of the form
\begin{eqnarray}\label{eq:KovacicPotentialR}
&V_r =& \frac14 \left( 2 \mathcal{B}_r' + \mathcal{B}_r^2 - 4 \mathcal{A}_r\right) \\
%&= &\frac{E^2}{256 \pi^2} \left( \frac{2 \alpha''(z)}{\alpha(z)} + \left( \frac{\alpha'(z)}{\alpha(z)} - \frac{\alpha'''(z)}{\alpha''(z)} \right)^2\right) \nonumber \\
& =&\frac{E^2}{256 \pi ^2 \alpha ^2 (\alpha'')^2} \Big(-3 \left(\alpha'\right)^2 \left(\alpha''\right)^2 - 2 \alpha \alpha ' \alpha '' \alpha''' + %\ldots \nonumber \\
%& &\ldots +
\alpha  \left(6 \left(\alpha ''\right)^3+5 \alpha  (\alpha''')^2- 4 \alpha \alpha'' \alpha'''' \right) \Big) \nonumber
\end{eqnarray}
If we now consider a function $\alpha(z)$ corresponding to a general background, $\alpha(z)$ will be a piece wise continuous polynomial along the $z$-direction of at most order 3 at every point. 
\begin{equation}\label{eq:GeneralSolutionForAlpha}
\alpha(z) = a_3 z^3 + a_2 z^2 + a_1 z + a_0
\end{equation}
Inserting this solution for $\alpha(z)$ into the potential in eq. (\ref{eq:KovacicPotentialR}) will allow us to analyse the pole structure of $V_r(\tau)$ corresponding to a general Massive IIA background, 
\begin{eqnarray}\label{eq:NVEequationRPotential}
V_r(\tau) = \frac{3 E^2}{2}\frac{ 3 a_3^4 E^6 \tau^6 + 24 a_2 a_3^3 E^5 \pi \tau^5 + \ldots }{(4 a_2 \pi + 3 a_3 E \tau)^2
   \left(64a_0 \pi^3 + 16 a_1 E \pi^2 \tau + 4 a_2 E^2 \pi \tau^2 + a_3 E^3 \tau^3 \right)^2}.
\end{eqnarray}
The dots in the numerator are quartic terms in $\tau$, the order of $V(\tau)$ at infinity is thus equal to two. Note that $V_r(\tau)$ has one pole of order two at $\tau = r_0 = - 4 a_2 \pi / 3 a_3 E$.  The other poles come from the cubic polynomial that is the second term in the denominator, one of the real roots of this cubic polynomial can coincide with the earlier pole $\tau = r_0$.

Let us first examine the possible roots coming from this cubic polynomial: A cubic polynomial of the form $a\tau^3 + b\tau^2 + c\tau + d = 0$ has three complex roots, the multiplicity of which can be obtained from the determinant
\begin{eqnarray}\label{eq:DeterminantCubicPolynomial}
&\Delta =& a^2 b^2 - 4 b^3 - 4a^3 c - 27c^2 + 18abc \nonumber\\
& =& -4096 \left(-a_1^2 a_2^2 + 4 a_1^3 a_3 - 18 a_0 a_1 a_2 a_3 + 
   a_0 \left(4 a_2^3 + 27 a_0 a_3^2\right)\right) E^6 \pi^6.
\end{eqnarray}
If $\Delta > 0$ the polynomial has three distinct real roots, for $\Delta < 0$ there is one real root and two complex conjugate roots. When $\Delta = 0$ there are roots with multiplicity larger than one, this can be either a single real root of multiplicity 3 or one real root of multiplicity 2 with another additional root. If we list all possible options we arrive at the following table where the order of the poles of $V_r(\tau)$ are listed in both the case when first pole $r_0$ does coincide with one of the real poles from the cubic term ($r_0 = r_1$), and when this does not happen. 
\begin{table}[h!]\centering
\label{table:zeros}
\begin{tabular}{l|l|l|l|l}
\cline{2-4}                        &  cubic poles                      & $r_0 \neq r_1$ & $r_0 = r_1$  & \\ \cline{1-4}
\multicolumn{1}{|l|}{$\Delta > 0$} &  $(\tau-r_1)(\tau-r_2)(\tau-r_3)$    & 2, 2, 2, 2     & 4, 2, 2      & \\ \cline{1-4}
\multicolumn{1}{|l|}{$\Delta = 0$} &  $(z-r_1)^3$                         & 2, 6           & 8            & \\ \cline{1-4}
\multicolumn{1}{|l|}{$\Delta = 0$} &  $(z-r_1)^2(z-r_2)$                  & 2, 4, 2        & 6, 2         & \\ \cline{1-4}
\multicolumn{1}{|l|}{$\Delta = 0$} &  $(z-r_1)(z-r_2)^2$                  & 2, 2, 4        & 4, 4         & \\ \cline{1-4}
\multicolumn{1}{|l|}{$\Delta < 0$} &  $(\tau-r_1)(\tau-c_2)(\tau-c_3)$    & 2, 2, 2, 2     & 4, 2, 2      & \\ \cline{1-4}
\end{tabular}
\caption{The orders of the poles of $V(\tau)$, depending on both the determinant $\Delta$ of the cubic polynomial in the denominator, and on whether the additional root $r_0$ coincides with one of the roots of the cubic polynomial or not.}
\end{table}
We see that in all cases all poles will be of even order, and that the resulting $V_r(\tau)$ thus might pass the first of Kovacic's criteria. These conditions are however necessary but not sufficient to guarantee the existence of Liouvillian solutions. Let us next turn to the string soliton moving along the $\varphi$ direction.\\

\subsection*{NVE for $\varphi$}
We now examine the equation of motion for $\ddot{\varphi}$. If we allow for small fluctuations in $\varphi(\tau) = 0 + \epsilon f(\tau)$, and we insert the solution for $z(\tau)$ from eq. (\ref{eq:solutionForZ}) while we now freeze the string along the other directions such that $\rho = \dot{\rho} = \ddot{\rho}$ we find for the NVE
\begin{eqnarray}\label{eq:NVEequationA}
& & \ddot{f}(\tau) + \mathcal{B}(\tau) \dot{f}(\tau) + \mathcal{A}(\tau) f(\tau) = 0 \nonumber\\
& & \qquad\qquad \mathcal{B}_f(\tau) = \left.\frac{f_1'(z)}{f_1(z)} \frac{E}{4\pi}\right|_{z_{sol}} = \left.\frac{E}{8\pi} \left( \frac{\alpha'}{\alpha} - \frac{\alpha'''}{\alpha''} \right)\right|_{z_{sol}} \\
& & \qquad\qquad \mathcal{A}_f(\tau) = \mu^2 \nonumber
\end{eqnarray}
Again, we see that when $f_1(z)$ is equal to a constant, the above NVE will reduce to the harmonic oscillator with solution $a(\tau) = \exp(i\mu \tau)$. 

When we allow for a more general warp factor where $f_1(z)$ is no longer constant, we can again use Kovacic's algorithm to determine if the above differential equation will still admit Liouvillian solutions. Inserting again for $\alpha(z)$ a general third order polynomial as given in eq. (\ref{eq:GeneralSolutionForAlpha}) gives us for the potential
\begin{eqnarray}
V_f(\tau) =\frac{ -36 a_3^4 E^8 \mu^2 \tau^8 - 384 a_2 a_3^3 E^7 \pi \mu^2 \tau^7 + \ldots }{4(4 a_2 \pi + 3 a_3 E \tau)^2
   \left(64a_0 \pi^3 + 16 a_1 E \pi^2 \tau + 4 a_2 E^2 \pi \tau^2 + a_3 E^3 \tau^3 \right)^2}
\end{eqnarray}
Notice that though the order of the numerator is different from $V_r(\tau)$, the pole structure in the denominator is identical to that in eq. (\ref{eq:NVEequationRPotential}). Since the numerator does now contain terms $\tau^8$, the order of $V(\tau)$ at infinity is zero, as $V_f(\tau) \sim \mu^2$ when $\tau \to \infty$. We thus see immediately that $V_f(\tau)$ fails to meet the first and third of Kovacic's criteria. Only when both $\Delta \neq 0$ and $r_0 \neq r_1$ all poles are of order 2, we can pass the second of Kovacic's criteria.

Requiring that $r_0$ is not a root of the cubic polynomial in the denominator implies, after inserting $\tau = r_0$, that
\begin{equation}\label{eq:NonIntegrabilityRelationC}
\boxed{c = 2 a_2^3 - 9 a_1 a_2 a_3 + 27 a_0 a_3^2}
\end{equation}
Here $c = 0$, when $r_0$ coincides with one of the roots of the cubic polynomial. We can use the expression for $c$ to simplify the expression for $\Delta$ in eq. (\ref{eq:DeterminantCubicPolynomial}).
\begin{equation}\label{eq:NonIntegrabilityRelationD}
\boxed{d = \frac{\Delta}{-4096 E^6 \pi^6} = a_1 \left( c - 27 a_0 a_3^2 \right) - a_1^2 \left( a_2^2 - 4a_1 a_3 \right)}
\end{equation}
We have thus found two constraints relating the constants $a_0$, $a_1$, $a_2$ and $a_3$ in a general solution for $\alpha(z)$ of the form in eq. (\ref{eq:GeneralSolutionForAlpha}), such that \textbf{when either $c = 0$ or $d = 0$, Kovacic's criteria guarantees the non-integrability of the string soliton}. 

From this, we can immediately conclude that for every function $\alpha(z)$ corresponding to a quiver diagram, on the first part of the $z$-interval the string-soliton is guaranteed to be non-integrable. This is because any quiver, starting with an $\mathrm{SU}(N)$ flavour group will have $\alpha(z) = -81\pi^2 N \left( \frac16 z^3 - a_1 z \right)$ for $z \in [0, 1]$, for which $c = 0$. The pole at $r_0$ will thus coincide with one of the roots of the cubic polynomial, giving us 2 poles of order two, and one pole at $r_0$ of order 4. The function $V_f(\tau)$ corresponding to this solution will fail to meet any of Kovacic's criteria and the NVE (\ref{eq:NVEequationA}) will have non-Liouvillian solutions.

\subsection*{NVE for $\chi$}
If we allow for small fluctuations in $\chi(\tau) = 0 + \epsilon x(\tau)$ and insert the solution in (\ref{eq:solutionForZ}) in the equation of motion for $\ddot{\chi}(\tau)$, we find for the NVE for $\ddot{x}(\tau)$
\begin{eqnarray}\label{eq:NVEequationX}
& & \ddot{x}(\tau) + \mathcal{B}_x(\tau) \dot{x}(\tau) + \mathcal{A}_x(\tau) x(\tau) = 0 \nonumber\\
& & \qquad\qquad \mathcal{B}_x(\tau) = \left.\frac{E f_3'(z)}{4\pi f_3(z)} \right|_{z_{sol}} = \left.\frac{E}{8\pi} \left( 3 \frac{\alpha'}{\alpha} + \frac{(\alpha'^2 + 2 \alpha\alpha'') \alpha'''}{(\alpha'^2-2\alpha\alpha'') \alpha''}\right) \right|_{z_{sol}}\\
& & \qquad\qquad \mathcal{A}_x(
\tau) =  \left.\left(\kappa^2 - \kappa \frac{E f_4'(z)}{4 \pi f_3(z)}\right)\right|_{z_{sol}} \\
& & \qquad\qquad\qquad\quad = \left.\left(\kappa^2 - \frac{E\kappa}{4\pi} \frac{1}{\sqrt{-2\alpha \alpha''}} \frac{6\alpha\alpha''^2 - 2\alpha\alpha'\alpha'''-3\alpha'^2\alpha''}{2 \alpha \alpha'' - \alpha'^2}\right)\right|_{z_{sol}} \nonumber
\end{eqnarray}
We now only consider fluctuations of the string along the $z$ and $\chi$-direction, which is the same situation that was considered in \cite{Nunez:2018ags}. Because the function $V_x(\tau)$ now involves $f_3(z)$ and $f_4(z)$ and their derivatives, it is far less obvious to see from this result that the string soliton will directly fail to be integrable when $f_1(z)$ is not equal to a constant (as we obtained from the NVE's for $\rho$ and $\varphi$). If we would again insert a general function $\alpha(z)$ of the form given in eq. (\ref{eq:GeneralSolutionForAlpha}), the resulting $V_x(\tau)$ will be a complicated sum of large fractions. We will omit the result here, but is it difficult to see from this result what functions $\alpha(z)$ would give rise to an integrable string soliton. For this reason the authors in \cite{Nunez:2018ags} did not make a general argument, but instead studied various examples. It is quite difficult to extract general expressions for these cases, as could be nicely done for the NVE's for $\rho$ and $a$, where we saw that the only backgrounds for which the string soliton could have a Liouvillian solution is when $f_1(z)$ is constant.

%%%%%%%%%%%%%%%%%%%%%%%%%%%%%%%%%%%%%%%%%%%%%%%
\section{Integrability on the Symmetric $\sigma$-model}\label{appendix:IntegrabilitySymmetricSpaces}
Classical Liouvillian integrability for a Hamiltonian dynamical system, or for a field theory, emerges over the existence of a flat Lax connection $\mathcal{L}$, i.e.
\begin{equation}
d{\mathcal{L}}+\mathcal{L}\wedge\mathcal{L}=0\label{LaxFlatnessAppendix},
\end{equation}
on the cotangent bundle $\mathcal{T}^*\mathcal{M}$ (phase space), together with the involution of all the analogous independent conserved quantities. Generally, though, there is no particular prescription for finding such a connection and one has to rely on their inspiration to address the problem.

However, given a 2-dimensional scalar field theory in a homogeneous space for a connected semisimple Lie group $G$, the action can be reformulated in terms of its underlying group structure as
\begin{equation}
S_{PCM}\equiv-\frac{\kappa^2}{\pi}\int\Tr\:j_aj^a\label{PCMactionAppendix},
\end{equation}
where the Lie-algebra-valued current $j\in\mathfrak{g}(G)$,
\begin{equation}
j_\pm\equiv g^{-1}\partial_\pm g=j_\pm^i\mathfrak{t}_i, \hspace{2cm} g\in G, \hspace{0.5cm}\mathfrak{t}_i\in\mathfrak{g},
\end{equation}\\
is defined over the group element $g=e^{X^i\mathfrak{t}_i}$, that is all the point transformations on the scalar field worldsheet, on the group manifold. This one-form current is by construction flat and its flatness condition, together with the equations of motion,
\begin{equation}
\begin{split}
\partial_+ j_- + \partial_- j_+ = 0,\\
\partial_+ j_- - \partial_- j_+ + \left[j_+ , j_- \right] = 0,
\end{split}\label{MaurerCartanEomAndFlatnessAppendix}
\end{equation}
can be combined in a parametrized Lax connection
\begin{equation}\label{LaxAppendix}
\mathcal{L}_\pm = \frac{j_\pm}{1\mp Z},
\end{equation}
where $Z\in\mathbb{C}$ is the spectral parameter, whose flatness condition, eq.(\ref{LaxFlatnessAppendix}), is equivalent to the equations of motion, eq.(\ref{MaurerCartanEomAndFlatnessAppendix}). Then, one also defines the holonomy of $\mathcal{L}$ for constant time, i.e. the monodromy
\begin{equation}
M(Z)=\mathcal{P}\exp\int\mathcal{L}
\end{equation}
which defines a parallel transport on the group manifold $\Sigma(G)$ and whose eigenvalues are conserved, which means that by expanding in $Z$ at infinity we can obtain an infinite set of conserved charges. This is known in the literature as the Principal Chiral Model (PCM), it exhibits a global $G_L\times G_R$ symmetry and it is obviously integrable.

Moreover, the $\sigma$-model (in the presence of a $B_2$ field) in a homogeneous space for a group $G$ can be represented by the Wess-Zumino-Witten (WZW) model as
\begin{equation}
S_{WZW,k}=\frac{k}{2\pi}\int_{\partial\mathcal{B}}\Tr\:j_aj^a+\frac{k}{6\pi}\int_{\mathcal{B}}\epsilon_{abc}\:\Tr\:j^aj^bj^c, \hspace{2cm}j\in\mathfrak{g}(G),\label{WZWactionAppendix}
\end{equation}
which exhibits an $G_{L,cur}\times G_{R,cur}$ current algebra symmetry, it is an exact CFT and thus integrable.\\
\\
The situation becomes even more elegant in the case of the non linear $\sigma$-model in a symmetric homogeneous space. Symmetric spaces are backgrounds with rich underlying group structure, which can be exploited in a natural way to make the integrability of the $\sigma$-model manifest. From the group theoretical point of view, a symmetric space is a \textit{coset space} $G/H$, where the isometry $G$ is a connected Lie group and the subgroup $H\subset G$ is its isotropy group. Then the $\sigma$-model (without a B-field) can be recast as a PCM with currents projected on the coset algebra. The WZW model on a symmetric coset, on the other hand, does not correspond to the $\sigma$-model on that space (except in the case of a group manifold) and exhibits alternative interpretations.\\
\\
In what follows we will illustrate the classical integrability of the string worldsheet on a symmetric space. To study this in more detail see \cite{Zarembo:2017muf}, for a more general review of integrability in the context of string theory  \cite{Arutyunov:2004yx} and AdS/CFT correspondence \cite{Beisert:2010jr}.\\

\subsection*{Integrability of AdS space}
The $\sigma$-model on AdS space is \textit{integrable}. We know this as a fact, since, as we illustrated above, the $\sigma$-model is integrable on every symmetric homogeneous space. Of course, an uneasy mind shall always ask for an explicit Lax formulation given a specific background, something that proves to be quite challenging as we climb higher in dimensions of the target space. The difficulty rests in the fact that finding the gauged group element (matrix) of the coset space becomes an involved task in higher dimensions.

Nevertheless, if one desires to make this portrait more delicate, they shall preserve the rich underlying group structure of the PCM, adopting at the same time a more geometric point of view.

In particular, one can realize the element of a group $G$ abstractly as
\begin{equation}
g\equiv\exp X^i\:\mathfrak{t}_i,
\end{equation}
where $\mathfrak{t}_i\in\mathfrak{g}(G)$ and $X^i$ parametrize the adjoint space, which produces another formulation of the PCM action as
\begin{equation}
S_{PCM}=-\frac{\kappa^2}{\pi}\int d^2\sigma\;\eta_{ij}\:{e^i}_\mu(X)\:{e^j}_\nu(X)\;\partial_+ X^\mu\partial_-X^\nu, \label{PCMaction}
\end{equation}
where $\eta_{ij}=\left<\mathfrak{t}_i,\mathfrak{t}_j\right>$ is the metric on the Lie algebra $\mathfrak{g}$, defined by $[\mathfrak{t}_i,\mathfrak{t}_j]={f_{ij}}^k\mathfrak{t}_k$, while the vielbein
\begin{equation}
{e^i}_\mu=\frac{\partial X^i}{\partial X^\mu},
\end{equation}
represents the relationship between the adjoint and the target space\footnote{$i$ runs in the adjoint space of $G$ while $\mu$ spans the target space dimensions. The vielbeins represent a relationship between different bases, i.e. they express an object in different frames. As such, this relationship can exist between any kind of spaces.}.

Therefore, in this context, the vielbeins ${e^i}_\mu$ represent the components of the symmetry transformations of $G$ or, equivalently, the \textit{Killing vectors} of the manifold at hand. Subsequently, the vielbein is realized as the Maurer-Cartan connection
\begin{equation}
j^i_\pm\equiv e^i_\pm={e^i}_\mu\;\partial_\pm X^\mu, \label{MaurerCartanVielbeinAppendix}
\end{equation}
where $j_\pm=j^i_\pm\mathfrak{t}_i$, and satisfies the structural flatness condition
\begin{equation}
\partial_\mu {e^i}_\nu-\partial_\nu {e^i}_\mu+{f^i}_{jk}\;{e^j}_\mu\;{e^k}_\nu=0. \label{MaurerFlatness}
\end{equation}
As in the standard case, this flatness identity together with the equations of motion of the PCM
\begin{equation}
\partial_+\left({e^i}_\mu\;\partial_-X^\mu\right)+\partial_-\left({e^i}_\mu\;\partial_+X^\mu\right)=0, \label{eomPCM}
\end{equation}
construct the Lax connection
\begin{equation}
\mathcal{L}_\pm=\frac{j_\pm}{1\mp Z}, \label{LaxPairVielbeinAppendix}
\end{equation}
where $Z\in\mathbb{C}$ is the spectral parameter, and whose flatness condition
\begin{equation}
\left[\partial_++\mathcal{L}_+,\partial_-+\mathcal{L}_-\right]=0, \label{LaxFlatnessVielbeinAppendix}
\end{equation}
is equivalent to equations of motion.\\
\\
Thus, we conclude that in order to specify a particular Lax connection for the $\sigma$-model on a symmetric space, one only needs the Killing vectors of the background manifold.\footnote{One could be naively troubled about the fact that a symmetric space has less degrees of freedom that the number of its Killing vectors, e.g. $S^2$ has two d.o.f. and three Killing vectors. In reality, the Killing vectors - the space isometries - are constrained by the metric and encode the actual degrees of freedom.}

The reader could argue that the Lax connection eq.(\ref{LaxPairVielbeinAppendix}) works only for the PCM on a group $G$, since it is not of the appropriate coset form, i.e. it doesn't project on separately the isotropy and coset algebras. However, this is not the case since, as we argued above, the Killing vectors are a special coset parametrization, constrained by the target space metric. In other words, as the Lax connection is defined up to a gauge transformation, one could gauge transform our Lax eq.(\ref{LaxPairVielbeinAppendix}) into a traditional coset Lax connection.

Next, finding the Killing vectors is, thankfully, a simple task for a symmetric space. This is because a symmetric space can always be realized as an embedding in a higher dimensional space, the former inheriting most of the isometries of the latter. A standard example is $S^2$ which inherits the SO$(3)$ isometries from $\mathbb{R}^3$ (but not the translations).

AdS$_n$ space is a hypersurface in $\mathbb{R}^{2,n-1}$ onto which only the Lorentz group is tangent. Therefore, the boosts and the rotations of $\mathbb{R}^{2,n-1}$
\begin{equation}
V_i\equiv {V_i}^A\partial_{Y^A}
\end{equation}
where $Y^A, A=0,...,n$ are the embedding coordinates which build the hypersurface
\begin{equation}
\eta_{AB}Y^AY^B=-l^2, \label{AdSembedding}
\end{equation}
with $\eta_{AB}=$diag$(-1,1,...,-1)$, are inherited into AdS$_n$ as the Killing vectors
\begin{equation}
\xi_i\equiv{\xi_i}^\mu\partial_\mu=g^{\mu\nu}\left(\frac{\partial Y^A}{\partial x^\nu}V_A\right)_i\partial_\mu,
\end{equation}
where $x^\mu, \mu=0,...,n-1$ are the AdS$_n$ coordinates and $g_{\mu\nu}$ its metric, while $i$ runs in the vector space. By choosing one of the solutions to eq.(\ref{AdSembedding}), like the global embedding
\begin{equation}
\begin{split}
Y^0&=l\cosh\rho\:\cos t,\\
Y^j&=l\sinh\rho\:\Omega^j, \hspace{2cm}j=1,...,n-1,\\
Y^n&=l\cosh\rho\:\sin t,
\end{split}
\end{equation}
where $\Omega^j$ are the Euclidean  coordinates for the unit sphere ($\Omega^j\Omega_j=1$), one can find each one of the $n(n+1)/2$ Killing vectors of AdS$_n$.\\
\\
It's worth emphasizing that the Killing vectors that are inherited into a symmetric space, through an embedding, are constrained by the metric tensor. This means that while their number (number of isometries) exceeds the dimension of the space, in reality they encode the actual degrees of freedom. In other words, the PCM metric
\begin{equation}
G_{\mu\nu}=\eta_{ij}\:{e^i}_\mu\:{e^j}_\nu\;\partial_+ X^\mu\partial_-X^\nu=\eta_{ij}\:{\xi^i}_\mu\:{\xi^j}_\nu\;\partial_+ X^\mu\partial_-X^\nu,
\end{equation}
matches the target space metric (it has not redundant degrees of freedom). Thus, while in a matrix realization of the PCM we would, traditionally, have to gauge the isotropy group $H$ out of the isometry group $G$ to obtain the element of the coset $G/H$, the Killing vectors constitute a natural environment to describe a symmetric space.\\
\\
Since we have identified the Killing vectors $\xi_i$ of the background space of the PCM with the vielbeins $e_i$ in eq.(\ref{PCMaction}), then one can explicitly check that the equations of motion of this action, eq.(\ref{eomPCM}), are equivalent to the standard equations of motion of the $\sigma$-model in the same background, as they should. Therefore, the Killing vectors can be used to build up an explicit Lax connection through equations (\ref{MaurerCartanVielbeinAppendix}) and (\ref{LaxPairVielbeinAppendix}), as promised.

\subsection*{The AdS$_3$ example}
While AdS$_n$ can give frustrating results as we climb up the ladder of $n$, AdS$_3$ constitutes a relatively compact example of the above methodology. The reader should not be worried about the special case of AdS$_3$, it being a group manifold. As we argued above, our construction holds for every symmetric coset and, in fact, it was also tested for higher dimensions, successfully as it should.

Choosing a global AdS$_3$ embedding in $\mathbb{R}^{2,2}$ as
\begin{equation}
\begin{split}
Y^0&=\cosh\rho\:\cos t,\\
Y^1&=\sinh\rho\:\cos t\sin\phi,\\
Y^2&=\sinh\rho\:\cos t\cos\phi,\\
Y^3&=\cosh\rho\:\sin t,
\end{split}\label{GlobalEmbedding}
\end{equation}
then the six corresponding Killing vectors are
\begin{equation}
\begin{split}
\xi_1&=\partial_t,\\
\xi_2&=\partial_\phi,\\
\xi_3&=\tanh\rho\:\sin t\sin\phi\:\partial_t+\coth\rho\:\cos t\cos\phi\:\partial_\phi+\cos t\sin\phi\:\partial_\rho,\\
\xi_4&=\tanh\rho\:\sin t\cos\phi\:\partial_t-\coth\rho\:\cos t\sin\phi\:\partial_\phi+\cos t\cos\phi\:\partial_\rho,\\
\xi_5&=\tanh\rho\:\cos t\sin\phi\:\partial_t+\coth\rho\:\sin t\cos\phi\:\partial_\phi+\sin t\sin\phi\:\partial_\rho,\\
\xi_6&=\tanh\rho\:\cos t\cos\phi\:\partial_t-\coth\rho\:\sin t\sin\phi\:\partial_\phi+\sin t\cos\phi\:\partial_\rho,\\\label{AdS3KillingsAppendix}
\end{split}
\end{equation}
where the curved indices of the components ${\xi_i}^\mu$ can be lowered, as usual, with the global AdS$_3$ metric $g_{\mu\nu}$. These Killing vectors $\xi_i$, as discussed before, are the vielbeins $e_i$ of the PCM action eq.(\ref{PCMaction}) that construct the flat current eq.(\ref{MaurerCartanVielbeinAppendix}), namely
\begin{equation}
j^i_\pm={\xi^i}_\mu\;\partial_\pm X^\mu, \label{MaurerCartanKillingAppendix}
\end{equation}
from which the Lax connection in eq.(\ref{LaxPairVielbeinAppendix}) is built as
\begin{equation}
\mathcal{L}^i_\pm=\frac{j^i_\pm}{1\mp Z}. \label{LaxPairKillingAppendix}
\end{equation}
The flatness eq.(\ref{LaxFlatnessVielbeinAppendix}) of the PCM Lax connection results in two sets of equations, the first being the flatness eq.(\ref{MaurerFlatness}) of the Maurer-Cartan current, which is a structural fact as it can be easily checked by the reader. This is an identity to be expected, since this flatness equation can be realized as just the Cartan's first structure equation applied on Killing vectors.

The second set of equations are the equations of motion eq.(\ref{eomPCM}) of the PCM, the necessary condition for an integrable model.

If one desires to further validate all the above, all they have to do is to secure the fact that the equations of motion of the PCM coincide with the equations of motion of the bosonic string, on AdS$_3$.

For that purpose, we use the AdS$_3$ Killing vectors, eq.(\ref{AdS3KillingsAppendix}), on the PCM equations of motion , eq.(\ref{eomPCM}), that is
\begin{equation}
\partial_+\left({\xi^i}_\mu\;\partial_-X^\mu\right)+\partial_-\left({\xi^i}_\mu\;\partial_+ X^\mu\right)=0. \label{eomPCMkillingAppendix}
\end{equation}
In particular, $\xi_1=e_1$ (which lifts to a boost in the $Y^0-Y^3$ plane of $\mathbb{R}^{2,2}$) gives
\begin{equation}
\cosh\rho\;\partial_+\partial_-t=-\sinh\rho\left(\partial_+\rho\:\partial_-t+\partial_+t\:\partial_-\rho\right), \label{EOMt}
\end{equation}
which is the correct equation of motion for $t$, while $\xi_2=e_2$ (which lifts to a rotation in the $Y^1-Y^2$ plane of $\mathbb{R}^{2,2}$) gives
\begin{equation}
\sinh\rho\;\partial_+\partial_-\phi=-\cosh\rho\left(\partial_+\rho\:\partial_-\phi+\partial_+\phi\:\partial_-\rho\right), \label{EOMphi}
\end{equation}
which is the correct equation of motion for $\phi$. Last but not least, $\xi_6=e_6$ (which lifts to a rotation in the $Y^2-Y^3$ plane of $\mathbb{R}^{2,2}$), supplemented with the above equations for $t$ and $\phi$, gives
\begin{equation}
\partial_+\partial_-\rho=\cosh\rho\:\sinh\rho\left(\partial_+\phi\:\partial_-\phi-\partial_+t\:\partial_-t\right), \label{EOMrho}
\end{equation}
which, of course, is the correct equation of motion for $\rho$.\\
\\
In accordance with what we have discussed so far, the fact that it took just three of the six Killing vectors of AdS$_3$ to deduce the equations of motion is just another manifestation of the actual degrees freedom encoded in the Killing vectors.
%%%%%%%%%%%%%%%%%%%%%%%%%%%%%%%%%%%%%%%%%%%%%%%%%%%%%%%

\newpage
\section{Numerical Analysis of the String on the Integrable AdS$_7$ Background}\label{numericsD8smeared}
In this Appendix, we will complement our analytical study of the integrability of the string worldsheet on the background (\ref{eq:SinAlphaZGeometryMetric}-\ref{eq:SinAlphaZGeometryRR}) with a numerical analysis, following \cite{Basu:2011di}. Our numerical analysis indeed confirms the integrability of the string worldsheet on this background. This underlines the reliability of the numerical methods used in \cite{Nunez:2018ags}, where these same numerical methods were used to show that the dynamics of string solitons on the more general quiver solutions discussed in Appendix \ref{sec:StringSolitons} were non-integrable. 

Here we will analyse the dynamics of a string soliton wrapping around the $\xi$-direction, and moving along the $\chi$ and $z$-directions of the internal space $\mathcal{M}_3$. This amounts to studying the numerical evolution of the last two equations of motion in eqs. (\ref{eq:EulerLagrangeEquations}), setting $\rho = \varphi = \mu = 0$. We study the solution where the function $\alpha(z)$ is given by
\begin{equation}\label{eq:AlphaSinZNumerics}
	\alpha(z) = -81\pi^2 \left[ A \sin \left( \frac{\pi z}{4} \right) + B \sin \left( \frac{\pi z}{2} \right) \right],
\end{equation}
where we let $B$ range from 0 (for which the dynamics of the string is integrable) to 1.

We will show that the dynamics becomes increasingly more chaotic as $B$ deviates from 0. We will first study how the string moves through the $(z, \chi)$-plane. Note that the `energy' of the classical string soliton - the integration constant $E$, that has to be tuned to satisfy the Virasoro constraint (\ref{eq:VirasoroConstraint}) - is given by 
\begin{equation}\label{eq:EnergyConstraint}
E^2 = f_1(z)^2 \left( \dot{\rho}^2 + \sinh^2 \rho \left( \dot{\varphi}^2 + \mu^2 \sin^2 \varphi \right) \right) + f_2(z)\dot{z}^2 + f_3(z)\left(\dot{\chi}^2 + \kappa^2 \sin^2 \chi \right).
\end{equation}
This energy minimises for the point $\rho = \varphi = \chi = 0$. %We have seen in Appendix \ref{sec:StringSolitons} that if we move the string away from the minima $\rho = \varphi = 0$ it will oscillate around it like an harmonic oscillator.
Here we will numerically study the dynamics of the string when we increase $\chi$ away from the stable point $\chi = 0$ and increase its energy. 

We show plots of various observables. The reader should compare them with the figures displayed in Section 4 of \cite{Nunez:2018ags}, where an analysis of strings on generic (non-integrable) backgrounds was performed.

In figure \ref{fig_lowE_chi0001_path} we see that if we start with an initial value that is very close to the poles of the 2-sphere ($\chi = 0$ and $\chi = \pi$) for the integrable background with $B = 0$, the string oscillates around this minimum. Every time the string hits the endpoints on the $z$-domain it flips to the other pole on the 2-sphere (indicated by the dashed grey lines) and moves back along the $z$-direction. As we increase the energy and allow the string to move further away from the poles, it starts to moving freely around the entire 2-sphere, see figure \ref{fig_lowE_chi01_path}. Note that even at high energies, the motion of the string remains quasi-periodic.
\begin{figure}[H]\label{fig:ChiZTrajectories}
{
 \centering
 \subfloat[\small $\chi(0) = 0.1$, $E \approx 6.58$, $t_{max} = 400$, $B=0$ \normalsize]{
   \label{fig_lowE_chi0001_path}
     \includegraphics[width=0.5\textwidth]{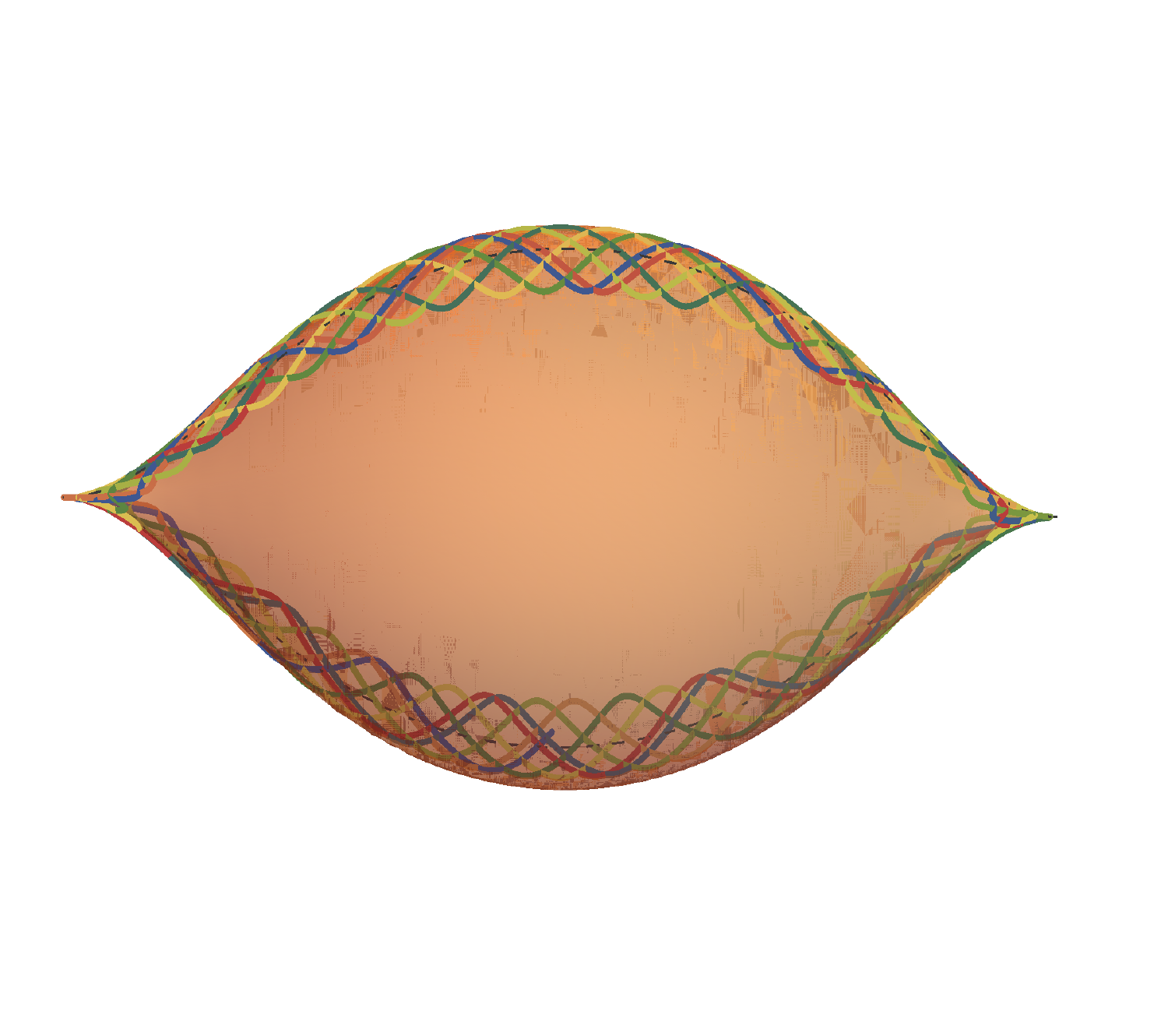}}
 \subfloat[\small $\chi(0) = 0.9$, $E \approx 43.33$, $t_{max} = 400$, $B=0$ \normalsize]{
   \label{fig_lowE_chi01_path}
    \includegraphics[width=0.5\textwidth]{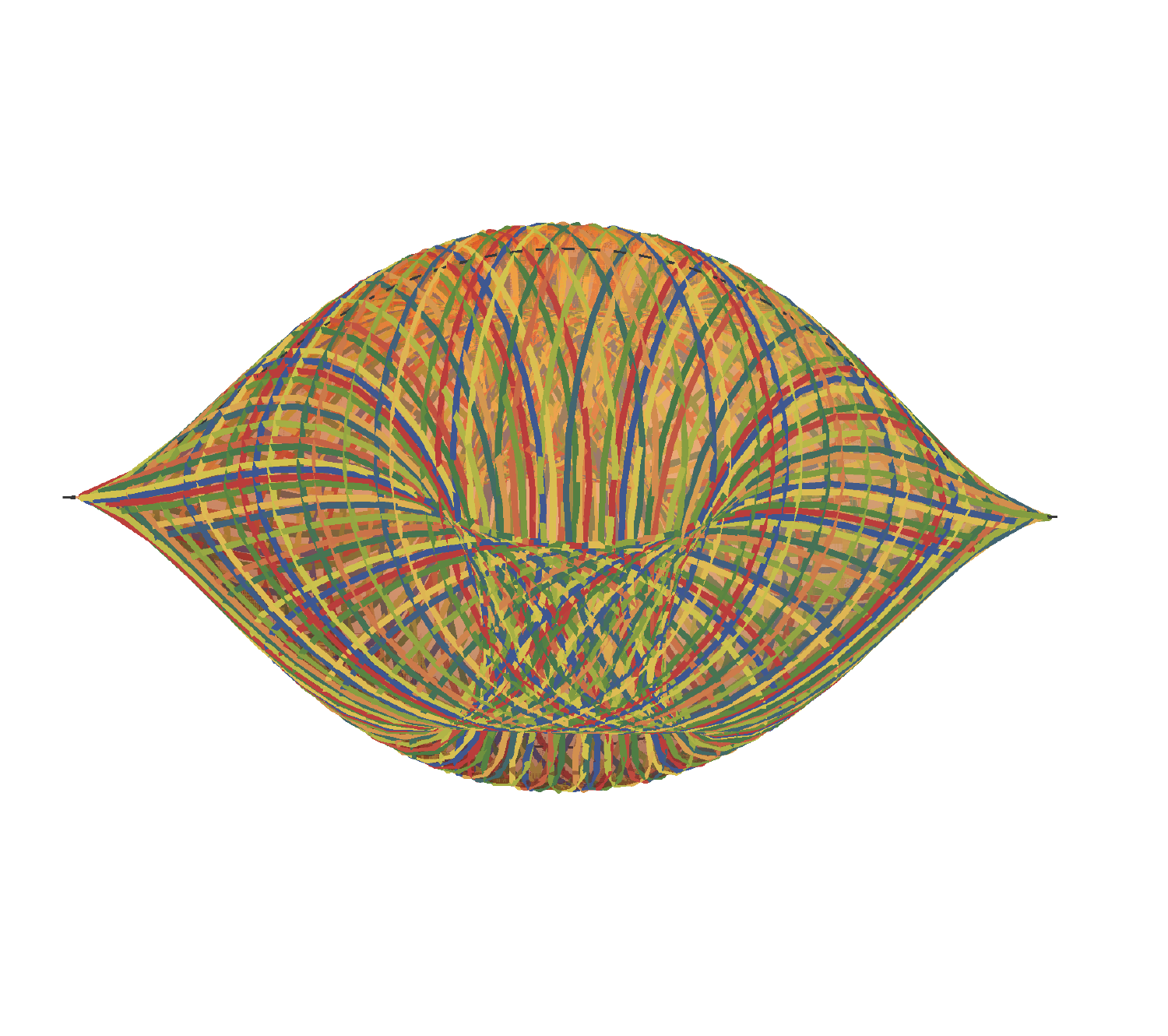}}\\    
 \centering
 \subfloat[\small $\chi(0) = 0.1$, $E \approx 7.19$, $t_{max} = 150$, $B=0.2$ \normalsize]{
   \label{fig_lowE_chi025_path}
     \includegraphics[width=0.5\textwidth]{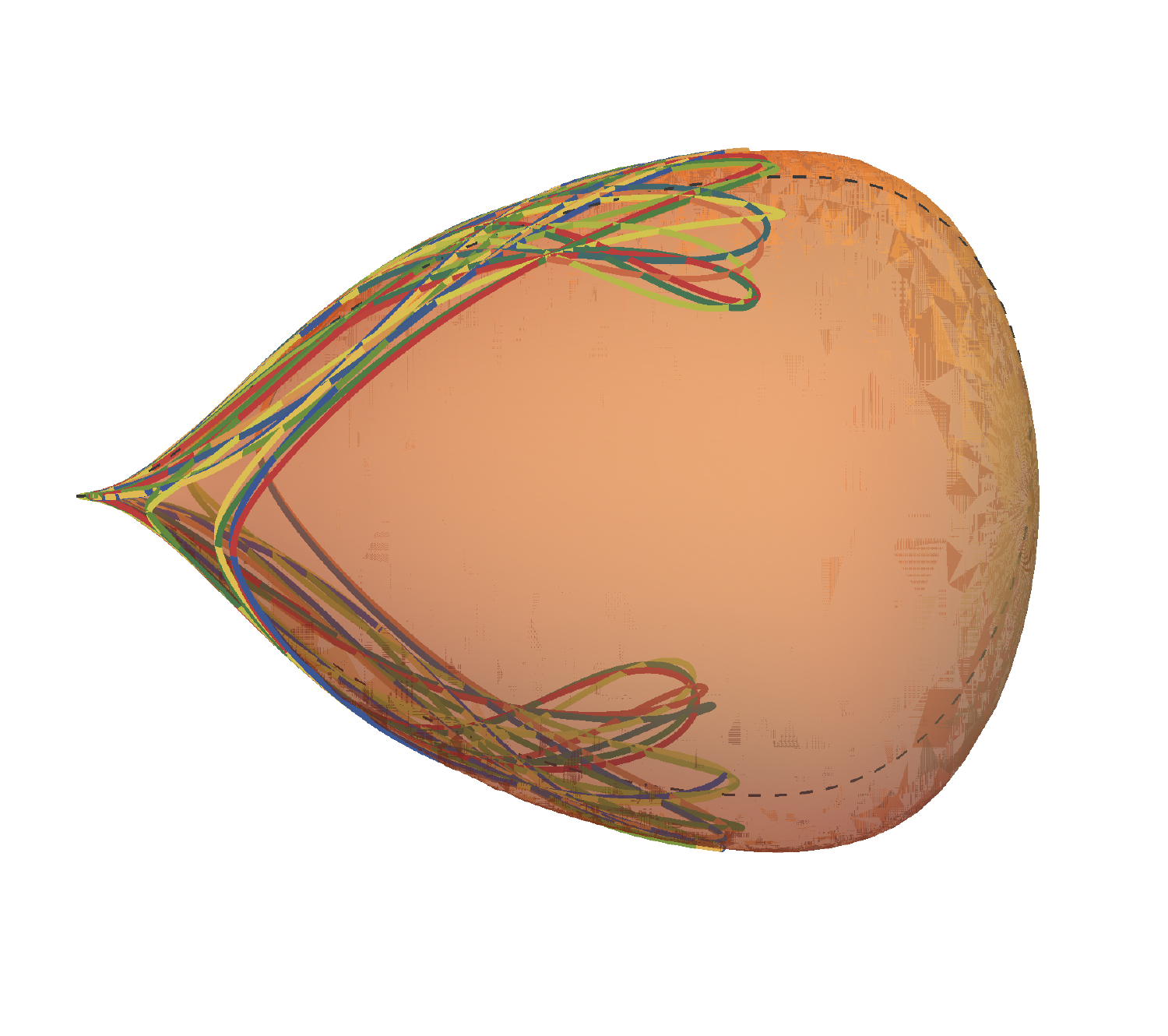}}
 \subfloat[\small $\chi(0) = 0.9$, $E \approx 48.95$, $t_{max} = 250$, $B=0.2$ \normalsize]{
   \label{fig_lowE_chi05_path}
    \includegraphics[width=0.5\textwidth]{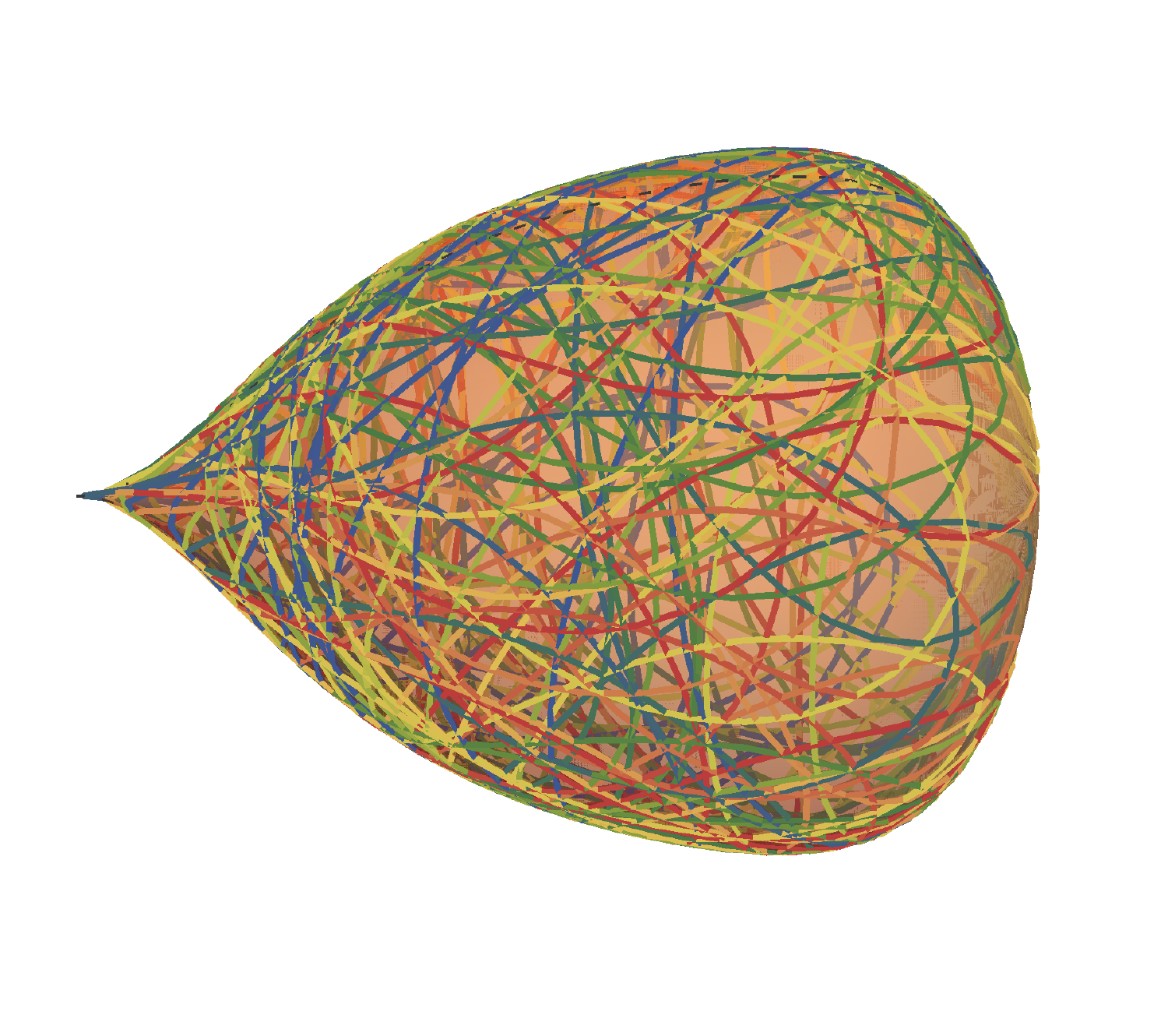}}
    \caption{Trajectories of the string on the internal space $\mathcal{M}^3$ for low and high energies (from left to right), for the embedding in eq. (\ref{eq:StringEmbedding}) with $\rho = \varphi = \lambda = 0$. The two images at the top correspond to the integrable background in eqs.(\ref{eq:SinAlphaZGeometryMetric}-\ref{eq:SinAlphaZGeometryRR}) with $B = 0$ in eq. (\ref{eq:AlphaSinZNumerics}), for those on the bottom $B = 0.2$. For the integrable background (in the top two images) the trajectories of the string soliton remain regular, even at high energies. We choose initial conditions $p_\chi(0) = 0, z(0) = 2, p_z(0) = 1$. The orange surface corresponds to the angle $\chi$ fibred over the $z$-interval with the warp-factor $f_3(z)$. The dashed line indicates the points where $\chi = 0, \pi$. Points on opposite sides of this line should be identified for fixed values of $z$ as $\chi \in [0, \pi]$.}\label{fig_lowE_path}
}
\end{figure}
As we go away from $B = 0$ in eq.(\ref{eq:AlphaSinZNumerics}) we are no longer considering the integrable background from eqs.(\ref{eq:SinAlphaZGeometryMetric}-\ref{eq:SinAlphaZGeometryRR}). The asymmetry along the $z$-direction of this background makes it harder for the string to probe the right side of the space. In figure \ref{fig_lowE_chi025_path} we see that again for small energies the string oscillates around the poles. Though it's motion around these poles does still look quite regular and quasiperiodic, it appears somewhat more disorderly that what we observed earlier for a string with roughly the same energy, moving on the integrable background. We see in figure \ref{fig_lowE_chi05_path} that as we now increase the energy for the string on the non-integrable background, its motion becomes chaotic.\\
\\
\textbf{Lyapunov Exponents} - To verify our intuition - that the trajectories for $B=0$ look regular while those for $B \neq0$ look chaotic, - we obtain the Lyapunov exponents corresponding to our initial conditions. One other typical characteristic of an integrable classical mechanical system is a vanishing Lyapunov exponent. The Lyapunov exponent is a measure of the sensitivity of the system to its initial conditions. Typically for a chaotic system, two nearby initial points will diverge during the dynamical evolution as
\begin{equation}
	\Delta F(x_i, p_i, t) \sim \Delta F(x_i, p_i, 0) e^{\lambda^i x_i + \tilde{\lambda}^i p_i},
\end{equation}
where the $\lambda$ are the Lyapunov exponents associated with the position directions in phase-space, and the $\tilde{\lambda}$ those associated with the momentum directions. Since our string soliton is a Hamiltonian system the initial volume in phase space will be conserved (as a consequence of Liouville's theorem). The Lyapunov exponents will therefore satisfy the additional constraint that their sum is equal to zero. The largest Lyapunov exponent (LLE) is typically used as an indicator to tell us how chaotic the dynamics of the system is.

We numerically estimate these largest Lyapunov exponents for the same low and high energy (left to right) initial conditions that we considered in figure \ref{fig_lowE_path}. The result is shown in figure \ref{fig_Lyapunov}. We see in figure \ref{LCE_massless} that the low energy dynamics for the string are indeed both not very chaotic, the LLE for the integrable background ($B=0$, in red) should asymptote to zero (with possibly some numerical noise making it slightly larger). The LLE for the non-integrable background ($B=0.2$, in blue) is a bit larger, telling us the low energy string on the non-integrable background is slightly more chaotic. This agrees with what we see in figure \ref{fig_lowE_chi0001_path} and \ref{fig_lowE_chi025_path}.

We see in figure \ref{LCE_quiver_1} that the dynamics for the high energy string still has an LLE of almost zero on the integrable background ($B=0$, in red), thus numerically confirming the absence of chaos for this case. On the non-integrable background ($B=0.2$, in blue) the value clearly asymptotes to a non-zero value $\lambda \approx 0.01$, larger than we saw for the low-energy string on the non-integrable background, clearly confirming its dynamics is chaotic. This agrees with what we see in figure \ref{fig_lowE_chi01_path} and \ref{fig_lowE_chi05_path}.
\begin{figure}[h!]
{
 \centering
 \subfloat[\small  LLE for $\chi(0) = 0.1$. \normalsize]{
   \label{LCE_massless}
     \includegraphics[width=0.5\textwidth]{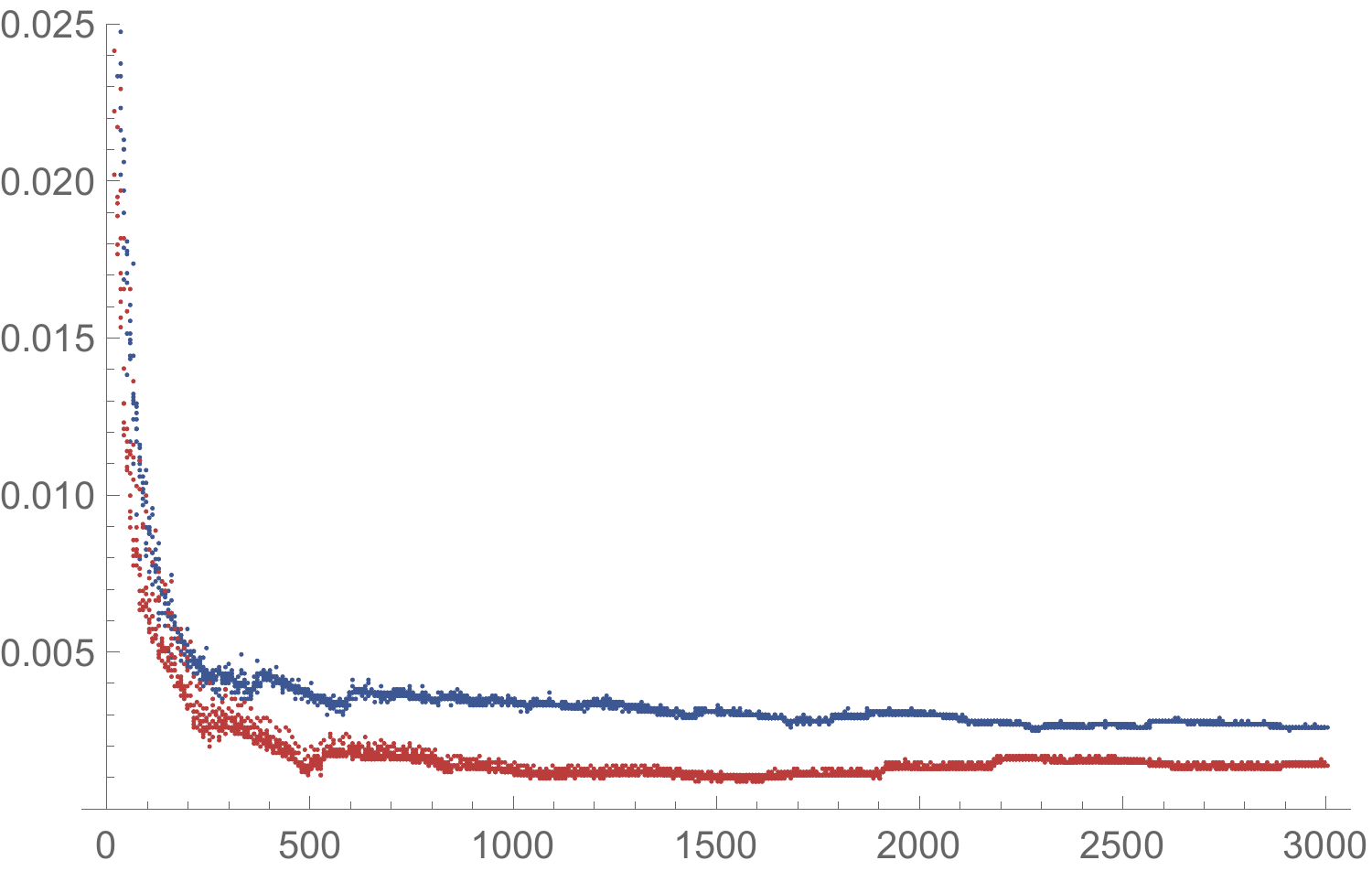}}
 \subfloat[\small LLE for $\chi(0) = 0.9$.\normalsize]{
   \label{LCE_quiver_1}
    \includegraphics[width=0.5\textwidth]{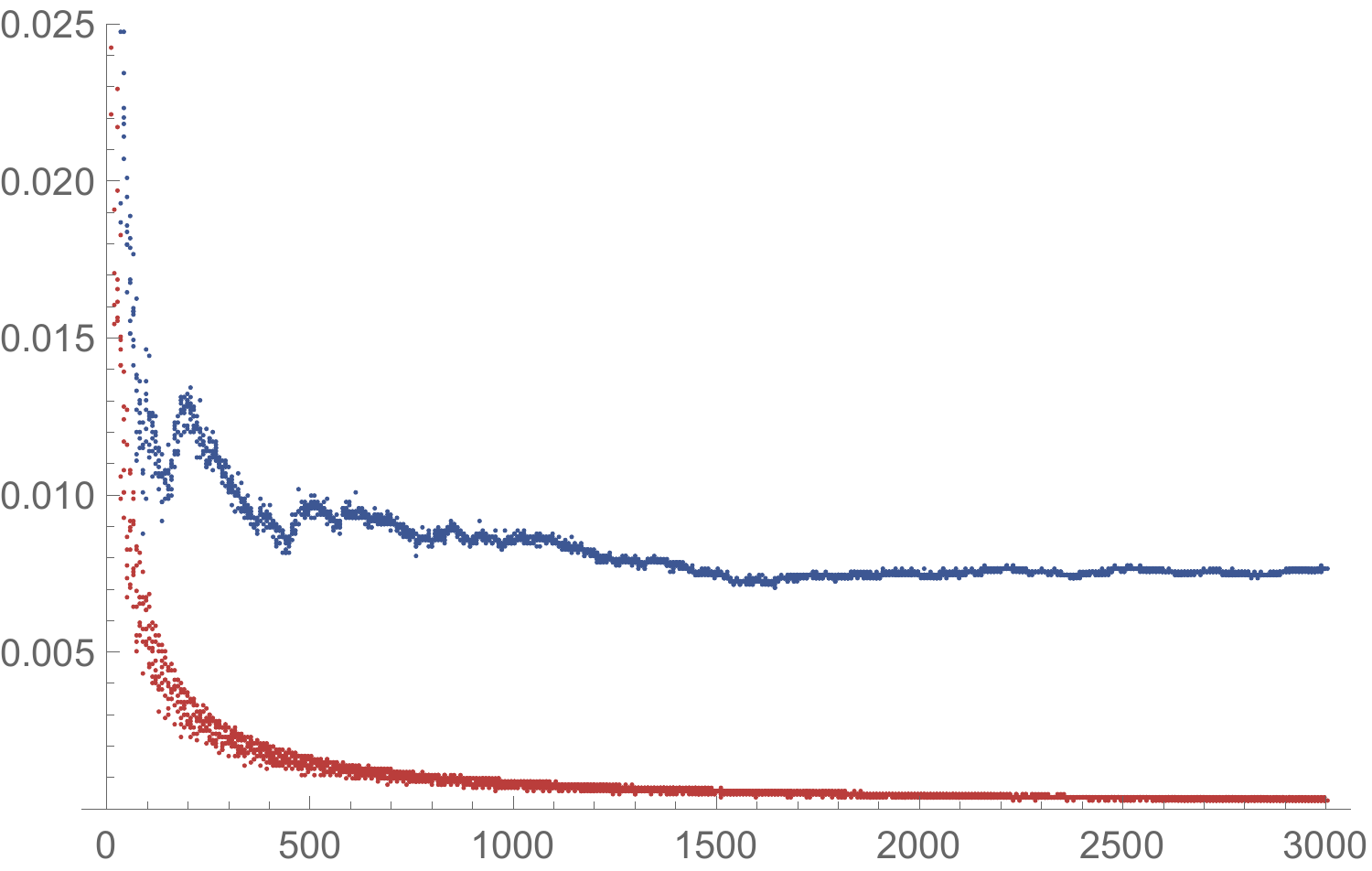}}\\  
\caption{Lyapunov exponents for low- (left) and high-energy (right) string configurations, using the same initial conditions as in figure \ref{fig_lowE_path}. We consider both the integrable background ($B=0$, in red) and on a non-integrable background ($B=0.2$, in blue). We find that the integrable background has a Lyapunov exponent that asymptotes to zero (indicating the absence of chaotic behaviour), while the LLE for the non-integrable background asymptotes to a finite value, indicating increasingly chaotic dynamics for higher energies. This agrees with what we see in figure \ref{fig_lowE_path}.}\label{fig_Lyapunov}
}
\end{figure}\\
\\\newpage
\noindent\textbf{Poincar\'{e} sections} - Another numerical tool we can use to examine if we are indeed dealing with an integrable system - where the dynamics is quasi-periodic - is to plot a Poincar\'{e} section. For this, we choose different initial conditions in the $(\chi, p_z)$-plane, that all correspond to the same energy (\ref{eq:EnergyConstraint}). We then run the numerical evolution for all these initial points and monitor the $(z, p_z)$-plane every time the trajectories pass through the point $\chi(t) = 0$. 

If the dynamics of the string soliton we are studying is integrable, this classical mechanical system with $2\times2$ degrees of freedom would have $2$ independent integrals of motion that are in involution (meaning their Poisson bracket vanishes). The trajectories of this system would then be confined to the surfaces of a series of embedded 2-dimensional KAM tori in the $(z, p_z, \chi, p_\chi)$ phase-space. We see this is exactly the case in figure \ref{PoincareB0} where we consider the integrable background ($B = 0$). As we increase the value of $B$ we lose the integrability of the dynamical system. This onset of chaos can clearly be seen in figure \ref{PoincareB0025}-\ref{PoincareB0065}, as more and more KAM tori break apart when we increase $B$ until there is no structure left and we have a purely chaotic system
\begin{figure}[h!]
{
 \centering
 \subfloat[\small  $B=0$ \normalsize]{
   \label{PoincareB0}
     \includegraphics[width=0.5\textwidth]{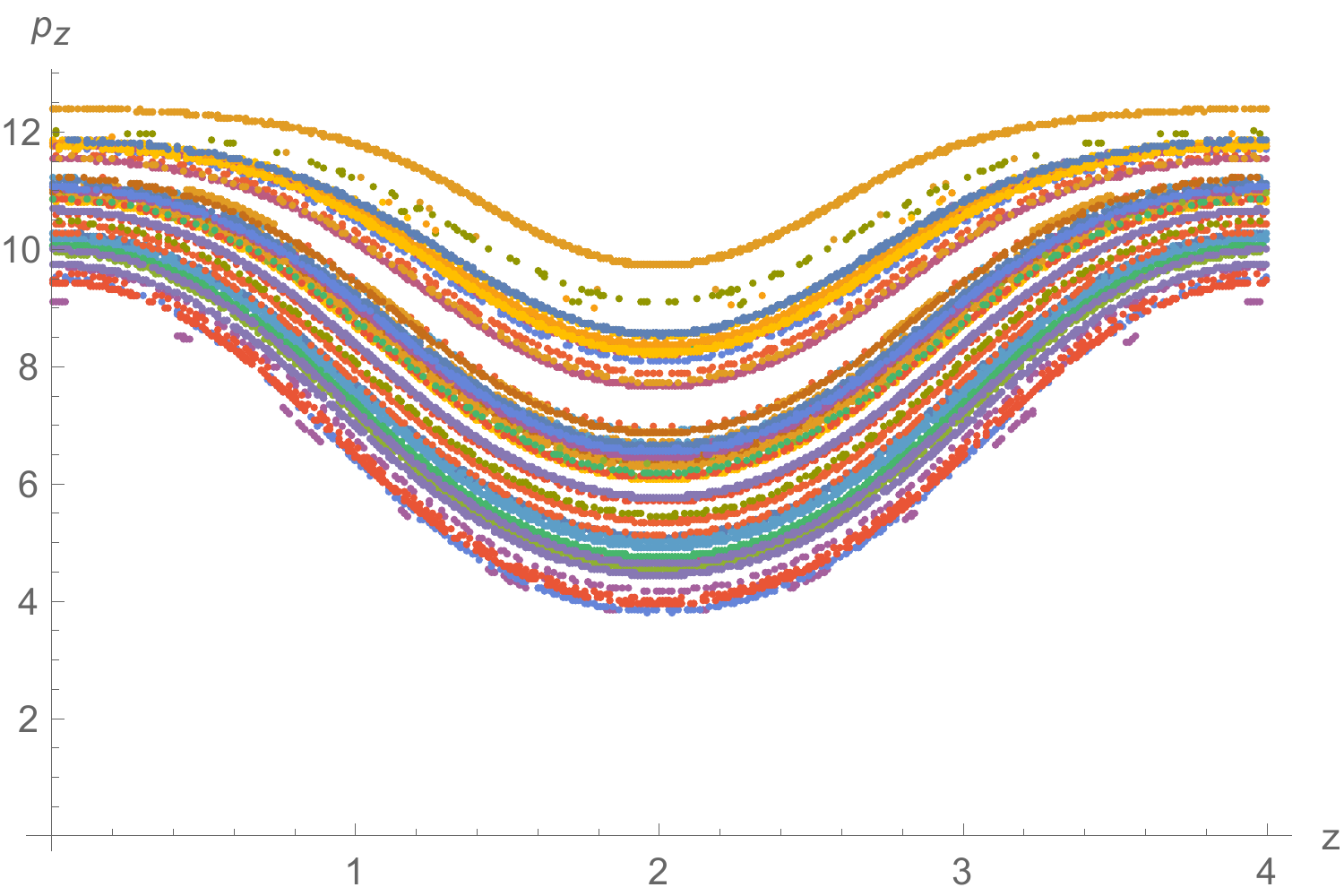}}
 \subfloat[\small  $B=0.025$ \normalsize]{
   \label{PoincareB0025}
    \includegraphics[width=0.5\textwidth]{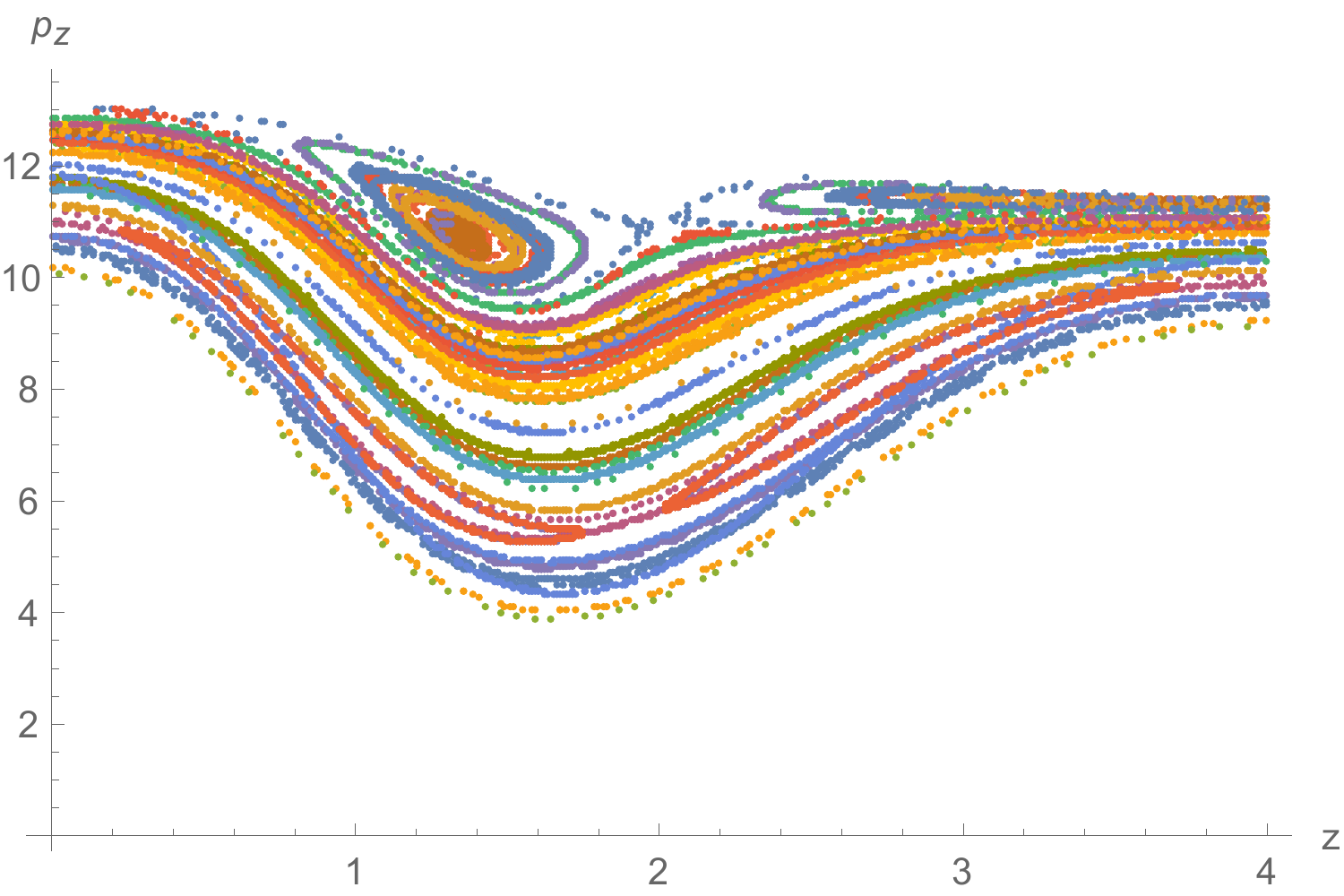}}\\
\subfloat[\small  $B=0.050$ \normalsize]{
   \label{PoincareB0050}
    \includegraphics[width=0.5\textwidth]{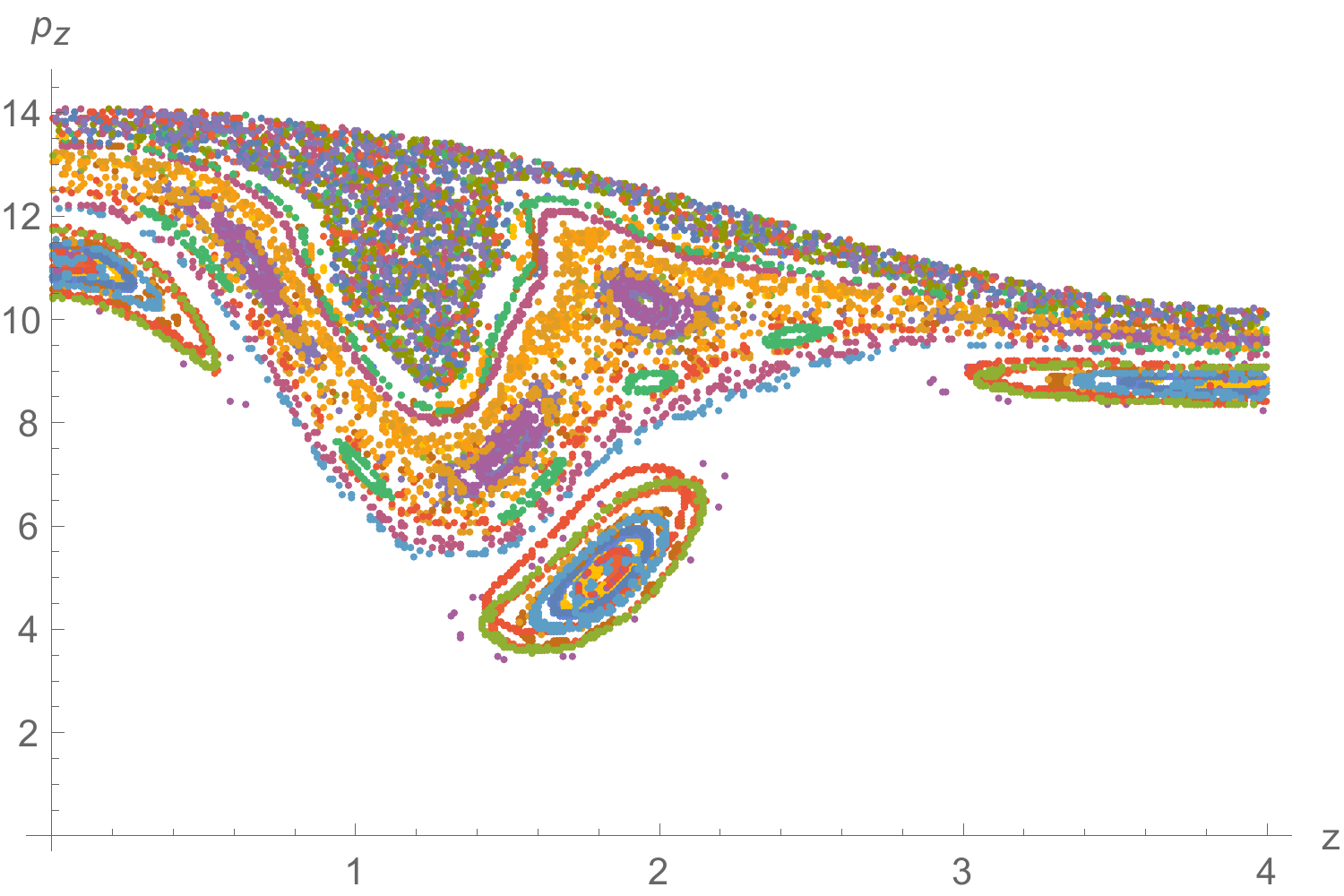}}
 \subfloat[\small $B=0.065$ \normalsize]{
   \label{PoincareB0065}
    \includegraphics[width=0.5\textwidth]{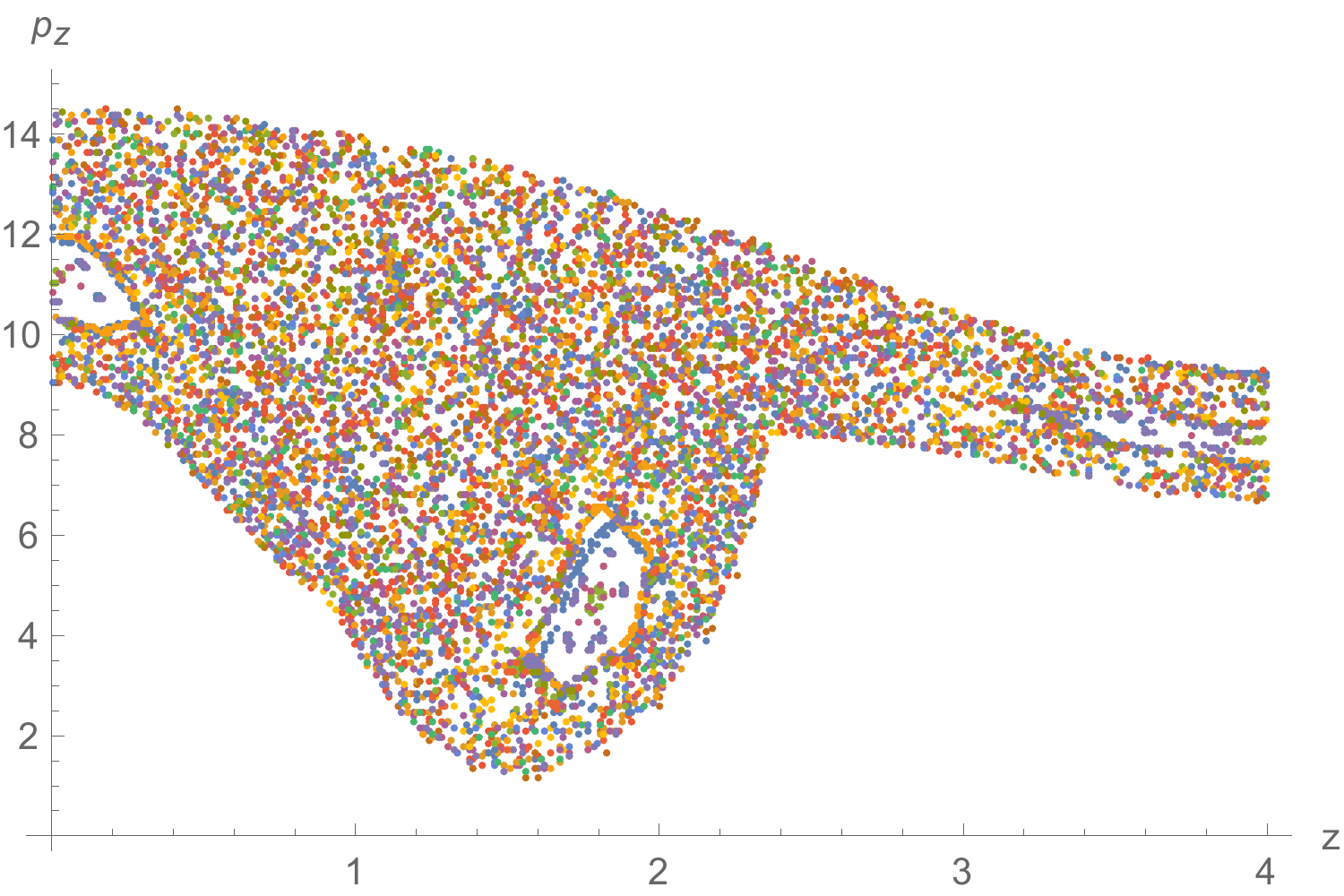}}\\  
\caption{Poincar\'{e} sections for $(z, p_z)$-plane at $\chi(t)=0$, for high energy string configurations ($E = 45$) on backgrounds with values of $B$. As we increase $B$, we clearly see the onset of chaos as more and more KAM tori break apart until there is no structure left and we have a purely chaotic system.}\label{fig_PoincarePlots}
}
\end{figure}\\
\\


\newpage
\begin{thebibliography}{99}

%\cite{Maldacena:1997re}
\bibitem{Maldacena:1997re} 
  J.~M.~Maldacena,
  %``The Large N limit of superconformal field theories and supergravity,''
  Int.\ J.\ Theor.\ Phys.\  {\bf 38}, 1113 (1999)
  [Adv.\ Theor.\ Math.\ Phys.\  {\bf 2}, 231 (1998)]
%  doi:10.1023/A:1026654312961, 10.4310/ATMP.1998.v2.n2.a1
  [hep-th/9711200].
  %%CITATION = doi:10.1023/A:1026654312961, 10.4310/ATMP.1998.v2.n2.a1;%%
  %14289 citations counted in INSPIRE as of 07 Jan 2019



%\cite{Witten:1995ex}
\bibitem{Witten:1995ex} 
  E.~Witten,
  %``String theory dynamics in various dimensions,''
  Nucl.\ Phys.\ B {\bf 443}, 85 (1995)
 % doi:10.1016/0550-3213(95)00158-O
  [hep-th/9503124].
  %%CITATION = doi:10.1016/0550-3213(95)00158-O;%%
  %2255 citations counted in INSPIRE as of 07 Jan 2019


%\cite{Seiberg:1996vs}
\bibitem{Seiberg:1996vs} 
  N.~Seiberg and E.~Witten,
  %``Comments on string dynamics in six-dimensions,''
  Nucl.\ Phys.\ B {\bf 471}, 121 (1996)
%  doi:10.1016/0550-3213(96)00189-7
  [hep-th/9603003].
  %%CITATION = doi:10.1016/0550-3213(96)00189-7;%%
  %392 citations counted in INSPIRE as of 07 Jan 2019

%\cite{Seiberg:1996vs}\cite{Blum:1997mm}
\bibitem{Blum:1997mm} 
  J.~D.~Blum and K.~A.~Intriligator,
  %``New phases of string theory and 6-D RG fixed points via branes at orbifold singularities,''
  Nucl.\ Phys.\ B {\bf 506}, 199 (1997)
%  doi:10.1016/S0550-3213(97)00449-5
  [hep-th/9705044].
  %%CITATION = doi:10.1016/S0550-3213(97)00449-5;%%
  %129 citations counted in INSPIRE as of 07 Jan 2019






%\cite{Seiberg:1996qx}
\bibitem{Seiberg:1996qx} 
  N.~Seiberg,
  %``Nontrivial fixed points of the renormalization group in six-dimensions,''
  Phys.\ Lett.\ B {\bf 390}, 169 (1997)
%  doi:10.1016/S0370-2693(96)01424-4
  [hep-th/9609161].
  %%CITATION = doi:10.1016/S0370-2693(96)01424-4;%%
  %165 citations counted in INSPIRE as of 22 Dec 2018
%\cite{Danielsson:1997kt}
\bibitem{Danielsson:1997kt} 
  U.~H.~Danielsson, G.~Ferretti, J.~Kalkkinen and P.~Stjernberg,
  %``Notes on supersymmetric gauge theories in five-dimensions and six-dimensions,''
  Phys.\ Lett.\ B {\bf 405}, 265 (1997)
 % doi:10.1016/S0370-2693(97)00645-X
  [hep-th/9703098].
  %%CITATION = doi:10.1016/S0370-2693(97)00645-X;%%
  %40 citations counted in INSPIRE as of 18 Jan 2019
%\cite{Hanany:1997gh}
\bibitem{Hanany:1997gh} 
  A.~Hanany and A.~Zaffaroni,
  %``Branes and six-dimensional supersymmetric theories,''
  Nucl.\ Phys.\ B {\bf 529}, 180 (1998)
 % doi:10.1016/S0550-3213(98)00355-1
  [hep-th/9712145].
  %%CITATION = doi:10.1016/S0550-3213(98)00355-1;%%
  %127 citations counted in INSPIRE as of 22 Dec 2018%\cite{Brunner:1997gf}
\bibitem{Brunner:1997gf} 
  I.~Brunner and A.~Karch,
  %``Branes at orbifolds versus Hanany Witten in six-dimensions,''
  JHEP {\bf 9803}, 003 (1998)
%  doi:10.1088/1126-6708/1998/03/003
  [hep-th/9712143].
  %%CITATION = doi:10.1088/1126-6708/1998/03/003;%%
  %117 citations counted in INSPIRE as of 22 Dec 2018
  
  %\cite{Ohmori:2014pca}
\bibitem{Ohmori:2014pca} 
  K.~Ohmori, H.~Shimizu and Y.~Tachikawa,
  %``Anomaly polynomial of E-string theories,''
  JHEP {\bf 1408}, 002 (2014)
 % doi:10.1007/JHEP08(2014)002
  [arXiv:1404.3887 [hep-th]].
  %%CITATION = doi:10.1007/JHEP08(2014)002;%%
  %39 citations counted in INSPIRE as of 07 Jan 2019
  
  %\cite{DelZotto:2014hpa}
\bibitem{DelZotto:2014hpa} 
  M.~Del Zotto, J.~J.~Heckman, A.~Tomasiello and C.~Vafa,
  %``6d Conformal Matter,''
  JHEP {\bf 1502}, 054 (2015)
%  doi:10.1007/JHEP02(2015)054
  [arXiv:1407.6359 [hep-th]].
  %%CITATION = doi:10.1007/JHEP02(2015)054;%%
  %136 citations counted in INSPIRE as of 07 Jan 2019
  %\cite{Cordova:2015fha}
\bibitem{Cordova:2015fha} 
  C.~Cordova, T.~T.~Dumitrescu and K.~Intriligator,
  %``Anomalies, renormalization group flows, and the a-theorem in six-dimensional (1, 0) theories,''
  JHEP {\bf 1610}, 080 (2016)
%  doi:10.1007/JHEP10(2016)080
  [arXiv:1506.03807 [hep-th]].
  %%CITATION = doi:10.1007/JHEP10(2016)080;%%
  %57 citations counted in INSPIRE as of 07 Jan 2019
  %\cite{Intriligator:2014eaa}
\bibitem{Intriligator:2014eaa} 
  K.~Intriligator,
  %``6d, $ \mathcal{N}=\left(1,\;0\right) $ Coulomb branch anomaly matching,''
  JHEP {\bf 1410}, 162 (2014)
  %doi:10.1007/JHEP10(2014)162
  [arXiv:1408.6745 [hep-th]].
  %%CITATION = doi:10.1007/JHEP10(2014)162;%%
  %45 citations counted in INSPIRE as of 18 Jan 2019
  %\cite{Ohmori:2014kda}
\bibitem{Ohmori:2014kda} 
  K.~Ohmori, H.~Shimizu, Y.~Tachikawa and K.~Yonekura,
  %``Anomaly polynomial of general 6d SCFTs,''
  PTEP {\bf 2014}, no. 10, 103B07 (2014)
%  doi:10.1093/ptep/ptu140
  [arXiv:1408.5572 [hep-th]].
  %%CITATION = doi:10.1093/ptep/ptu140;%%
  %89 citations counted in INSPIRE as of 18 Jan 2019
  
  %\cite{Beccaria:2015ypa}
\bibitem{Beccaria:2015ypa} 
  M.~Beccaria and A.~A.~Tseytlin,
  %``Conformal anomaly c-coefficients of superconformal 6d theories,''
  JHEP {\bf 1601}, 001 (2016)
%  doi:10.1007/JHEP01(2016)001
  [arXiv:1510.02685 [hep-th]].
  %%CITATION = doi:10.1007/JHEP01(2016)001;%%
  %26 citations counted in INSPIRE as of 18 Jan 2019
  %\cite{Bhardwaj:2015xxa}
\bibitem{Bhardwaj:2015xxa} 
  L.~Bhardwaj,
  %``Classification of 6d $ \mathcal{N}=\left(1,0\right) $ gauge theories,''
  JHEP {\bf 1511}, 002 (2015)
%  doi:10.1007/JHEP11(2015)002
  [arXiv:1502.06594 [hep-th]].
  %%CITATION = doi:10.1007/JHEP11(2015)002;%%
  %63 citations counted in INSPIRE as of 18 Jan 2019
  %\cite{Bandos:2013jva}
\bibitem{Bandos:2013jva} 
  I.~Bandos, H.~Samtleben and D.~Sorokin,
  %``Duality-symmetric actions for non-Abelian tensor fields,''
  Phys.\ Rev.\ D {\bf 88}, no. 2, 025024 (2013)
  %doi:10.1103/PhysRevD.88.025024
  [arXiv:1305.1304 [hep-th]].
  %%CITATION = doi:10.1103/PhysRevD.88.025024;%%
  %22 citations counted in INSPIRE as of 18 Jan 2019
  %\cite{Chang:2018xmx}
\bibitem{Chang:2018xmx} 
  C.~M.~Chang,
  %``5d and 6d SCFTs Have No Weak Coupling Limit,''
  arXiv:1810.04169 [hep-th].
  %%CITATION = ARXIV:1810.04169;%%
  %2 citations counted in INSPIRE as of 18 Jan 2019
  
%\cite{Gaiotto:2014lca}
\bibitem{Gaiotto:2014lca} 
  D.~Gaiotto and A.~Tomasiello,
  %``Holography for (1,0) theories in six dimensions,''
  JHEP {\bf 1412}, 003 (2014)
%  doi:10.1007/JHEP12(2014)003
  [arXiv:1404.0711 [hep-th]].
  %%CITATION = doi:10.1007/JHEP12(2014)003;%%
  %80 citations counted in INSPIRE as of 22 Dec 2018

%\cite{Apruzzi:2014qva}
\bibitem{Apruzzi:2014qva} 
  F.~Apruzzi, M.~Fazzi, A.~Passias, D.~Rosa and A.~Tomasiello,
  %``AdS$_{6}$ solutions of type II supergravity,''
  JHEP {\bf 1411}, 099 (2014)
  Erratum: [JHEP {\bf 1505}, 012 (2015)]
%  doi:10.1007/JHEP11(2014)099, 10.1007/JHEP05(2015)012
  [arXiv:1406.0852 [hep-th]].
  %%CITATION = doi:10.1007/JHEP11(2014)099, 10.1007/JHEP05(2015)012;%%
  %62 citations counted in INSPIRE as of 22 Dec 2018
  
  %\cite{Apruzzi:2015wna}
\bibitem{Apruzzi:2015wna} 
  F.~Apruzzi, M.~Fazzi, A.~Passias, A.~Rota and A.~Tomasiello,
  %``Six-Dimensional Superconformal Theories and their Compactifications from Type IIA Supergravity,''
  Phys.\ Rev.\ Lett.\  {\bf 115}, no. 6, 061601 (2015)
 % doi:10.1103/PhysRevLett.115.061601
  [arXiv:1502.06616 [hep-th]].
  %%CITATION = doi:10.1103/PhysRevLett.115.061601;%%
  %38 citations counted in INSPIRE as of 22 Dec 2018
  %\cite{Passias:2015gya}
\bibitem{Passias:2015gya} 
  A.~Passias, A.~Rota and A.~Tomasiello,
  %``Universal consistent truncation for 6d/7d gauge/gravity duals,''
  JHEP {\bf 1510}, 187 (2015)
%  doi:10.1007/JHEP10(2015)187
  [arXiv:1506.05462 [hep-th]].
  %%CITATION = doi:10.1007/JHEP10(2015)187;%%
  %29 citations counted in INSPIRE as of 22 Dec 2018

%\cite{Apruzzi:2017nck}
\bibitem{Apruzzi:2017nck} 
  F.~Apruzzi and M.~Fazzi,
  %``AdS$_{7}$/CFT$_{6}$ with orientifolds,''
  JHEP {\bf 1801}, 124 (2018)
 % doi:10.1007/JHEP01(2018)124
  [arXiv:1712.03235 [hep-th]].
  %%CITATION = doi:10.1007/JHEP01(2018)124;%%
  %9 citations counted in INSPIRE as of 07 Jan 2019

%\cite{Macpherson:2016xwk}
\bibitem{Macpherson:2016xwk} 
  N.~T.~Macpherson and A.~Tomasiello,
  %``Minimal flux Minkowski classification,''
  JHEP {\bf 1709}, 126 (2017)
  doi:10.1007/JHEP09(2017)126
  [arXiv:1612.06885 [hep-th]].
  %%CITATION = doi:10.1007/JHEP09(2017)126;%%
  %14 citations counted in INSPIRE as of 28 Mar 2019

%\cite{Bobev:2016phc}
\bibitem{Bobev:2016phc} 
  N.~Bobev, G.~Dibitetto, F.~F.~Gautason and B.~Truijen,
  %``Holography, Brane Intersections and Six-dimensional SCFTs,''
  JHEP {\bf 1702}, 116 (2017)
  doi:10.1007/JHEP02(2017)116
  [arXiv:1612.06324 [hep-th]].
  %%CITATION = doi:10.1007/JHEP02(2017)116;%%
  %10 citations counted in INSPIRE as of 18 Jan 2019


%\cite{Cremonesi:2015bld}
\bibitem{Cremonesi:2015bld} 
  S.~Cremonesi and A.~Tomasiello,
  %``6d holographic anomaly match as a continuum limit,''
  JHEP {\bf 1605}, 031 (2016)
%  doi:10.1007/JHEP05(2016)031
  [arXiv:1512.02225 [hep-th]].
  %%CITATION = doi:10.1007/JHEP05(2016)031;%%
  %30 citations counted in INSPIRE as of 22 Dec 2018

%\cite{Hanany:1996ie}
\bibitem{Hanany:1996ie} 
  A.~Hanany and E.~Witten,
  %``Type IIB superstrings, BPS monopoles, and three-dimensional gauge dynamics,''
  Nucl.\ Phys.\ B {\bf 492}, 152 (1997)
%  doi:10.1016/S0550-3213(97)00157-0, 10.1016/S0550-3213(97)80030-2
  [hep-th/9611230].
  %%CITATION = doi:10.1016/S0550-3213(97)00157-0, 10.1016/S0550-3213(97)80030-2;%%
  %1050 citations counted in INSPIRE as of 22 Dec 2018




%\cite{Nahm:1977tg}
\bibitem{Nahm:1977tg} 
  W.~Nahm,
  %``Supersymmetries and their Representations,''
  Nucl.\ Phys.\ B {\bf 135}, 149 (1978).
 % doi:10.1016/0550-3213(78)90218-3
  %%CITATION = doi:10.1016/0550-3213(78)90218-3;%%
  %682 citations counted in INSPIRE as of 27 Dec 2018%\cite{Minwalla:1997ka}
%\bibitem{Minwalla:1997ka} 
  S.~Minwalla,
  %``Restrictions imposed by superconformal invariance on quantum field theories,''
  Adv.\ Theor.\ Math.\ Phys.\  {\bf 2}, 783 (1998)
  %doi:10.4310/ATMP.1998.v2.n4.a4
  [hep-th/9712074].
  %%CITATION = doi:10.4310/ATMP.1998.v2.n4.a4;%%
  %292 citations counted in INSPIRE as of 27 Dec 2018














%\cite{Nunez:2019gbg}
\bibitem{Nunez:2019gbg} 
  C.~Nunez, D.~Roychowdhury, S.~Speziali and S.~Zacarias,
  %``Holographic Aspects of Four Dimensional ${\cal N }=2$ SCFTs and their Marginal Deformations,''
  arXiv:1901.02888 [hep-th].
  %%CITATION = ARXIV:1901.02888;%%











%\cite{Nunez:2018ags}
\bibitem{Nunez:2018ags} 
  C.~Nunez, J.~M.~Penin, D.~Roychowdhury and J.~van Gorsel,
  %``The non-Integrability of Strings in Massive Type IIA and their Holographic duals,''
  JHEP {\bf 1806}, 078 (2018)
%  doi:10.1007/JHEP06(2018)078
  [arXiv:1802.04269 [hep-th]].
  %%CITATION = doi:10.1007/JHEP06(2018)078;%%
  %2 citations counted in INSPIRE as of 21 Oct 2018
  
  
  
  %\cite{Macpherson:2014eza}
\bibitem{Macpherson:2014eza} 
%\bibitem{Kol:2014nqa} 
  U.~Kol, C.~Nunez, D.~Schofield, J.~Sonnenschein and M.~Warschawski,
  %``Confinement, Phase Transitions and non-Locality in the Entanglement Entropy,''
  JHEP {\bf 1406}, 005 (2014)
%  doi:10.1007/JHEP06(2014)005
  [arXiv:1403.2721 [hep-th]].
  %%CITATION = doi:10.1007/JHEP06(2014)005;%%
  %30 citations counted in INSPIRE as of 06 Jan 2019
  N.~T.~Macpherson, C.~Nunez, L.~A.~Pando Zayas, V.~G.~J.~Rodgers and C.~A.~Whiting,
  %``Type IIB supergravity solutions with AdS$_{5}$ from Abelian and non-Abelian T dualities,''
  JHEP {\bf 1502}, 040 (2015)
 % doi:10.1007/JHEP02(2015)040
  [arXiv:1410.2650 [hep-th]].
  %%CITATION = doi:10.1007/JHEP02(2015)040;%%
  %33 citations counted in INSPIRE as of 06 Jan 2019%\cite{Kol:2014nqa}


   
  %\cite{Basu:2011di}
\bibitem{Basu:2011di} 
  P.~Basu and L.~A.~Pando Zayas,
  %``Chaos rules out integrability of strings on AdS$_5 \times T^{1,1}$,''
  Phys.\ Lett.\ B {\bf 700}, 243 (2011)
 % doi:10.1016/j.physletb.2011.04.063
  [arXiv:1103.4107 [hep-th]].
  %%CITATION = doi:10.1016/j.physletb.2011.04.063;%%
  %66 citations counted in INSPIRE as of 30 Oct 2018
 %\cite{Basu:2011di}
%\cite{Basu:2011fw}
\bibitem{Basu:2011fw} 
  P.~Basu and L.~A.~Pando Zayas,
  %``Analytic Non-integrability in String Theory,''
  Phys.\ Rev.\ D {\bf 84}, 046006 (2011)
 % doi:10.1103/PhysRevD.84.046006
  [arXiv:1105.2540 [hep-th]].
  %%CITATION = doi:10.1103/PhysRevD.84.046006;%%
  %40 citations counted in INSPIRE as of 30 Oct 2018
  
  %\cite{Nunez:2018ags}  
  %\cite{Nunez:2018qcj}
\bibitem{Nunez:2018qcj} 
  C.~Nunez, D.~Roychowdhury and D.~C.~Thompson,
  %``Integrability and non-integrability in $ \mathcal{N}=2 $ SCFTs and their holographic backgrounds,''
  JHEP {\bf 1807}, 044 (2018)
%  doi:10.1007/JHEP07(2018)044
  [arXiv:1804.08621 [hep-th]].
  %%CITATION = doi:10.1007/JHEP07(2018)044;%%
  %2 citations counted in INSPIRE as of 23 Dec 2018


  
  %\cite{Sfetsos:2013wia}
\bibitem{Sfetsos:2013wia} 
  K.~Sfetsos,
  %``Integrable interpolations: From exact CFTs to non-Abelian T-duals,''
  Nucl.\ Phys.\ B {\bf 880}, 225 (2014)
%  doi:10.1016/j.nuclphysb.2014.01.004
  [arXiv:1312.4560 [hep-th]].
  %%CITATION = doi:10.1016/j.nuclphysb.2014.01.004;%%
  %110 citations counted in INSPIRE as of 21 Oct 2018
  
  %\cite{Driezen:2018glg}
\bibitem{Driezen:2018glg} 
%\cite{Sfetsos:2014cea}
%\bibitem{Sfetsos:2014cea} 
  K.~Sfetsos and D.~C.~Thompson,
  %``Spacetimes for $\lambda$-deformations,''
  JHEP {\bf 1412}, 164 (2014)
  %doi:10.1007/JHEP12(2014)164
  [arXiv:1410.1886 [hep-th]].
  %%CITATION = doi:10.1007/JHEP12(2014)164;%%
  %62 citations counted in INSPIRE as of 21 Jan 2019
  S.~Driezen, A.~Sevrin and D.~C.~Thompson,
  %``D-branes in $\lambda$-deformations,''
  JHEP {\bf 1809}, 015 (2018)
%  doi:10.1007/JHEP09(2018)015
  [arXiv:1806.10712 [hep-th]].
  %%CITATION = doi:10.1007/JHEP09(2018)015;%%
  %5 citations counted in INSPIRE as of 25 Dec 2018
  
  
  
  
  
  
  %\cite{Sfetsos:2010uq}
\bibitem{Sfetsos:2010uq} 
  K.~Sfetsos and D.~C.~Thompson,
  %``On non-abelian T-dual geometries with Ramond fluxes,''
  Nucl.\ Phys.\ B {\bf 846}, 21 (2011)
  %doi:10.1016/j.nuclphysb.2010.12.013
  [arXiv:1012.1320 [hep-th]].
  %%CITATION = doi:10.1016/j.nuclphysb.2010.12.013;%%
  %107 citations counted in INSPIRE as of 17 Jan 2019
  %\cite{Lozano:2016kum}
%\bibitem{Lozano:2016kum} 
  Y.~Lozano and C.~Nunez,
  %``Field theory aspects of non-Abelian T-duality and $ \mathcal{N}  =$ 2 linear quivers,''
  JHEP {\bf 1605}, 107 (2016)
  %doi:10.1007/JHEP05(2016)107
  [arXiv:1603.04440 [hep-th]].
  %%CITATION = doi:10.1007/JHEP05(2016)107;%%
  %28 citations counted in INSPIRE as of 18 Jan 2019
  
  
  %\cite{Lunin:2005jy}
\bibitem{Lunin:2005jy} 
  O.~Lunin and J.~M.~Maldacena,
  %``Deforming field theories with U(1) x U(1) global symmetry and their gravity duals,''
  JHEP {\bf 0505}, 033 (2005)
  %doi:10.1088/1126-6708/2005/05/033
  [hep-th/0502086].
  %%CITATION = doi:10.1088/1126-6708/2005/05/033;%%
  %530 citations counted in INSPIRE as of 17 Jan 2019
  
  %\cite{Wulff:2017lxh}
\bibitem{Wulff:2017lxh} 
  L.~Wulff,
  %``Condition on Ramond-Ramond fluxes for factorization of worldsheet scattering in anti–de Sitter space,''
  Phys.\ Rev.\ D {\bf 96}, no. 10, 101901 (2017)
  %doi:10.1103/PhysRevD.96.101901
  [arXiv:1708.09673 [hep-th]].
  %%CITATION = doi:10.1103/PhysRevD.96.101901;%%
  %4 citations counted in INSPIRE as of 17 Jan 2019%\cite{Wulff:2017vhv}
%\bibitem{Wulff:2017vhv} 
  L.~Wulff,
  %``Classifying integrable symmetric space strings via factorized scattering,''
  JHEP {\bf 1802}, 106 (2018)
  %doi:10.1007/JHEP02(2018)106
  [arXiv:1711.00296 [hep-th]].
  %%CITATION = doi:10.1007/JHEP02(2018)106;%%
  %1 citations counted in INSPIRE as of 17 Jan 2019
  
  
  
  
  %\cite{Maldacena:1998im}
\bibitem{Maldacena:1998im} 
  J.~M.~Maldacena,
  %``Wilson loops in large N field theories,''
  Phys.\ Rev.\ Lett.\  {\bf 80}, 4859 (1998)
%  doi:10.1103/PhysRevLett.80.4859
  [hep-th/9803002].
  %%CITATION = doi:10.1103/PhysRevLett.80.4859;%%
  %1561 citations counted in INSPIRE as of 07 Jan 2019
  
  
 %\cite{Nunez:2010sf}
\bibitem{Nunez:2010sf} 
  C.~Nunez, A.~Paredes and A.~V.~Ramallo,
  %``Unquenched Flavor in the Gauge/Gravity Correspondence,''
  Adv.\ High Energy Phys.\  {\bf 2010}, 196714 (2010)
  %doi:10.1155/2010/196714
  [arXiv:1002.1088 [hep-th]].
  %%CITATION = doi:10.1155/2010/196714;%%
  %116 citations counted in INSPIRE as of 24 Jan 2019 %\cite{Casero:2006pt}
%\bibitem{Casero:2006pt} 
  R.~Casero, C.~Nunez and A.~Paredes,
  %``Towards the string dual of N=1 SQCD-like theories,''
  Phys.\ Rev.\ D {\bf 73}, 086005 (2006)
 % doi:10.1103/PhysRevD.73.086005
  [hep-th/0602027].
  %%CITATION = doi:10.1103/PhysRevD.73.086005;%%
  %204 citations counted in INSPIRE as of 24 Jan 2019
  %\cite{Koerber:2007hd}
%\bibitem{Koerber:2007hd} 
  P.~Koerber and D.~Tsimpis,
  %``Supersymmetric sources, integrability and generalized-structure compactifications,''
  JHEP {\bf 0708}, 082 (2007)
  %doi:10.1088/1126-6708/2007/08/082
  [arXiv:0706.1244 [hep-th]].
  %%CITATION = doi:10.1088/1126-6708/2007/08/082;%%
  %113 citations counted in INSPIRE as of 24 Jan 2019%\cite{Barranco:2013fza}
%\bibitem{Barranco:2013fza} 
  A.~Barranco, J.~Gaillard, N.~T.~Macpherson, C.~Nunez and D.~C.~Thompson,
  %``G-structures and Flavouring non-Abelian T-duality,''
  JHEP {\bf 1308}, 018 (2013)
%  doi:10.1007/JHEP08(2013)018
  [arXiv:1305.7229 [hep-th]].
  %%CITATION = doi:10.1007/JHEP08(2013)018;%%
  %41 citations counted in INSPIRE as of 24 Jan 2019
  
  
  
  
  
\bibitem{Kovacic:jsc} 
  J.~J.~Kovacic,
  %``An algorithm for solving second order linear homogeneous differential equations,''
  J. Symbolic Computation (1986)
  {\bf 2}, 3-43.
  
  
%\cite{Torrielli:2016ufi}
\bibitem{Torrielli:2016ufi} 
  A.~Torrielli,
  %``Lectures on Classical Integrability,''
  J.\ Phys.\ A {\bf 49}, no. 32, 323001 (2016)
%  doi:10.1088/1751-8113/49/32/323001
  [arXiv:1606.02946 [hep-th]].
  %%CITATION = doi:10.1088/1751-8113/49/32/323001;%%
  %15 citations counted in INSPIRE as of 15 Jan 2019

%\cite{Zarembo:2017muf}
\bibitem{Zarembo:2017muf} 
  K.~Zarembo,
  %``Integrability in Sigma-Models,''
  arXiv:1712.07725 [hep-th].
  %%CITATION = ARXIV:1712.07725;%%
  %2 citations counted in INSPIRE as of 18 Jan 2019


  %\cite{Arutyunov:2004yx}
\bibitem{Arutyunov:2004yx} 
  G.~Arutyunov and S.~Frolov,
  %``Integrable Hamiltonian for classical strings on AdS(5) x S**5,''
  JHEP {\bf 0502}, 059 (2005)
  %doi:10.1088/1126-6708/2005/02/059
  [hep-th/0411089].
  %%CITATION = doi:10.1088/1126-6708/2005/02/059;%%
  %159 citations counted in INSPIRE as of 15 Jan 2019
  
%\cite{Beisert:2010jr}
\bibitem{Beisert:2010jr} 
  N.~Beisert {\it et al.},
  %``Review of AdS/CFT Integrability: An Overview,''
  Lett.\ Math.\ Phys.\  {\bf 99}, 3 (2012)
%  doi:10.1007/s11005-011-0529-2
  [arXiv:1012.3982 [hep-th]].
  %%CITATION = doi:10.1007/s11005-011-0529-2;%%
  %842 citations counted in INSPIRE as of 15 Jan 2019
\end{thebibliography}
\end{document}